\documentclass{article}
\usepackage{graphicx} 
\usepackage{subfigure}
\usepackage{bm}
\usepackage{amsmath}
\usepackage{amssymb}
\usepackage{mathtools}
\usepackage{xcolor} 
\usepackage{hyperref} 
\usepackage{authblk}
\usepackage{orcidlink}

\title{{Generalization of Stoney's equation for flexoelectric thin films on elastic substrates}\footnote{Notice: This manuscript has been coauthored by UT-Battelle, LLC, under Contract No. DE-AC0500OR22725 with the U.S. Department of Energy. The United States Government retains and the publisher, by accepting the article for publication, acknowledges that the United States Government retains a non-exclusive, paid-up, irrevocable, world-wide license to publish or reproduce the published form of this manuscript, or allow others to do so, for the United States Government purposes. The Department of Energy will provide public access to these results of federally sponsored research in accordance with the DOE Public Access Plan (\href{http://energy.gov/downloads/doe-public-access-plan}{http://energy.gov/downloads/doe-public-access-plan}).}}
\author[1]{Swarnava Ghosh\orcidlink{0000-0003-3800-5264}}

\affil[1]{National Center for Computational Sciences, Oak Ridge National Laboratory, TN 37830\\Email: ghoshs@ornl.gov}
\begin{document}

\maketitle

\begin{abstract}
When a thin film is deposited on an incompatible elastic substrate, the film develops an elastic mismatch strain, causing the film-substrate system to bend. Stoney's equation relates the curvature of the bent film-substrate system with the residual stress developed in the film, and can be used to infer film properties from curvature measurements. Certain materials exhibit electromechanical coupling, such as piezoelectricity and flexoelectricity, which can alter the curvature and strains. In this work, we generalize Stoney's equation to include flexoelectric and piezoelectric effects in the film. Considering both open and closed circuit configurations, as well as uniform and non-uniform film properties, we compare different cases of electromechanical coupling and discuss their influence on curvature, strains, and electric polarization in the film.  

\end{abstract}


\section{Introduction}
Consider a thin film deposited on an incompatible elastic substrate. Because of the incompatibility, the film develops an elastic mismatch, causing the film-substrate system to bend \cite{Freund:Book}. Stoney, in his seminal work \cite{Stoney}, derived an expression relating the curvature of the film-substrate system to the substrate height, substrate thickness, and the residual force in the film. Stoney’s equation serves as a fundamental relation for calculating thin-film properties based on experimentally measured substrate curvature\cite{Freund:1999,janssen2009celebrating}. Notwithstanding its popularity, Stoney’s relation ignores the properties of the film, assumes infinitesimal displacements, and neglects the variation of stress through the film thickness, thereby often leading to errors in the estimation of film properties from curvature measurements \cite{Freund:Book}. 

Several attempts have been made to extend Stoney's equation. In particular, Freund and co-workers \cite{Freund:1999} derived an expression for curvature and stretching strain that accounts for the properties of both the film and substrate. Feng and co-workers \cite{feng2007stoney} extended it to the case where the thin film-substrate system is subjected to nonuniform misfit strain with nonuniform substrate thickness. Using the principle of energy minimization, Pureza and collaborators \cite{pureza2009enhancing} derived a three-dimensional system of equations to model stresses in thin films deposited on plane substrates that are much thicker than the film. Stoney's equation has also been extended to account for different biaxial stresses in thin films \cite{chen2015modified}, elastically anisotropic substrates \cite{janssen2009celebrating,injeti2016extending}, non-uniformly distributed incremental stress in films \cite{qiang2021extension}, large deformation bending of the film-substrate system\cite{finot1997large,liu2021modified,masters1993geometrically,schicker2016stress,injeti2021modified}, and piezoelectric films on elastic substrates \cite{mccartney2014,zhou2017stoney}. 

While these developments are encouraging, they ignore the possibility that certain materials exhibit coupling between strain gradients and electric polarization. This electromechanical phenomenon will influence the curvature and strains of the film-substrate system. In this work, we extend Freund's \cite{Freund:1999} modification of Stoney's equation to include the effects of coupling between strain gradient, strain, and electric polarization. 

Flexoelectricity is the coupling between strain gradients and electric polarization, and it is an important and technologically relevant electromechanical effect\cite{Yudin2013Review,Tripathy2021Review,Codony2021Review,Nguyen2013Review,Wang2016Review}. Unlike piezoelectricity, the coupling between electric polarization and strain, which is observed in non-centrosymmetric materials, flexoelectricity is a universal property of all dielectrics \cite{Sharma2007,Codony2021Review}. However, the flexoelectric effect is significant at small length scales where strain gradients are appreciable, and diminishes with the increase in the size of the material \cite{Sharma2007,Abdollahi:2014,Codony2021Review}. Because of this, nanoscale materials such as two-dimensional materials display several interesting features due to flexoelectricity\cite{kumar2021,kumar2025,codony2021,Kothari2018,Kothari2019}. The flexoelectric effect has also been observed in soft matter \cite{Deng2014,Grasinger2021,codony:Finite}, liquid crystals \cite{Rahmati2025} and biological materials \cite{Deng2019,Gao2008,Witt2023}.  In the remainder of this section, we provide a brief overview of some of the classic as well as contemporaneous work related to the flourishing field of flexoelectricity.

In the continuum theory of flexoelectricity, the flexoelectric tensor is sixth-order, coupling the strain gradient with electric polarization. In crystalline materials, the underlying symmetry of the lattice determines the symmetry of the flexoelectric tensor. The classification of all possible rotational symmetries, as well as the number of independent material parameters required to specify flexoelectric tensors, have been investigated in reference \cite{Le2011}. Interestingly, the strain gradient elasticity theory of Mindlin \cite{Mindlin1965} can provide insights into the theory of flexoelectricity \cite{MaoPurohit2014}. Shen and Hu \cite{Shen2010} extended the theory of flexoelectricity to include surface stress and surface polarization. Flexoelectricity can influence the electroelastic response of bending of piezoelectric nanobeams \cite{Yan2013}.

In materials with defects, the flexoelectric effect is more prominent near the defects due to high strain gradients and decays away at distances larger than the flexoelectric length scale \cite{Mao2015}. Interestingly, flexoelectricity-induced deformation in ferroelectrics gives rise to asymmetric mechanical properties \cite{Lun2022}. It is surprising to note that in flexural sensors and actuators, the effective piezoelectric effect may not only be enhanced but can also be decreased due to flexoelectricity, and this constructive or destructive interplay is size-dependent \cite{Abdollahi2015}. In composites, large strain gradients near nanoscale inhomogenities can be used to tailor flexoelectric response \cite{Sharma2007}.
 
Flexoelectricity has a profound effect on the electromechanical response of nanomaterials because of their small sizes. Accurate quantification of flexoelectric constants of nanomaterials using first-principles methods such as Density Functional Theory \cite{hohenberg1964,kohn1965} can be challenging due to the large computational cost involved. To resolve this, a novel formulation incorporating the spatial symmetry of nanostructures into Density Functional Theory has been developed \cite{Banerjee2016,Ghosh2019} using a real-space formalism\cite{ghosh2017a,ghosh2017b}, and has been successfully used to calculate the flexoelectric constants of two-dimensional materials \cite{kumar2021,kumar2025,codony2021}. 

In two-dimensional materials, flexoelectricity can introduce some surprising effects. For example, multi-layer graphene nanosheets exhibit curvature-localizing and subcritical buckling mode that produces shallow-kink corrugation, which has been attributed to the quantum flexoelectric effect\cite{Kothari2018}. A thermodynamic framework incorporating a non-local interaction model of polarization with flexoelectricity–dielectricity coupling has been developed to study this phenomenon\cite{Kothari2019}.

Recently, the thriving research in non-linear continuum mechanics of soft matter has also motivated the study of flexoelectricity in these materials \cite{Deng2014,Grasinger2021,codony:Finite}. The mechanisms and underlying theory of flexoelectricity in elastomers have been explored by Grasinger and co-workers\cite{Grasinger2021}. Their analysis shows that a giant flexoelectric effect can emerge in incompressible soft matter when pre-stretched in the direction of the strain gradient. Furthermore, piezoelectricity can also emerge by poling the chains of a polymer network. However, the flexoelectric effect diminishes with poling and thus, piezoelectricity comes at a cost of flexoelectricity. Their seminal work paves the way for the development of flexoelectric effect-based soft robotics and soft actuators. Following this line of investigation, a static and dynamic theory for flexoelectricity in photoactive liquid crystal elastomers is developed in \cite{Rahmati2025}, which shows that the coupling between light, deformation and polarization is maximized at an optimal size-scale. A theoretical framework for flexoelectricity in soft materials using nonlinear continuum mechanics is derived in \cite{Deng2014}. Codony and co-workers \cite{codony:Finite} have also derived a finite deformation theory to model of bending and buckling of flexoelectric beams. A framework to model dielectric rods using finite deformation continuum theory, for both direct and converse flexoelectric effects, using the flexoelectric strain-gradient special Cosserat rod model, is presented in \cite{Mishra2025}. 

Flexoelectricity plays a critical role in the electromechanical response of thin films and layered materials. For example, epitaxially grown thin films of oxides can exhibit flexoelectric effect several orders of magnitude larger than those observed in conventional bulk solids due to interfacial strain gradients \cite{Lee2012}. An odd number of non-piezoelectric but flexoelectric layers stacked into a superlattice can be used to create apparent piezoelectricity \cite{Sharma2010}. In ferroelectric nanofilms,  domain switching \cite{Chen2018} and evolution of domain patterns are attributed to flexoelectricity \cite{Chen2015}. In a two-dimensional suspended membrane, flexoelectric effects influence the bending response and can be modeled by extending von Kármán plate theory \cite{Song2024}. 

These advances motivate us to investigate its influence on the curvature and strains of a film-substrate bilayer system when the film exhibits flexoelectricity or piezoelectricity.

The remainder of this paper is organized as follows. In Section \ref{Sec:Constitutive} we provide an overview of the constitutive laws of flexoelectricity, followed by a discussion on the film-substrate system in Section \ref{Sec:System}. Next, considering different electromechanical cases, the analytical expressions of curvature, stretching strain, and electrical polarization of the film-substrate system with uniform properties and a discussion comparing them are presented in Section \ref{Sec:Uniform}. Next, we extend this analysis to the non-uniform film properties in Section \ref{Sec:Nonuniform}, and finally, we provide concluding remarks in Section \ref{Sec:Conclusion}

\section{Constitutive relations in flexoelectricity}\label{Sec:Constitutive}
Consider a body occupying a domain $\Omega$. The enthalpy density of the body, considering elastic, dielectric, piezoelectric, and flexoelectric contributions, is \cite{codony:JAP}.
\begin{equation}
    \psi ({\bm {\varepsilon}}, \nabla {\bm {\varepsilon}}, {\bf E}, {\bf \nabla E}) = \psi_{elastic} ({\bm{\varepsilon}})+ \psi_{di}({\bf E}) + \psi_{piezo} ({\bf E}, {\bm {\varepsilon}}) +  \psi_{flexo} ({\bm {\varepsilon}}, \nabla {\bm {\varepsilon}}, {\bf E}, {\bf \nabla E}) \,\,.
\end{equation}
where ${\bm {\varepsilon}}$ is the strain in the body and ${\bf E}$ is the electric field. We assume the body to be linearly elastic. The elastic, dielectric, and piezoelectric contributions to the total enthalpy density are
\begin{eqnarray}\label{Eq:EnthalpyElDiPiezo}
& & \psi_{elastic} ({\bm{\varepsilon}}) = \frac{1}{2} C_{ijkl} \varepsilon_{ij} \varepsilon_{kl} \\
& & \psi_{di} ({\bf E}) = -\frac{1}{2} k_{ij} E_i E_j \,\,, \\
& & \psi_{piezo} ({\bm {\varepsilon}}, {\bf E}) = -e_{ikl} E_i \varepsilon_{kl} \,\,,
\end{eqnarray}
where $C_{ijkl}$ is the elastic modulus tensor, $k_{ij}$ is the permittivity tensor, and $e_{ikl}$ is the piezoelectric tensor. This formulation can also be extended to a non-linear regime by employing an appropriate stored energy density \cite{codony:Finite}. 

In this work, we consider two variants of the flexoelectric enthalpy density. Specifically, \emph{direct} flexoelectricity (i.e., voltage from applied deformation) and \emph{converse} flexoelectricity (i.e., deformation from applied voltage). The flexoelectric enthalpy density for the direct case is
\begin{equation}\label{Eq:EnthalpyDirect}
    \psi_{flexo}^{Dir} ( \nabla {\bm \varepsilon}, {\bf E}) = -\mu_{lijk} \varepsilon_{ij,k} E_l \,\,,
\end{equation}
and the flexoelectric enthalpy density for the converse case is
\begin{equation}\label{Eq:EnthalpyConv}
    \psi_{flexo}^{Conv} ( {\bm \varepsilon}, \nabla {\bf E}) = \mu_{lijk} \varepsilon_{ij} E_{l,k} \,\,,
\end{equation}
where $\mu_{lijk}$ is the flexocoupling tensor.

The total free enthalpy of the system is
\begin{equation}
    \Pi [\bm{u},\phi] = \int_{\Omega}  \psi (\bm{u},\phi) \mathrm{d} \Omega - W^{ext} \,\,.
\end{equation}
where the displacement field $\bm{u}$ is related to the strain as ${\bm{\varepsilon}} = \frac{1}{2} (\nabla \bm{u} + (\nabla \bm{u})^T)$, $W^{ext}$ is the external work, and the electrostatic potential $\phi$ is related to the electric field by
\begin{equation}
    {\bf E} = -\nabla \phi \,\,.
\end{equation}
The corresponding variational principle is \cite{codony:JAP}
\begin{equation}
    (\bm{u}^*,\phi^*) = \arg \min_{\bm{u}} \max_{\phi} \Pi [\bm{u},\phi]\,\,.
\end{equation}
In the absence of external charges and body forces, the Euler-Lagrange equations for this variational principle are the equilibrium equations given by 
\begin{equation}
    \nabla \cdot {\bm{\sigma}} = 0 \,\, \text{in} \,\, \Omega \,\,,
\end{equation}
and 
\begin{equation}
    \nabla \cdot {\bf{D}}  = 0 \,\, \text{in} \,\, \Omega \,\,,
\end{equation}
where ${\bm{\sigma}}$ is the stress tensor and ${\bf{D}}$ is the electric displacement vector.

The stress tensor is given by
\begin{eqnarray}\label{Eqn:stress}
    {\bm{\sigma}} (\bm{u},\phi) &=& \frac{d \psi (\bm{u},\phi) }{d \bm{\varepsilon}} =  \left[ \frac{\partial \psi }{\partial \bm{\varepsilon}}  - \nabla \cdot \frac{\partial \psi }{\partial \nabla  \bm{\varepsilon}} \right] (\bm{u},\phi) \nonumber \\
    \sigma_{ij} &=& C_{ijlk} \varepsilon_{kl} - e_{kij} E_{k} + \mu_{lijk} E_{l,k} \,\,,
\end{eqnarray}
and the electric displacement is 
\begin{eqnarray}\label{Eqn:displacement}
    {\bf{D}} (\bm{u},\phi) &=& - \frac{d \psi (\bm{u},\phi) }{d \bf{E}} =  \left[ -\frac{\partial \psi }{\partial \bf{E}}  + \nabla \cdot \frac{\partial \psi }{\partial \nabla  \bf{E}} \right] (\bm{u},\phi) \nonumber \\
    D_{i} &=& e_{ikl} \varepsilon_{kl} + k_{ij} E_{j} + \mu_{ijkl} \varepsilon_{ij,l} \,\,.
\end{eqnarray}
These expressions for the stress tensor and the electric displacement hold irrespective of the choice of the flexoelectric enthalpy density \cite{codony:JAP}.

The electric polarization is
\begin{eqnarray}\label{Eqn:polarization}
    {\bf{P}} = e_0 {\bf{E}} + {\bf{D}} \,\,,
\end{eqnarray}
where $e_0$ is the electrical permittivity of free space. The polarization is calculated once the electric field and electric displacement are known.

\section{Thin flexoelectric film on an elastic substrate}\label{Sec:System}
\begin{figure}[h!]
\centering
\includegraphics[keepaspectratio=true,width=0.75\textwidth]{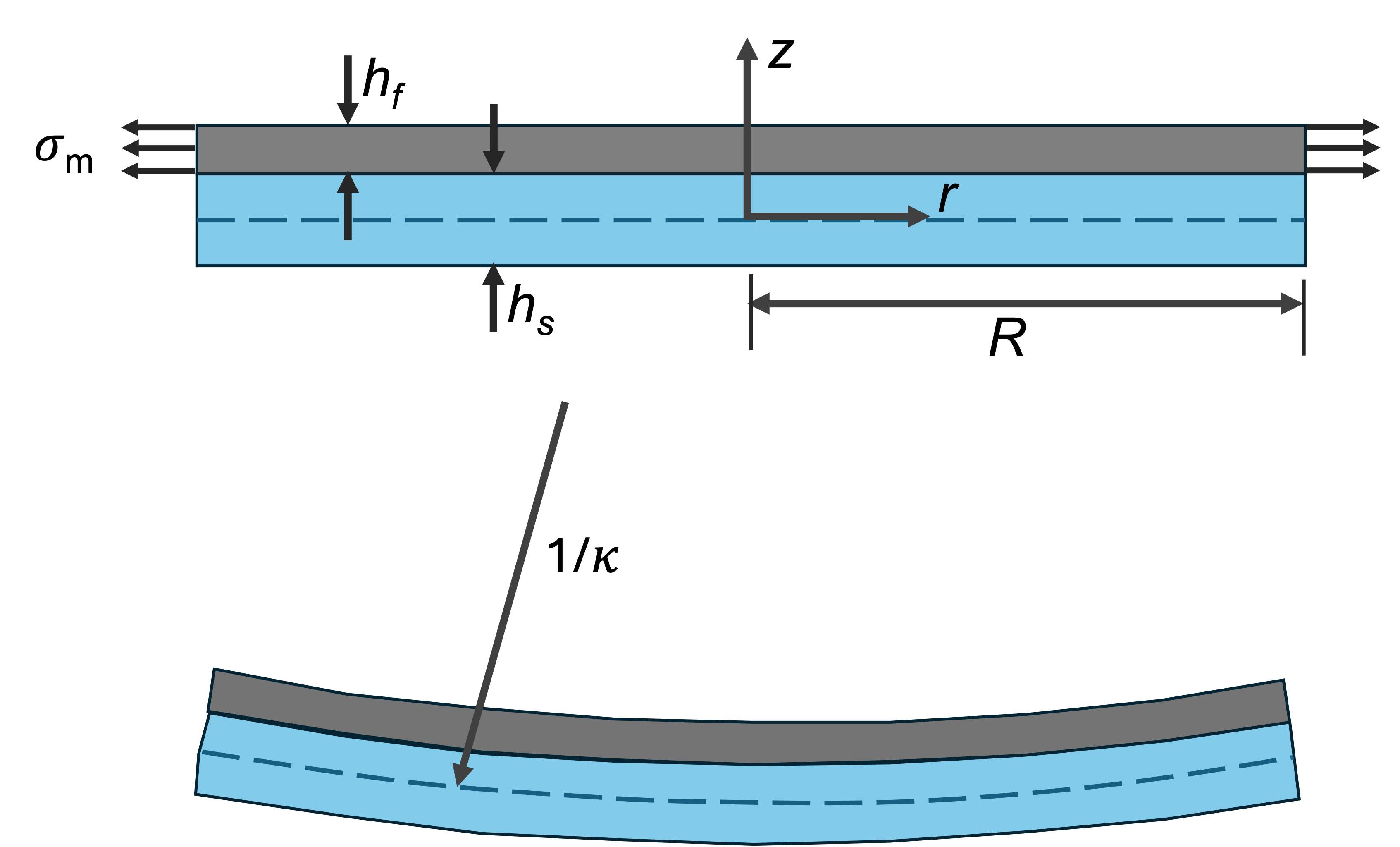}\label{Fig:geometry}
{\caption{Top figure shows a film deposited on a flexible substrate. The film has an elastic mismatch and is held in a flat configuration by an applied stress $\sigma_m$. Upon removal of this applied stress, the film-substrate system deforms as shown in the bottom.}\label{Fig:geometry}}
\end{figure}

Consider a thin film of dielectric material and thickness $h_f$ on an elastic substrate of uniform thickness $h_s$. The film is assumed to exhibit electromechanical coupling, such as piezoelectricity, flexoelectricity, or a combination of both. The substrate is circular with radius $R$ such that $h_f+h_s << R$. Due to the film's and substrate's incompatibility, the film has an elastic mismatch strain $\varepsilon_m$, which is assumed to correspond to an isotropic in-plane expansion or contraction of the interface. The cause of the elastic mismatch strain in the film can arise from various phenomena such as epitaxial mismatch, thermal expansion, chemical reaction, moisture absorption, and phase transformations \cite{Freund:Book,Freund:1999}. The incompatible mismatch strain introduces stresses that cause the film-substrate system to bend. This effect is schematically shown in Figure \ref{Fig:geometry}.

Because of the symmetric nature of the problem, we choose a cylindrical coordinate system, as depicted in Figure \ref{Fig:geometry}. The coordinate system is defined such that the origin lies at the center of the substrate midplane, the $z$-axis is normal to the substrate face, the midplane is located at $z=0$, and the film–substrate interface is at $z=\frac{h_s}{2}$.


Following Freund \cite{Freund:Book,Freund:1999}, we assume the following for the film-substrate bi-layer system. First, the deformation is within the range of small-strain theory, the material of the film and substrate is isotropic, and the deformation of the substrate follows Kirchhoff's theory of thin plates. This implies that the normal stress component is zero ($\sigma_{zz}=0$), a line that is initially straight and perpendicular to the mid-plane of the substrate remains straight after deformation, and thus the shear strains $\varepsilon_{rz}=\varepsilon_{\theta z} =0$. Furthermore, the deformation is axially symmetric, so all fields are independent of $\theta$, and hence the shear strain $\varepsilon_{r\theta}=0$. As the film-substrate deforms,  the bending of the midplane is approximately a spherical surface, and the curvature $\kappa$ is uniform throughout. The stretching strain $\varepsilon_0$ of the midplane surface of the substrate is uniform along the midplane. Thus, the strains in the radial and angular directions along the midplane are equal to the stretching strain (i.e. $\varepsilon_{rr}(r,0)$ = $\varepsilon_{\theta\theta} (r,0)$ = $\varepsilon_0$). We neglected localized edge effects around the boundary. 

The strains $\varepsilon_{rr}$ and $\varepsilon_{\theta \theta}$ are related to the displacement field $u_r$ as 
\begin{equation}\label{Eq:displacementfield}
    \varepsilon_{rr} = \frac{\partial{u_r}}{\partial r} \,\,\, \text{and} \,\, \varepsilon_{\theta \theta} = \frac{{u_r}}{r} \,\,.
\end{equation}

As both the film and the substrate are transversely isotropic, the non-zero components of the elastic moduli are $c_{1111}$, $c_{1122}$, $c_{1133}$, $c_{3333}$ and $c_{2323}$ \cite{Book:Anand&Govindjee}. We express these in Voigt notation as $C_{11}$, $C_{12}$, $C_{13}$, $C_{33}$, $C_{44}$ respectively. 

When piezoelectric effect in the film is considered, it is assumed to be piezoelectrically polarized in the $z$-direction. The film is also isotropically flexoelectric. Piezoelectric and flexoelectric effects are absent in the substrate. Therefore, the only non-zero components of the piezoelectric tensor, and their two-index notation are $e_{113}$=$e_{223}$=$e_{15}$, $e_{311}$ = $e_{322}$ = $e_{31}$ and $e_{333}$ = $e_{33}$. The permittivity tensor $k_{ij}$ is diagonal and the non-zero components are $k_{11}$, $k_{22}$ and $k_{33}$. Finally, due to isotropy, the only non-zero components of the flexocoupling tensor are $\mu_{1111}$, $\mu_{1221}$ and $\mu_{1212}$ which in a two-index notation is denoted by $\mu_{11}$, $\mu_{12}$ and $\mu_{44}$, respectively \cite{codony:JAP}. The first two are the longitudinal and transversal components, and the third is due to shear, which is typically neglected due to it being poorly characterized \cite{Abdollahi:2014}. 

The electrostatic potential $\phi$ and the radial $E_r$ and axial $E_z$ components of the electric field in the film are related as 
\begin{equation}\label{Eq:ElectricFieldPotential}
E_r = - \frac{\partial \phi}{\partial r} \,\,\,, \text{and} \,\, E_z = - \frac{\partial \phi}{\partial z} \,\,, 
\end{equation}
and the electric field in the $\theta$-direction is zero due to axial symmetry. 

Given this problem of a thin film on an elastic substrate, we seek to calculate the stretching strain $\varepsilon_0^{st}$ and the curvature $\kappa_{st}$. If the thickness of the film is small compared to the thickness of the substrate, then the change in film stress due to substrate deformation is negligible. Furthermore, when the dielectric, piezoelectric, and flexoelectric effects of the film are neglected, and the effective stiffness of the bilayer system is assumed to depend only on the stiffness of the substrate, the stretching  strain as derived by Stoney \cite{Freund:Book,Stoney} is 
\begin{equation}\label{Eq:stoneyepsilon}
\varepsilon_0^{st} = -\varepsilon_m \frac{h_f M_f}{h_s M_s}\,\,,
\end{equation}
and the curvature $\kappa^{st}$ is given by
\begin{equation}\label{Eq:stoneykappa}
    \kappa^{st} = 6\varepsilon_m \frac{h_f M_f}{h_s^2 M_s}\,\,.
\end{equation}

Freund and co-workers \cite{Freund:1999} further extended this analysis for the case when the stiffness of the film is non-negligible. They derived analytical expressions of the stretching strain $\varepsilon_0^{st}$ and curvature $\kappa_{st}$, and showed that these quantities depend on the ratio of the stiffness and thickness of the bilayer. In this paper, we are interested in the problem when additional flexoelectric and piezoelectric effects are present in the film. In the following Section \ref{Sec:Uniform}, we discuss the case when the properties of the film are uniform. In Section \ref{Sec:Nonuniform}, we further extend our analysis to non-uniform properties of the film.


\section{Uniform elastic mismatch strain and elastic properties}\label{Sec:Uniform}
\begin{figure}[h!]\centering
\centering
\includegraphics[keepaspectratio=true,width=0.75\textwidth]{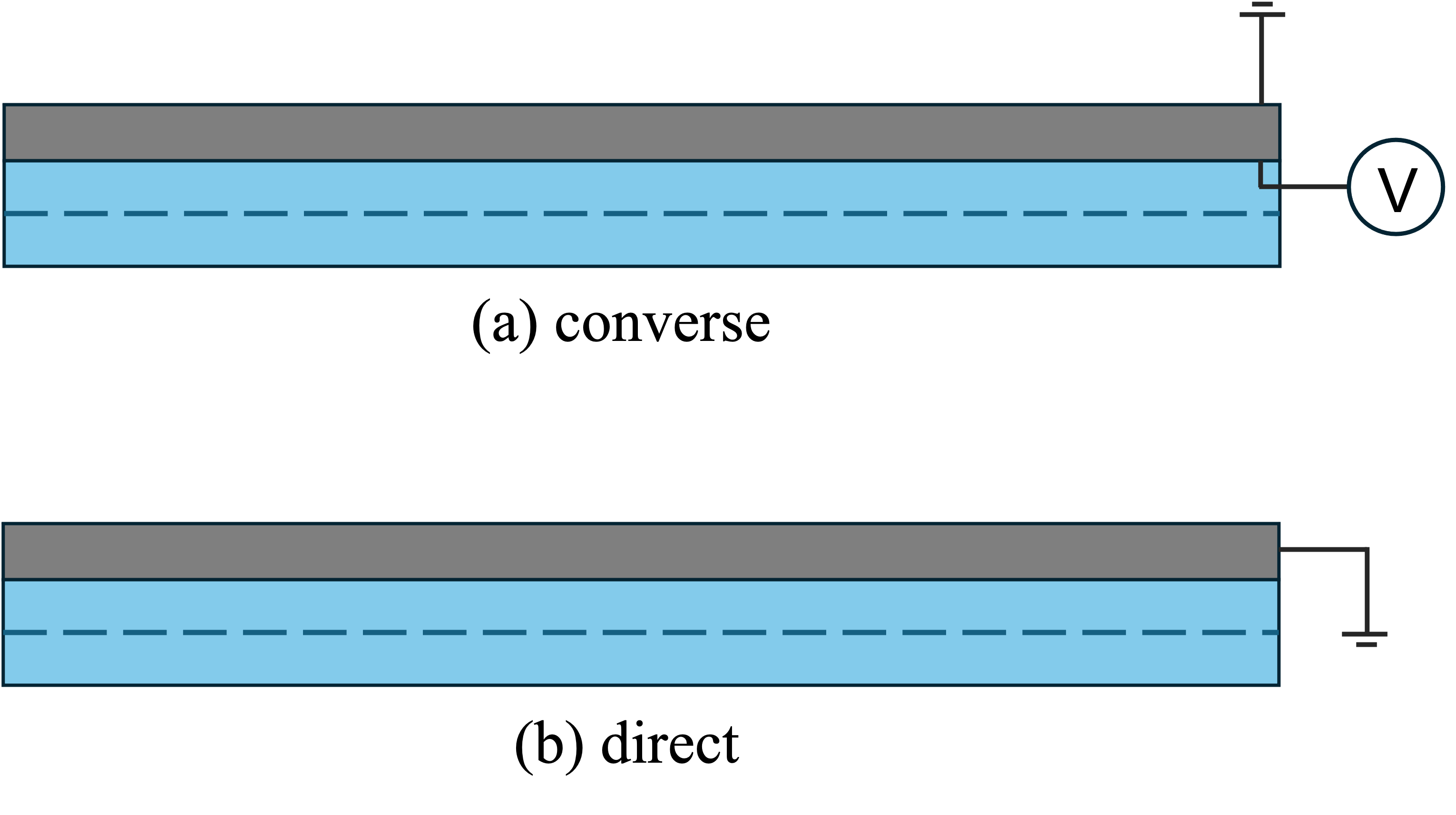}\label{Fig:convdir}
{\caption{The film substrate system for (a) converse case with applied voltage $V$ across the height of the film and (b) direct case without any applied voltage is shown. }\label{Fig:convdir}}
\end{figure}
We consider a film-substrate bilayer system where the elastic mismatch strain $\varepsilon_m$ and the elastic moduli of the film and the substrate are constant. We analyze two cases. In the first case, a voltage of $V$ is applied across the film layer. This is called a \emph{closed circuit} configuration as depicted in Figure \ref{Fig:convdir}a, and corresponds to \emph{converse} flexoelectric effect. In the second case, there is no external voltage, and this is called an \emph{open circuit} configuration is shown in Figure \ref{Fig:convdir}b, and corresponds to \emph{direct} flexoelectric effect.

Since the deformation is assumed to be small and the radial and out-of-plane deformations are uncoupled, the radial and transverse displacement fields of the mid-plane of the substrate are related to the stretching  strain $\varepsilon_0$ and the mean curvature $\kappa$ as \cite{Freund:Book,Freund:1999} 
\begin{equation}
    u_r(r,0) = \varepsilon_0 r \,\,\,, \text{and} \,\, u_z(r,0) = \frac{1}{2}\kappa r^2 \,\,,
\end{equation}
and the strains in the bilayer are
\begin{equation}\label{Eqn:strainbilayer}
    \varepsilon_{rr}=\varepsilon_{\theta\theta} = \begin{cases} 
    \varepsilon_0 - \kappa z + \varepsilon_m \,\, ,  \frac{h_s}{2} < z < \frac{h_s}{2}+ h_f\\
    \varepsilon_0 - \kappa z \,\,,  \,\,\,\,\,\,\,\,\, -\frac{h_s}{2} < z < \frac{h_s}{2} \,\,. \\
    \end{cases}
\end{equation}

Using Equation \ref{Eqn:stress}, the non-zero components of the stress tensor in the film are
\begin{eqnarray} \label{Eq:sigmafilm}
& &\sigma_{rr} = C_{11}^f \varepsilon_{rr} + C_{12}^f \varepsilon_{\theta \theta} - e_{31} E_z + \mu_{12} \frac{\partial E_{z}}{\partial z} + \mu_{11} \frac{\partial E_{r}}{\partial r} \,\,, \\
& &\sigma_{\theta \theta} = C_{12}^f \varepsilon_{rr} + C_{11}^f \varepsilon_{\theta \theta} - e_{31} E_z + \mu_{12} \frac{\partial E_{z}}{\partial z} + \mu_{12} \frac{\partial E_{r}}{\partial r} \,\,, \\
& &\sigma_{r z} = -e_{15} E_r \,\,,
\end{eqnarray}
and the non-zero components of the stress tensor in the substrate are
\begin{eqnarray} 
& &\sigma_{rr} = C_{11}^s \varepsilon_{rr} + C_{12}^s \varepsilon_{\theta \theta} \,\,,\\
& &\sigma_{\theta \theta} = C_{12}^s \varepsilon_{rr} + C_{11}^s \varepsilon_{\theta \theta} \,\,. \label{Eq:sigmasubstrate}
\end{eqnarray}
In Equations \ref{Eq:sigmafilm} - \ref{Eq:sigmasubstrate}, the superscript $f$ denote the elastic constant of the film and $s$ denote the elastic constants of the substrate.

Again, using Equation \ref{Eqn:displacement}, the non-zero components of the electric displacement vector in the film are
\begin{eqnarray}\label{Eq:disr}
& & D_r = k_{11} E_r + \mu_{11} \frac{\partial \varepsilon_{rr}}{\partial r} + \mu_{12} \frac{\partial \varepsilon_{\theta \theta}}{\partial r}\,\,, \\
& & D_z = e_{31} \varepsilon_{rr} + e_{31} \varepsilon_{\theta\theta} + k_{33} E_z + \mu_{12}\frac{\partial \varepsilon_{rr}}{\partial z} + \mu_{12}\frac{\partial \varepsilon_{\theta \theta}}{\partial z} \,\,,\label{Eq:disz}
\end{eqnarray}
and the electric displacement in the substrate is zero due to the absence of piezoelectricity and flexoelectricity.

Having obtained the expression of stress and electric displacement in the film substrate system, the equilibrium equation $\nabla \cdot {\bm{\sigma}}=0$ for stress in cylindrical coordinates is
\begin{eqnarray}\label{Eqn:sigmacylindrical}
& & \frac{\partial \sigma_{rr}}{\partial r} + \frac{\partial \sigma_{rz}}{\partial z} + \frac{1}{r}(\sigma_{rr} -\sigma_{\theta \theta}) = 0 \,\,, \\
& & \frac{\partial \sigma_{rz}}{\partial r} + \frac{\sigma_{rz}}{r} = 0 \,\,, 
\end{eqnarray}
and the equilibrium equation $\nabla \cdot {\bf{D}}=0$ for electric displacement in cylindrical coordinates
\begin{equation}\label{Eqn:displacementcylindrical}
    \frac{\partial D_{r}}{\partial r} +  \frac{\partial D_{z}}{\partial z} + \frac{D_{r}}{r} = 0 \,\,.
\end{equation}

Next, using strains (Equation \ref{Eqn:strainbilayer}), stresses (Equations \ref{Eq:sigmafilm} -\ref{Eq:sigmasubstrate}) and electric displacement  (Equation \ref{Eq:disr}-\ref{Eq:disz}), respectively, in the equilibrium Equations \ref{Eqn:sigmacylindrical} - \ref{Eqn:displacementcylindrical}, we arrive at the following set of equations for the electrostatic potential in the film
\begin{eqnarray}\label{Eq:potential}
   \frac{\partial ^2 \phi}{\partial r \partial z} = 0 \,\,,\,\,\,
    \frac{\partial ^2 \phi}{\partial r^2} + \frac{1}{r} \frac{\partial \phi}{\partial r} = 0 \,\,\text{and}\,\,\, 
   k_{33} \frac{\partial ^2 \phi}{\partial z^2} + 2e_{31} \kappa = 0 \,\,.
\end{eqnarray}

We first consider the closed circuit case (and correspondingly the converse flexoelectric effect) as shown in Figure \ref{Fig:convdir}a. Using the boundary conditions $\phi(\frac{h_s}{2}) = V$ and $\phi(h_f+\frac{h_s}{2})=0$, for solving Equation \ref{Eq:potential} and then using Equation \ref{Eq:ElectricFieldPotential}, we obtain the following solution for the electric field
\begin{eqnarray}\label{Eq:ElectricFleldConv}
&&    E_z^{Conv}(z) = \frac{2e_{31}}{k_{33}} \kappa \left( z - \frac{h_f+h_s}{2} \right) + \frac{V}{h_f} \,\,,\nonumber\\
 &&   E_r^{Conv}(r) = 0 \,\,.
\end{eqnarray}

Next, we consider the open circuit case (and correspondingly the direct flexoelectric effect) as shown in Figure \ref{Fig:convdir}b. For this case, the electric displacement in the normal direction is zero ${\bf{D} \cdot {\bf{e}}_z} = 0$, where $\bf{e}_z$ is the unit normal vector over the smooth surface of the boundary in the $z-$direction, and we obtain
\begin{eqnarray}\label{Eq:ElectricFleldDir}
    & & E_z^{Dir}(z) = \frac{2 e_{31}}{k_{33}} \left[ \kappa \left( z + \frac{\mu_{12}}{e_{31 }} \right) - (\varepsilon_{0} + \varepsilon_{m}) \right] \,\,, \nonumber \\
    & & E_r^{Dir}(r) = 0 \,\,.
\end{eqnarray}

It is easy to see that the derivative of the electric fields is the same irrespective of the choice of flexoelectricity
\begin{equation}
    E_{z,z}^{Conv} = E_{z,z}^{Dir}= \frac{2 e_{31}}{k_{33}}\kappa \,\,.
\end{equation}

We use the electric field from Equations \ref{Eq:ElectricFleldConv} - \ref{Eq:ElectricFleldDir} and strains from Equation \ref{Eqn:strainbilayer} to calculate the enthalpy densities Equations \ref{Eq:EnthalpyElDiPiezo} - \ref{Eq:EnthalpyConv} for the bilayer system. The enthalpy density due to elasticity is 
\begin{eqnarray}\label{Eq:elasticenthalpydensity}
& &   \psi_{elastic} ({\bm{\varepsilon}}) = \begin{cases} 
    M^f (\varepsilon_{rr})^2 = M^f (\varepsilon_0 - \kappa z + \varepsilon_m)^2 \,\, ,  \frac{h_s}{2} < z < \frac{h_s}{2}+ h_f\\
    M^s (\varepsilon_{rr})^2 = M^f (\varepsilon_0 - \kappa z )^2 \,\,, -\frac{h_s}{2} < z < \frac{h_s}{2} \\
\end{cases} 
\end{eqnarray}
where $ M^f = C_{11}^f + C_{12}^f $ and $ M^s = C_{11}^s + C_{12}^s $ are the bending modulus of the film and the substrate, respectively.

Notice that the piezoelectric, dielectric, and flexoelectric contributions depend on the electric field, and hence depends on the nature of electrical boundary condition i.e. open circuit (direct) or closed circuit (converse). The contribution of dielectricity to the enthalpy density in the film for the converse case is
\begin{eqnarray}
& &\psi_{di}^{Conv} ({\bf E}) = -\frac{1}{2} k_{33} \left(E_z^{Conv}\right)^2 =   -\frac{1}{2} k_{33} \left[ \frac{2e_{31}}{k_{33}} \kappa \left( z - \frac{h_f+h_s}{2} \right) + \frac{V}{h_f}  \right]^2 \,\,.
\end{eqnarray}
and for the direct case
\begin{eqnarray}
& &\psi_{di}^{Dir} ({\bf E}) = -\frac{1}{2} k_{33} \left(E_z^{Dir}\right)^2 =   - e_{31} \left[ \kappa \left( z + \frac{\mu_{12}}{e_{31 }} \right) - (\varepsilon_{0} + \varepsilon_{m}) \right]^2  \,\,, 
\end{eqnarray}

The contribution of piezoelectricity to the enthalpy density for the converse case is
\begin{eqnarray}
& &\psi_{piezo}^{Conv} ({\bm {\varepsilon}}, {\bf E}) = -2e_{31}E_z^{Conv} \varepsilon_{rr} = -2e_{31} \left[  \frac{2e_{31}}{k_{33}} \kappa \left( z - \frac{h_f+h_s}{2} \right) + \frac{V}{h_f} \right] \left( \varepsilon_0 - \kappa z + \varepsilon_m \right) \,\,.
\end{eqnarray}
and for the direct case, we obtain
\begin{eqnarray}
& &\psi_{piezo}^{Dir} ({\bm {\varepsilon}}, {\bf E}) = -2e_{31}E_z^{Dir} \varepsilon_{rr} = -2e_{31} \left[  \frac{2 e_{31}}{k_{33}} \left\{ \kappa \left( z + \frac{\mu_{12}}{e_{31 }} \right) - (\varepsilon_{0} + \varepsilon_{m}) \right\} \right] \left( \varepsilon_0 - \kappa z + \varepsilon_m \right) \,\,,
\end{eqnarray}


The flexoelectric enthalpy densities for the film-substrate system for the converse case is 
\begin{eqnarray}
& & \psi_{flexo}^{Conv} ({\bm \varepsilon},  \nabla {\bf E}) = 2\mu_{12}  \varepsilon_{rr} 
 \frac{\partial  E_z^{Conv}}{\partial z} = \frac{4}{k_{33}} \mu_{12} e_{31} \kappa (\varepsilon_0 - \kappa z +\varepsilon_m) \,\,.
\end{eqnarray}
and for the direct case
\begin{eqnarray}
 & & \psi_{flexo}^{Dir} ( \nabla {\bm \varepsilon}, {\bf E}) = -2\mu_{12} \frac{\partial \varepsilon_{rr}}{\partial z} E_z^{Dir} =  \frac{4}{k_{33}} \mu_{12} e_{31} \kappa \left[ \kappa\left( z+ \frac{\mu_{12}}{e_{31}}\right) -(\varepsilon_0 + \varepsilon_m) \right] \,\,.
\end{eqnarray}

Next, we evaluate the total enthalpy of the film-substrate system. The total enthalpy of the system is the sum of elastic, dielectric, piezoelectric, and flexoelectric contributions, and is evaluated by integrating the densities over the volume of the film-substrate system. As dielectric, piezoelectric, and flexoelectric effects in the substrate are absent, it suffices to integrate over the volume of the film for dielectric, piezoelectric, and flexoelectric contributions.

The contribution of elasticity to the total enthalpy is given by
\begin{eqnarray}
&& \Pi_{elastic} (\varepsilon_0,\kappa) = \int_{0}^R \int_{-\frac{h_s}{2}}^{\frac{h_s}{2}+h_f} \psi_{elastic} ({\bm{\varepsilon}}) \,\, 2\pi r \,\mathrm{d} r \mathrm{d} z 
= \pi R^2 \biggl[ M_s h_s \left( \varepsilon_0^2 + \frac{k^2h_s^2}{12}\right) \nonumber \\ && + M_f h_f \biggl\{ \varepsilon_0^2 + \varepsilon_m^2 + 2\varepsilon_0 \varepsilon_m  - \kappa(\varepsilon_0+\varepsilon_m)(h_s + h_f) + \frac{k^2 h_s^2}{4} + \frac{k^2 h_s h_f}{2} + \frac{k^2 h_f^2}{3} \biggr\} \biggr] \,\,.
\end{eqnarray}
This expression for the elastic contribution is the same as that obtained by Freund and co-workers \cite{Freund:1999,Freund:Book}.

For the closed circuit configuration and correspondingly the converse effect, the dielectric, piezoelectric, and flexoelectric contributions to the total enthalpy are
\begin{eqnarray}
\Pi_{di}^{Conv}(\kappa) = \int_{0}^R \int_{\frac{h_s}{2}}^{\frac{h_s}{2}+h_f} \psi_{di}^{Conv} ({\bf E}) \,\, 2\pi r \, \mathrm{d}r \mathrm{d}z 
= -\frac{\pi R^2k_{33}h_f}{6} \left[ \left( \frac{e_{31}h_f}{k_{33}} \right)^2\kappa^2+ \frac{3V^2}{h_f^2} \right] \,\,,
\end{eqnarray}
\begin{eqnarray}
\Pi_{piezo}^{Conv}(\varepsilon_0,\kappa) &=& \int_{0}^R \int_{\frac{h_s}{2}}^{\frac{h_s}{2}+h_f} \psi_{piezo}^{Conv} ({\bm {\varepsilon}}, {\bf E}) \,\, 2\pi r \,  \mathrm{d}r \mathrm{d}z \,\,\nonumber \\
 &=& \pi R^2 e_{31}\left[ (h_s+h_f)V\kappa + \frac{e_{31}}{3k_{33}} h_f^3 \kappa^2 - 2 (\varepsilon_0+\varepsilon_m)V \right]\,\,,
\end{eqnarray}
and
\begin{eqnarray}
      \Pi_{flexo}^{Conv}(\varepsilon_0,\kappa) &=& \int_{0}^R \int_{\frac{h_s}{2}}^{\frac{h_s}{2}+h_f} \psi_{flexo}^{Conv} ({\bm \varepsilon}, \nabla {\bf E}) \,\, 2\pi r \, \mathrm{d}r \mathrm{d}z \,\,\nonumber \\
    &=& -\biggl( \frac{2 \pi R^2 e_{31} \mu_{12} h_f }{ k_{33}} \biggr) \biggl[ 
(h_f +  h_s) \kappa ^2 - 
   2 (\varepsilon_0 + \varepsilon_m) \kappa \biggr]   \,.
\end{eqnarray}

For the open circuit configuration and correspondingly the direct flexoelectric effect, the dielectric, piezoelectric, and flexoelectric contributions to the total enthalpy are
\begin{eqnarray}
\Pi_{di}^{Dir}(\varepsilon_0,\kappa) &=& \int_{0}^R \int_{\frac{h_s}{2}}^{\frac{h_s}{2}+h_f} \psi_{di}^{Dir} ({\bf E}) \,\, 2\pi r \, \mathrm{d}r \mathrm{d}z 
\,\,\nonumber \\ &=& -\frac{\pi  R^2 h_f}{6 k_{33}} \biggl[e_{31}^2 \biggl\{ 4 \kappa ^2 h_f^2+6 \kappa  h_f \biggl(\kappa  h_s-2 (\varepsilon _0+\varepsilon _m)\biggr)+3
\biggl(\kappa  h_s-2 (\varepsilon _0+\varepsilon _m )\biggr){}^2\biggr\} \nonumber \\&& + 12 \kappa e_{31} \mu _{12} \biggl\{\kappa  (h_f+ h_s) -2 (\varepsilon _0+\varepsilon
_m)\biggr\}+12 \kappa ^2 \mu _{12}^2\biggr]\,\,,
\end{eqnarray}
\begin{eqnarray}
\Pi_{piezo}^{Dir}(\varepsilon_0,\kappa) &=& \int_{0}^R \int_{\frac{h_s}{2}}^{\frac{h_s}{2}+h_f} \psi_{piezo}^{Dir} ({\bm {\varepsilon}}, {\bf E}) \,\, 2\pi r \,  \mathrm{d}r \mathrm{d}z \,\,\nonumber \\
 &=&  \frac{1}{3 k_{33}}\pi  R^2 e_{31} h_f \biggl[e_{31} \biggl\{4 \kappa ^2 h_f^2  +6 \kappa  h_f \biggl(\kappa  h_s-2 (\varepsilon _0+\varepsilon
_m)\biggr)+3 \biggl(\kappa  h_s-2 (\varepsilon _0+\varepsilon _m)\biggr){}^2\biggr\}\nonumber \\ && +6 \kappa  \biggl(\kappa  h_f+\kappa  h_s-2 (\varepsilon
_0+\varepsilon _m)\biggr) \mu _{12}\biggr] \,\,,
\end{eqnarray}
and
\begin{eqnarray}
      \Pi_{flexo}^{Dir}(\varepsilon_0,\kappa) &=& \int_{0}^R \int_{\frac{h_s}{2}}^{\frac{h_s}{2}+h_f} \psi_{flexo}^{Dir} (\nabla  {\bm \varepsilon}, {\bf E}) \,\, 2\pi r \, \mathrm{d}r \mathrm{d}z \,\,\nonumber \\
    &=& \biggl( \frac{2 \pi  R^2 e_{31}\mu _{12} h_f}{k_{33}} \biggr)  \biggl[ \kappa^2  (h_f+ h_s) +\kappa^2 \biggl(\frac{2\mu _{12}}{e_{31}}\biggr) - 2 \kappa \left(\varepsilon _0+\varepsilon _m\right)
 \biggr] \,.
\end{eqnarray}


The total enthalpy of the system is the sum of the elastic, dielectric, piezoelectric, and flexoelectric contributions
\begin{equation}\label{Eq:TotalEnthalpy}
    \Pi^{(\cdot)}(\varepsilon_0, \kappa) = \Pi_{elastic}(\varepsilon_0, \kappa) + \Pi_{di}^{(\cdot)}(\varepsilon_0,\kappa) + \Pi_{piezo}^{(\cdot)} (\varepsilon_0, \kappa) + \Pi_{flexo}^{(\cdot)} (\varepsilon_0, \kappa)\,\,,
\end{equation}
where the superscript $(\cdot)$ denotes the converse or the direct cases. 

Note that within the class of all admissible deformations, the deformation of the mid-plane of the film-substrate system is the one that minimizes the total enthalpy of the system and hence the total enthalpy is stationary with $\varepsilon_0$ and $\kappa$ \cite{Freund:1999} 
\begin{equation}\label{Eq:simultaneous}
    \frac{\partial \Pi^{(\cdot)}}{\partial \varepsilon_0} = 0 \,\,\, \text{and} \,\, \frac{\partial \Pi^{(\cdot)}}{\partial \kappa} = 0 \,\,.
\end{equation}
These two conditions give a set of equations which can be solved for $\varepsilon_0$ and $\kappa$. 

\subsection{Expressions for stretching  strain, curvature, and electric polarization}
In this section, we present analytical expressions of $\varepsilon_0$ and $\kappa$ for different electromechanical effects. 

\paragraph{Case I: both piezoelectric and flexoelectric effects are present} We present the expressions of the stretching strain and curvature for the film-substrate system when both piezoelectric and flexoelectric effects are present in the film. The stretching strain and curvature are respectively normalized by Stoney's approximations $\varepsilon_0^{st}$ and $\kappa^{st}$ as defined in Equations \ref{Eq:stoneyepsilon} and \ref{Eq:stoneykappa}, respectively. The stretching  strain $\varepsilon_{0,pf}^{Conv}$ for the converse flexoelectric case is
\begin{eqnarray}\label{Eq:strain:conv:pf}
 & &  \frac{\varepsilon_{0,pf}^{Conv}}{\varepsilon_0^{st}}  = \frac{{\splitfrac{h_s M_s \biggl\{-2 V e_{31}^3 h_f^3 k_{33}-V e_{31} k_{33}^2 (h_f^3 M_f+h_s^3 M_s )+h_f k_{33}^2 M_f (h_f^3M_f+h_s^3 M_s ) \varepsilon_m \mathstrut} { +2 e_{31}^2 h_f \biggl(h_f^3 k_{33} M_f \varepsilon_m+6 V h_s k_{33} \mu _{12}+6 h_f \mu _{12} (V k_{33}-4 \varepsilon_m \mu _{12})\biggr)\biggr\}\mathstrut}}}{{\splitfrac{{h_f M_f \varepsilon_m \biggl\{k_{33}^2 (h_f^4 M_f^2+4 h_f^3 h_s M_f M_s+6 h_f^2 h_s^2 M_f M_s+4 h_f h_s^3 M_f M_s+h_s^4 M_s^2 )\mathstrut}}{-24 e_{31} h_f h_s (h_f+h_s ) k_{33} M_s \mu_{12}+2 e_{31}^2 h_f^2 (h_f^2 k_{33} M_f+h_f h_s k_{33}M_s-24 \mu _{12}^2 ) \biggr\}  \mathstrut}}}
\end{eqnarray}
and the curvature $\kappa_{pf}^{Conv}$ for the converse case is
\begin{eqnarray}\label{Eq:kappa:conv:pf}
    \frac{\kappa_{pf}^{Conv}}{\kappa^{st}} = \frac{{\splitfrac{-h_s^2 k_{33} M_s \biggl\{-h_f h_s (h_f+h_s) k_{33} M_f M_s \varepsilon_m+4 V e_{31}^2 h_f \mu_{12}\mathstrut}{+e_{31}h_s M_s \biggl(V h_s k_{33}+h_f (V k_{33}+4 \varepsilon _m \mu _{12})\biggr)\biggr\}\mathstrut}}}{{\splitfrac{h_f M_f \varepsilon_m \biggl\{ k_{33}^2 (h_f^4M_f^2+4 h_f^3 h_s M_f M_s+6 h_f^2 h_s^2 M_f M_s+4 h_f h_s^3 M_f M_s+h_s^4 M_s^2)\mathstrut}{-24 e_{31} h_f h_s \left(h_f+h_s\right) k_{33} M_s \mu _{12}+2e_{31}^2 h_f^2 (h_f^2 k_{33} M_f+h_f h_s k_{33} M_s-24 \mu _{12}^2)\biggr\}\mathstrut}}}\,\,.
\end{eqnarray}
For the direct case, we obtain the stretching strain $\varepsilon_{0,pf}^{Dir}$ as
\begin{eqnarray}\label{Eq:strain:dir:pf}
&& \frac{\varepsilon_{0,pf}^{Dir}}{\varepsilon_0^{st}} \nonumber \\ && = 
\frac{{h_s M_s \biggl\{4 e_{31}^4 h_f^3+2 e_{31}^2 k_{33} (2 h_f^3 M_f+h_s^3 M_s)+k_{33} M_f (h_f^3 k_{33} M_f+h_s^3
k_{33} M_s+24 h_f \mu_{12}^2) \biggr\}}}
{\splitfrac{M_f \biggl\{ 4 e_{31}^4 h_f^4+4 e_{31}^2 h_f k_{33} (h_f^3 M_f+2 h_f^2 h_s M_s+3 h_f h_s^2
M_s+2 h_s^3 M_s)+24 e_{31} h_f h_s (h_f+h_s) k_{33} M_s \mu_{12}\mathstrut}{+k_{33} \biggl(h_f^4 k_{33} M_f^2+4 h_f^3 h_s k_{33} M_f M_s+h_s^4k_{33} M_s^2+6 h_f^2 M_f (h_s^2 k_{33} M_s+4 \mu _{12}^2)+4 h_f h_s M_s (h_s^2 k_{33} M_f+6 \mu _{12}^2) \biggr) \biggr\}\mathstrut}} \nonumber \\
\end{eqnarray}
and the curvature $\kappa_{pf}^{Dir}$ as
\begin{eqnarray}\label{Eq:kappa:dir:pf}
&& \frac{\kappa_{pf}^{Dir}}{\kappa^{st}} \nonumber \\ && = \frac{h_s^3 k_{33} M_s^2 \biggl\{2 e_{31}^2 (h_f+h_s)+(h_f+h_s) k_{33} M_f+4 e_{31} \mu _{12}\biggr\}}
{\splitfrac{M_f \biggl\{4 e_{31}^4 h_f^4+4 e_{31}^2 h_f k_{33} (h_f^3 M_f+2 h_f^2 h_s M_s+3 h_f h_s^2 M_s+2 h_s^3 M_s)+24 e_{31} h_f h_s (h_f+h_s)k_{33} M_s \mu _{12}\mathstrut}{+k_{33} \biggl(h_f^4 k_{33} M_f^2+4 h_f^3 h_s k_{33} M_f M_s+h_s^4 k_{33} M_s^2+6 h_f^2 M_f (h_s^2k_{33} M_s+4 \mu _{12}^2)+4h_f h_s M_s (h_s^2 k_{33} M_f+6 \mu _{12}^2) \biggr) \biggr\}}} \nonumber \\ \,\,.
\end{eqnarray}

\paragraph{Case II: piezoelectric, dielectric and flexoelectric effects are absent} In the absence of piezoelectric and flexoelectric effects, The expressions of stretching strain and curvature of the film-substrate system, can be obtained from Equations \ref{Eq:strain:conv:pf}, \ref{Eq:kappa:conv:pf} or Equations \ref{Eq:strain:dir:pf}, \ref{Eq:kappa:dir:pf} by setting $e_{31} = 0$, and $\mu_{12} = 0$. In this case, the stretching strain $\varepsilon_0^*$ and the curvature $\kappa^*$ are solely due to elastic effects of the film and substrate, and have been previously reported by Freund and co-workers \cite{Freund:Book,Freund:1999}. These are
\begin{eqnarray}
    \frac{\varepsilon_0^*}{\varepsilon_0^{st}} = \frac{h_s M_s (h_f^3 M_f+h_s^3 M_s)}{h_f^4 M_f^2+4 h_f^3 h_s M_f M_s+6 h_f^2 h_s^2 M_f M_s+4 h_f h_s^3 M_f M_s+h_s^4 M_s^2} \,\,,
\end{eqnarray}
and
\begin{eqnarray}
    \frac{\kappa^*}{\kappa_0^{st}} = \frac{h_s^3 M_s^2 \left(h_f+h_s\right)}{h_f^4 M_f^2+4 h_f^3 h_s M_f M_s+6 h_f^2 h_s^2 M_f M_s+4 h_f h_s^3 M_f M_s+h_s^4 M_s^2} \,\,.
\end{eqnarray}

\paragraph{Case III: flexoelectric effect is absent} The expressions of stretching strain and curvature of the film-substrate system in the absence of flexoelectric effect are obtained by setting $\mu_{12} = 0$ in Equations \ref{Eq:strain:conv:pf} - \ref{Eq:kappa:dir:pf}. In this case, the stretching strain and the curvature are due to the combined effects of the film's piezoelectricity, dielectricity, and elasticity of the film and substrate.

\paragraph{Case IV: piezoelectric effect is absent} Lastly, we discuss when the piezoelectric effect in the film is absent. In this case, the stretching strain and curvature of the system are only influenced by the flexoelectric effects in the film and the combined elastic effects of the film and substrate. We can obtain the curvatures and the stretching strains by setting $e_{31}=0$ in Equations \ref{Eq:strain:conv:pf} - \ref{Eq:kappa:dir:pf}. Interestingly, for the converse case, the stretching  strain and the curvature are the same as those when only elasticity is considered (Case II) 
\begin{eqnarray}
    \varepsilon_{0,f}^{Conv} = \varepsilon_0^* \,\,, \kappa_{f}^{Conv} = \kappa^* \,\,.
\end{eqnarray}

\paragraph{Electric polarization in the film} Once the curvature $\kappa$, the stretching strains $\varepsilon_0$, the strains in the film $\varepsilon_{rr}$ and $\varepsilon_{\theta\theta}$ are calculated, we calculate the electric field using Equation \ref{Eq:ElectricFleldConv} for the converse case and Equation \ref{Eq:ElectricFleldDir} for the direct case. We use these to calculate the electric displacement in the film, and calculate the electric polarization using Equation \ref{Eqn:polarization}. 

For the converse case, the polarization in the $z$-direction is
\begin{eqnarray}\label{Eq:PolarizationConv}
    && P_z^{Conv} (z) = (e_0 + k_{33}) E_z^{Conv} (z) + 2 e_{31} \varepsilon_{rr}(z) + 2\mu_{12} \frac{\partial \varepsilon_{rr}}{\partial z} \,\,, \nonumber \\
     &=& (e_0 + k_{33})\Bigg[  \frac{2e_{31}}{k_{33}} \kappa \left( z - \frac{h_f+h_s}{2} \right) + \frac{V}{h_f} \Bigg ] + 2 e_{31} (\varepsilon_0 - \kappa z + \varepsilon_m ) - 2\mu_{12} \kappa \,\,, 
\end{eqnarray}
and for the direct case
\begin{eqnarray}\label{Eq:PolarizationDir}
    && P_z^{Dir} (z) = (e_0 + k_{33}) E_z^{Dir} (z) + 2 e_{31} \varepsilon_{rr}(z) + 2\mu_{12} \frac{\partial \varepsilon_{rr}}{\partial z} \,\,, \nonumber \\
     &=& (e_0 + k_{33})\left( \frac{2 e_{31}}{k_{33}}\right)\left[ \kappa \left( z + \frac{\mu_{12}}{e_{31 }} \right) - (\varepsilon_{0} + \varepsilon_{m}) \right]  + 2 e_{31} (\varepsilon_0 - \kappa z + \varepsilon_m ) - 2\mu_{12} \kappa \,\,.
\end{eqnarray}
The polarization in the radial and angular directions is zero.

From Equations \ref{Eq:PolarizationConv} and \ref{Eq:PolarizationDir}, we see that the polarization varies linearly across the thickness of the film. Also, notice that the polarization not only depends on the electromechanical constants, but also on the geometry, curvature, and stretching strain. The dependence of polarization on the stiffness $M_s$ and $M_f$, and the thickness $h_f$ and $h_s$ of the film and the substrate is non-linear because the stretching strain $\varepsilon_0$ and the curvature $\kappa$ depend on these parameters in a non-linear fashion.



\subsubsection{Discussion}
As seen from the equations presented in the previous section, the thickness ratio $h_f/h_s$ and the stiffness ratio $M_f/M_s$ influence the stretching strain and curvature. In this section, we discuss this using the material constants and parameters listed in Table \ref{parametertable}. 


\begin{table}[!h]
\centering
\caption{Materials constants and parameters}
\label{parametertable}
\begin{tabular}{ll}
\hline
constants and parameters & value \\
\hline
$\mu_{12}$ & 1 $\mu$C/m \cite{Abdollahi:2014}\\
$e_{31}$ & -4.4 C/m$^2$ \cite{Abdollahi:2014}\\
$k_{11}$ & 11 nC/Vm \cite{Abdollahi:2014}\\
$k_{33}$ & 12.48 nC/Vm \cite{Abdollahi:2014}\\
$\chi_{33}$ & 1408 \cite{Abdollahi:2014}\\
$e_0$ & 8.854$\times10^{-12}$ C/V-m \cite{Yan2013}\\
$h_s$ & 100 $\mu$ m \\
$M_s$ & 180 GPa \cite{Freund:1999} \\
$\varepsilon_m$ & -0.0082 \cite{Freund:1999}\\
$V$ & 0.05 Volt \\
\hline
\end{tabular}
\vspace*{-4pt}
\end{table}

\begin{figure}[h!]
\centering
\subfigure[$\frac{\varepsilon_0^*}{\varepsilon_0^{st}}$]{\includegraphics[keepaspectratio=true,width=0.32\textwidth]{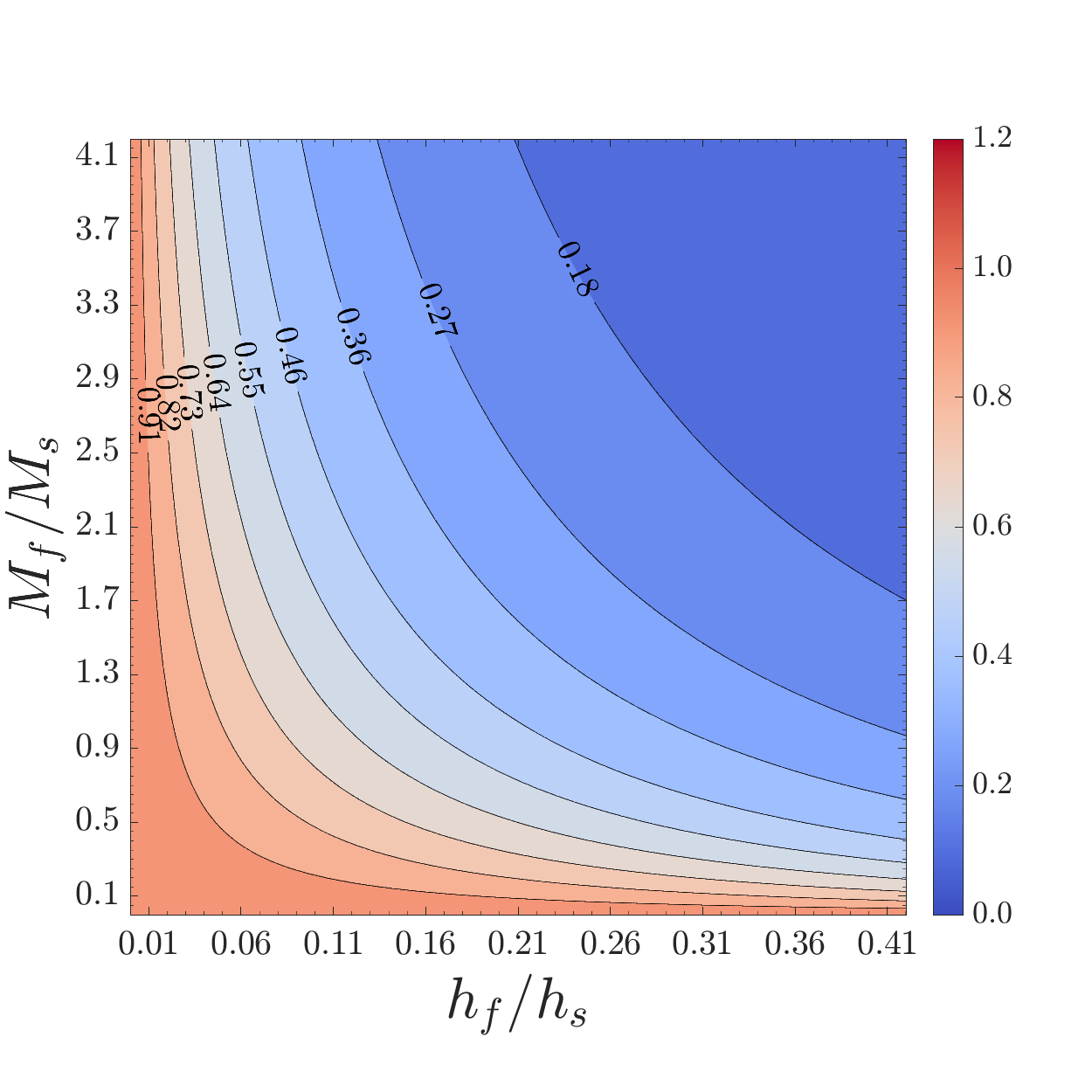}\label{Fig:Uni:e0:Star}}
\subfigure[$\frac{\varepsilon_{0,pf}^{Conv}}{\varepsilon_0^{st}}$]{\includegraphics[keepaspectratio=true,width=0.32\textwidth]{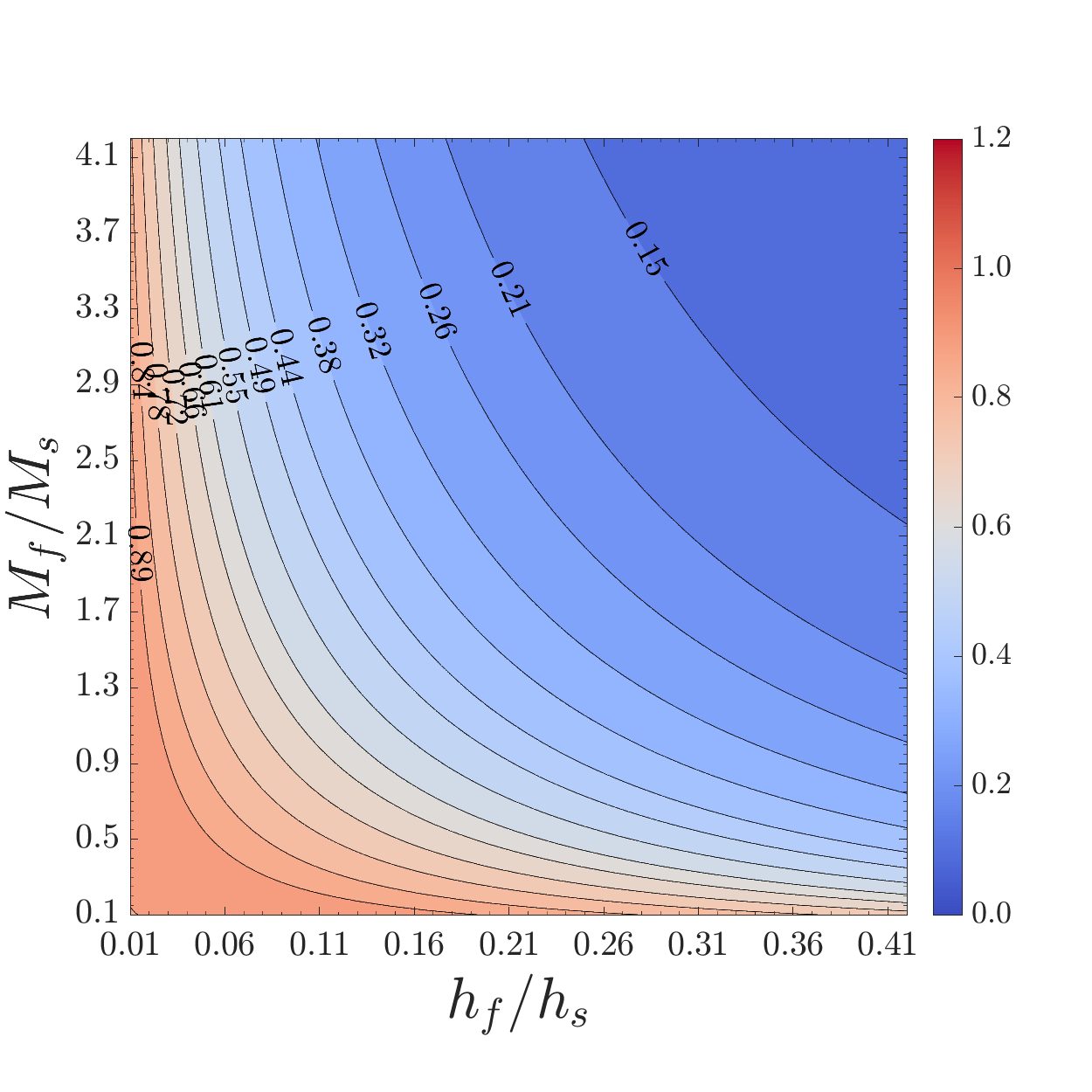}\label{Fig:Uni:e0:Conv}}
\subfigure[$\frac{\varepsilon_{0,pf}^{Dir}}{\varepsilon_0^{st}}$]{\includegraphics[keepaspectratio=true,width=0.32\textwidth]{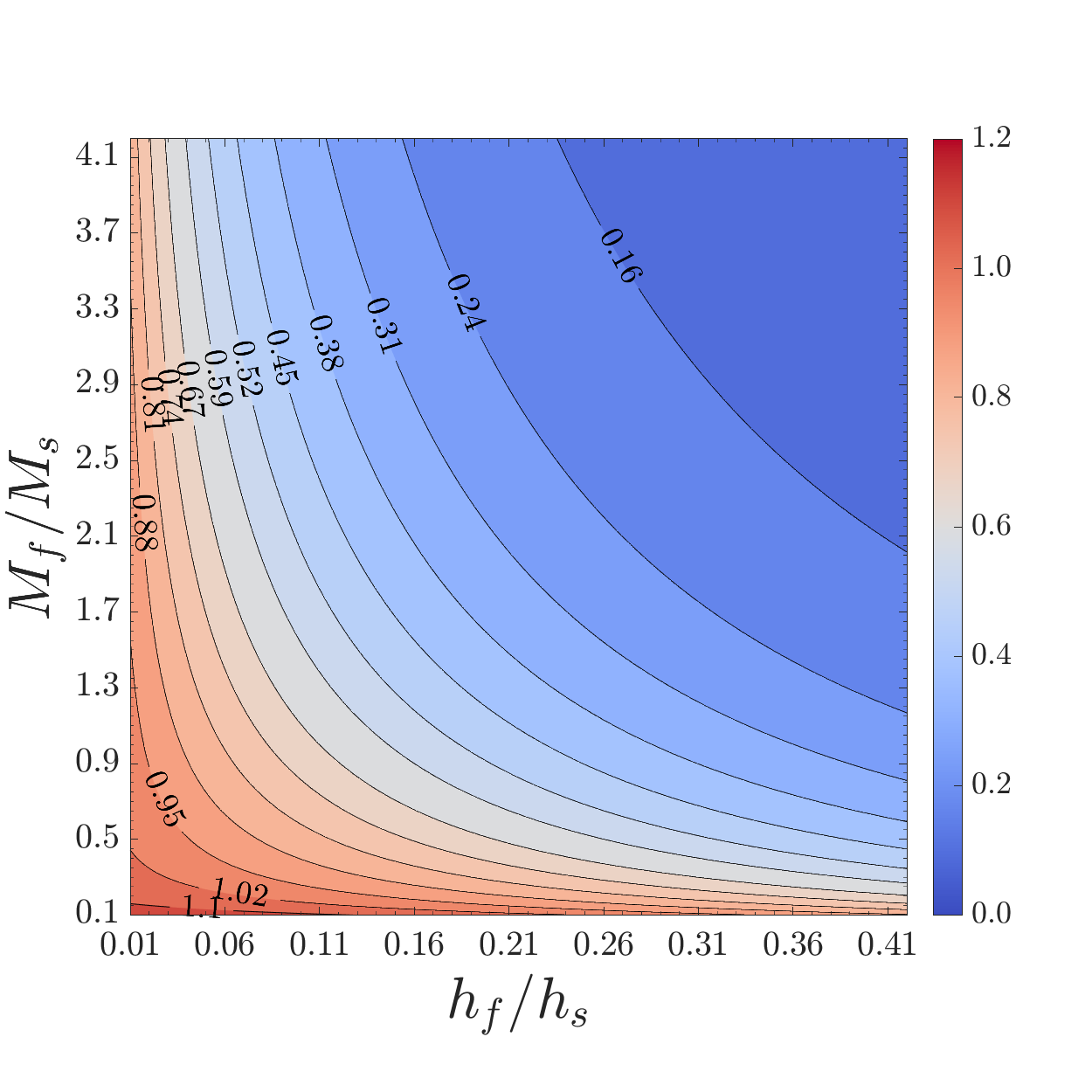}\label{Fig:Uni:e0:Dir}}\\
\subfigure[$\frac{\kappa^*}{\kappa^{st}}$]{\includegraphics[keepaspectratio=true,width=0.32\textwidth]{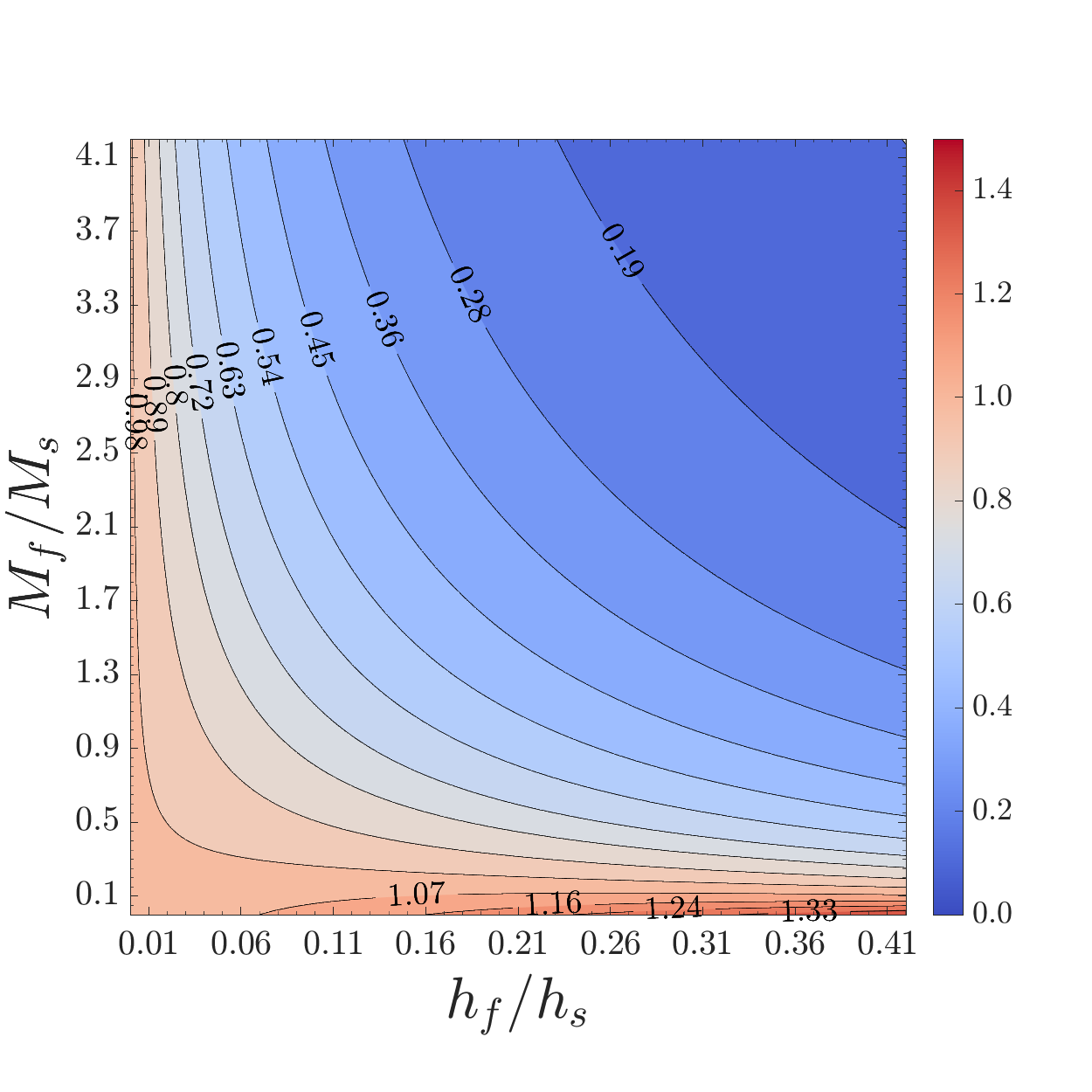}\label{Fig:Uni:Curv:Star}}
\subfigure[$\frac{\kappa_{pf}^{Conv}}{\kappa^{st}}$]{\includegraphics[keepaspectratio=true,width=0.32\textwidth]{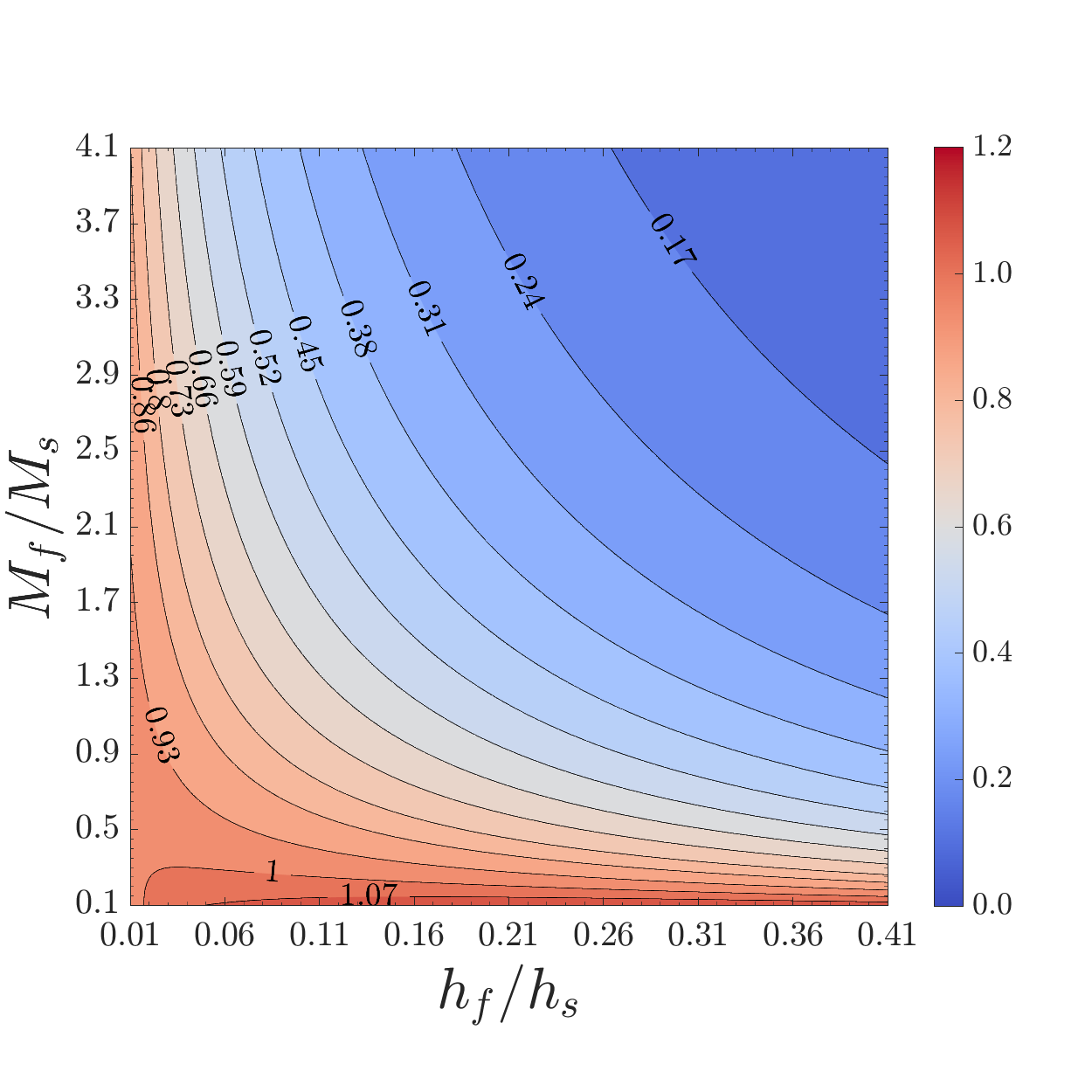}\label{Fig:Uni:Curv:Conv}}
\subfigure[$\frac{\kappa_{pf}^{Dir}}{\kappa^{st}}$]{\includegraphics[keepaspectratio=true,width=0.32\textwidth]{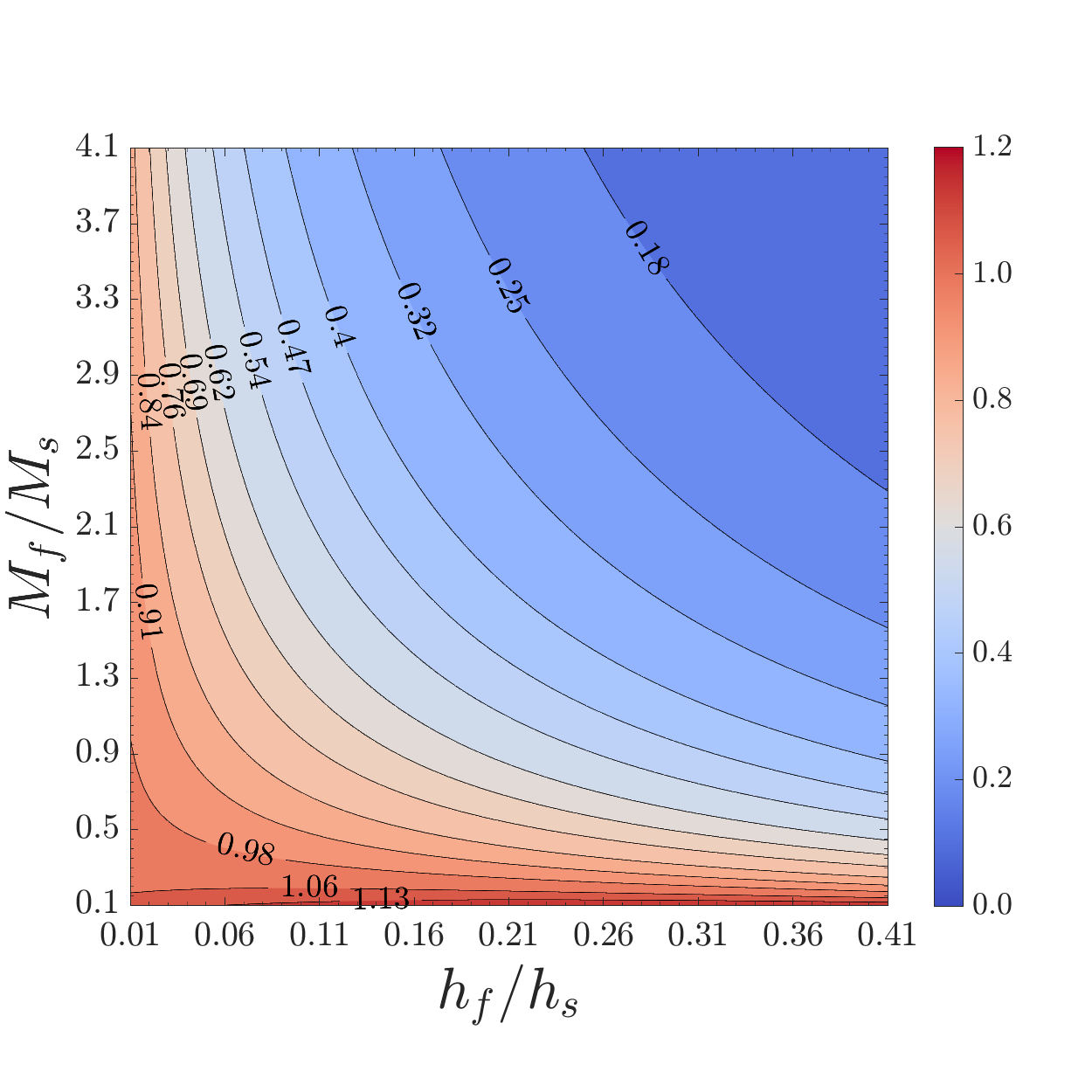}\label{Fig:Uni:Curv:Dir}}
{\caption{Normalized stretching strains and curvatures of the film-substrate system for various ratios of film-substrate thickness $h_f/h_s$ and stiffness $M_f/M_s$. (a) show the stretching strain and (d) the curvature when only elastic effects are considered (case II). (b), (c) show the stretching strains and (e), (f) the curvatures for converse and direct effects when both piezoelectric and flexoelectric effects are present (case I). }\label{Fig:e0kappa}}
\end{figure}

In Figure \ref{Fig:e0kappa}, we show the variation of the stretching strains and curvatures with varying ratios of $h_f/h_s$ and $M_f/M_s$. In Figure \ref{Fig:Uni:e0:Star}, stretching strain, and in Figure \ref{Fig:Uni:Curv:Star} normalized curvature when only elasticity is present are shown. The normalized stretching strains in Figures \ref{Fig:Uni:e0:Conv} and \ref{Fig:Uni:e0:Dir} and the normalized curvatures in Figures \ref{Fig:Uni:Curv:Conv} and \ref{Fig:Uni:Curv:Dir} show the converse and direct cases when both piezoelectric and flexoelectric effects are present. In these figures, the contour line with the value of $1$ gives the range of values of $h_f/h_s$ and $M_f/M_s$ for which Stoney's approximation is valid. We see that when the ratios $h_f/h_s$ and $M_f/M_s$ are increased, Stoney's formula overestimates stretching strains and curvatures. 

\begin{figure}[h!]\centering
\subfigure[$M_f/M_s=0.1$]{\includegraphics[keepaspectratio=true,width=0.30\textwidth]{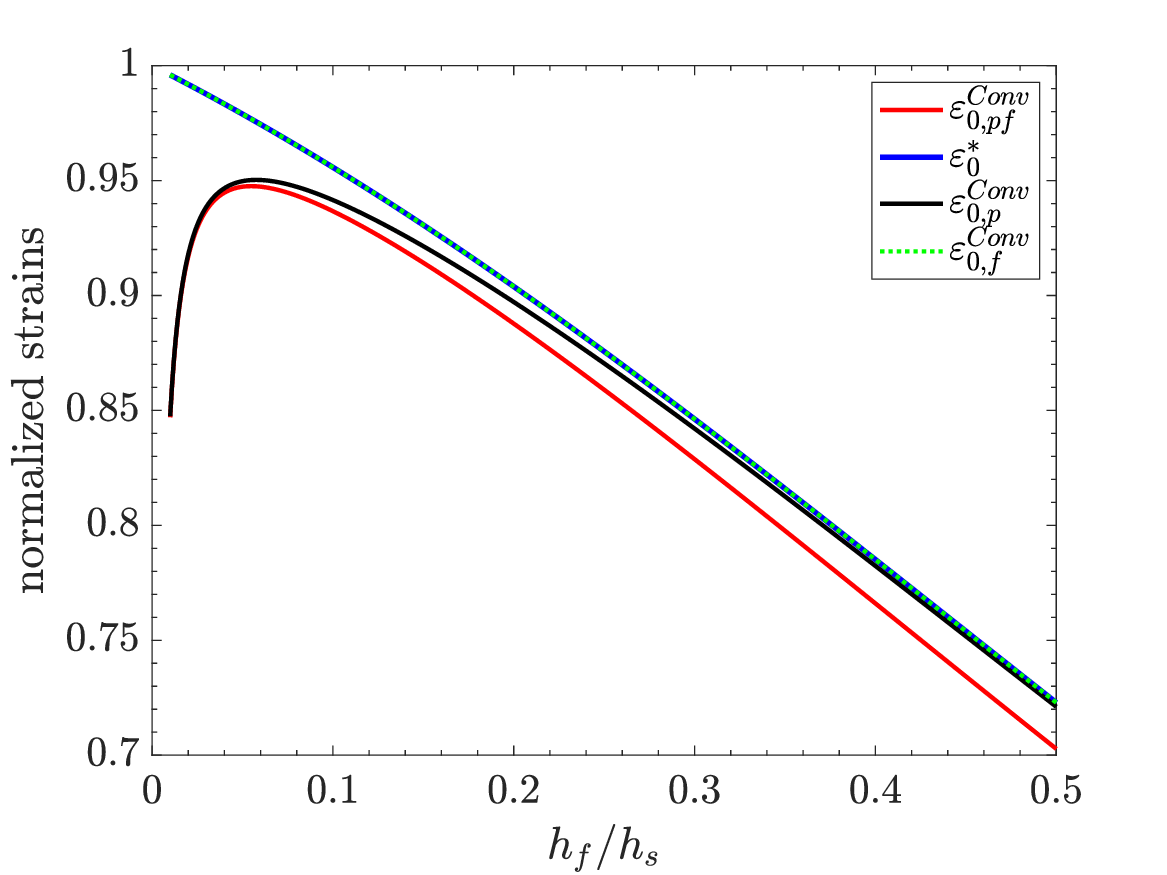}}
\subfigure[$M_f/M_s=1.0$]{\includegraphics[keepaspectratio=true,width=0.30\textwidth]{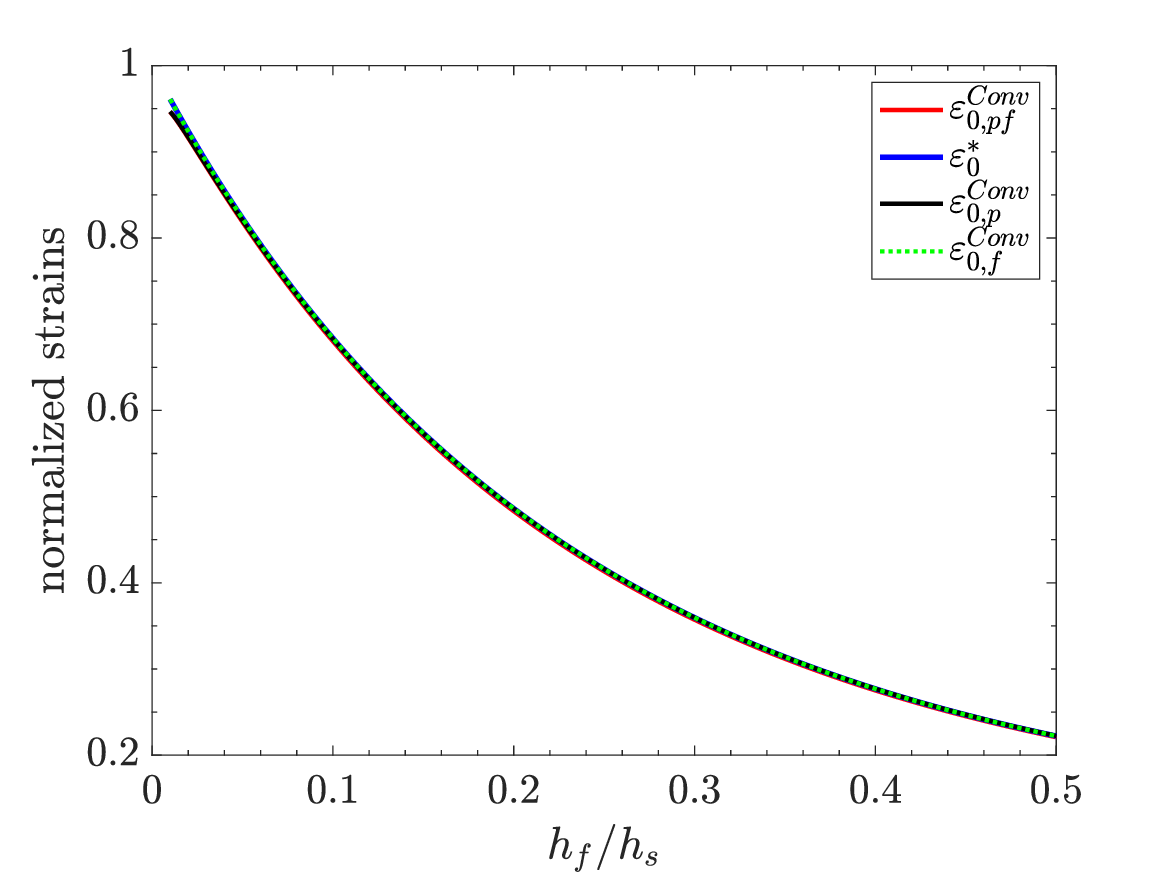}}
\subfigure[$M_f/M_s=1.5$]{\includegraphics[keepaspectratio=true,width=0.30\textwidth]{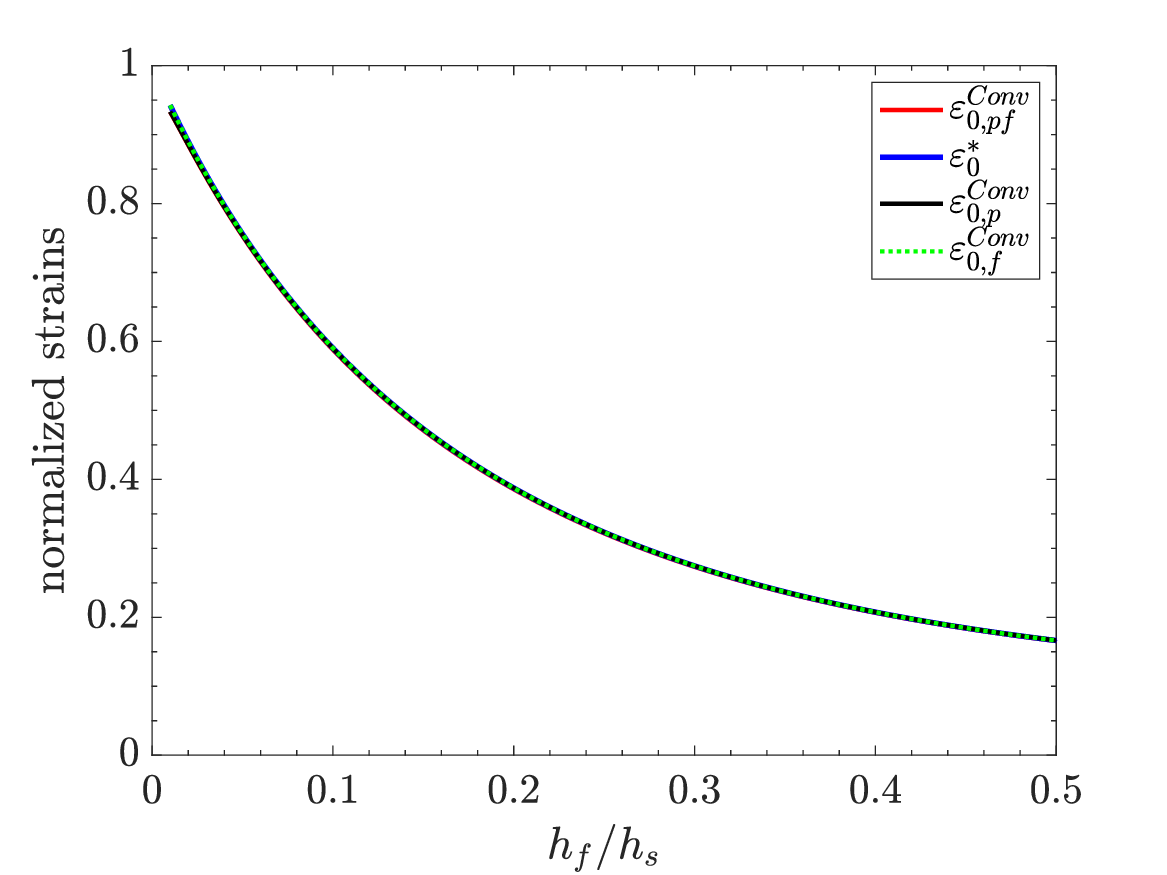}}
\subfigure[$M_f/M_s=0.1$]{\includegraphics[keepaspectratio=true,width=0.30\textwidth]{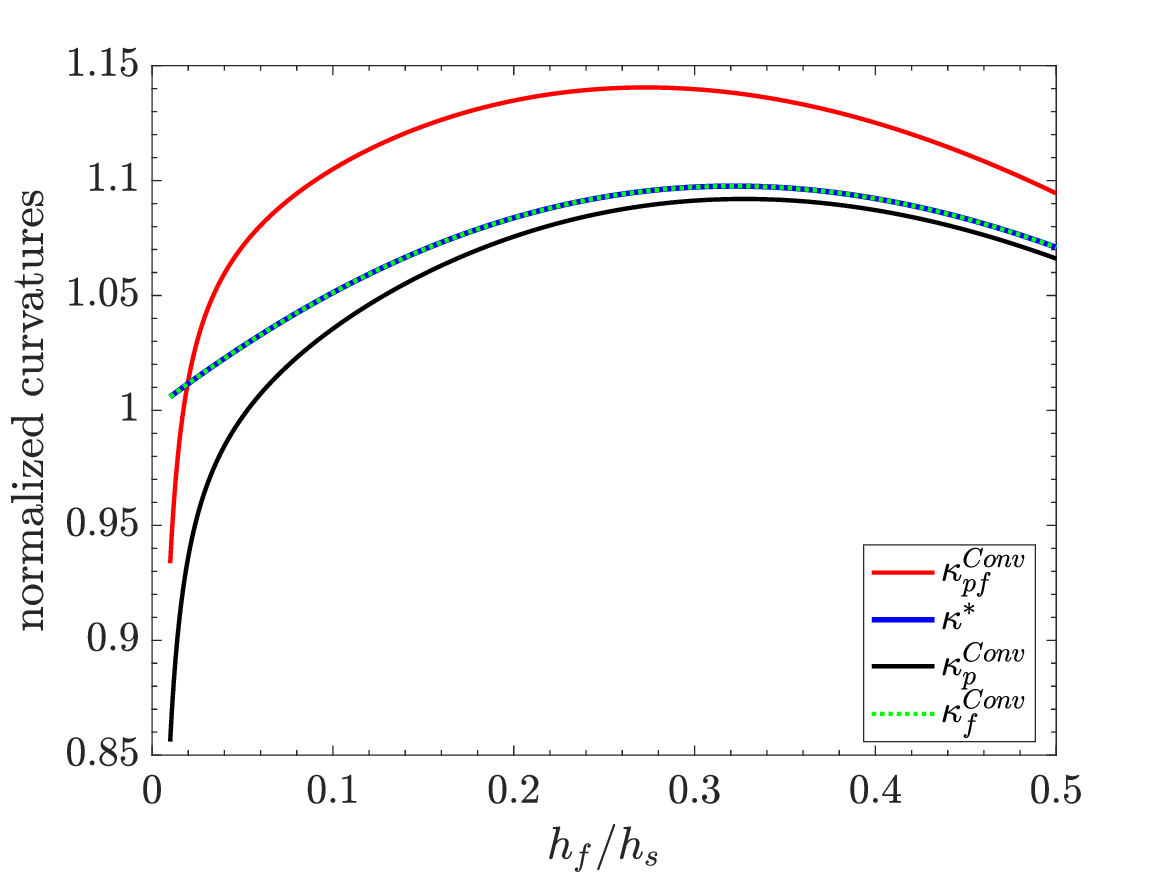}}
\subfigure[$M_f/M_s=1.0$]{\includegraphics[keepaspectratio=true,width=0.30\textwidth]{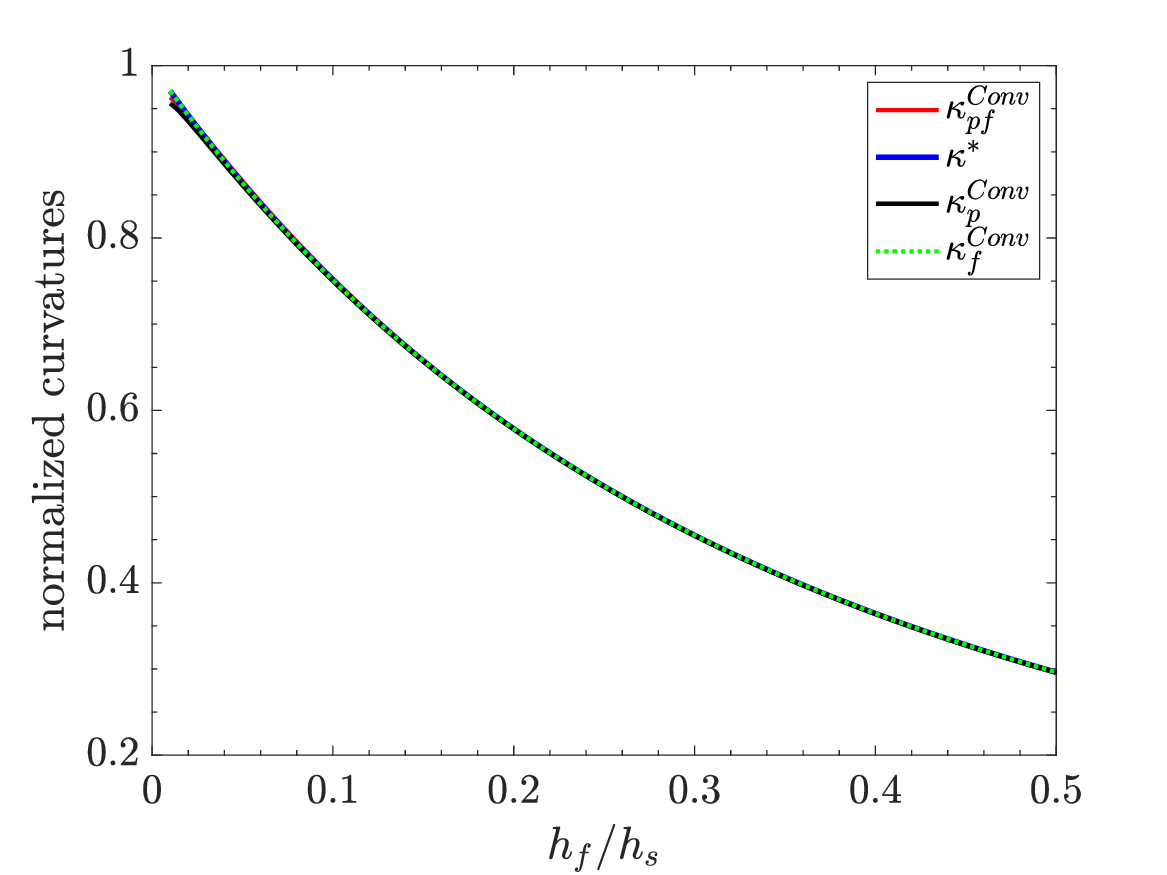}}
\subfigure[$M_f/M_s=1.5$]{\includegraphics[keepaspectratio=true,width=0.30\textwidth]{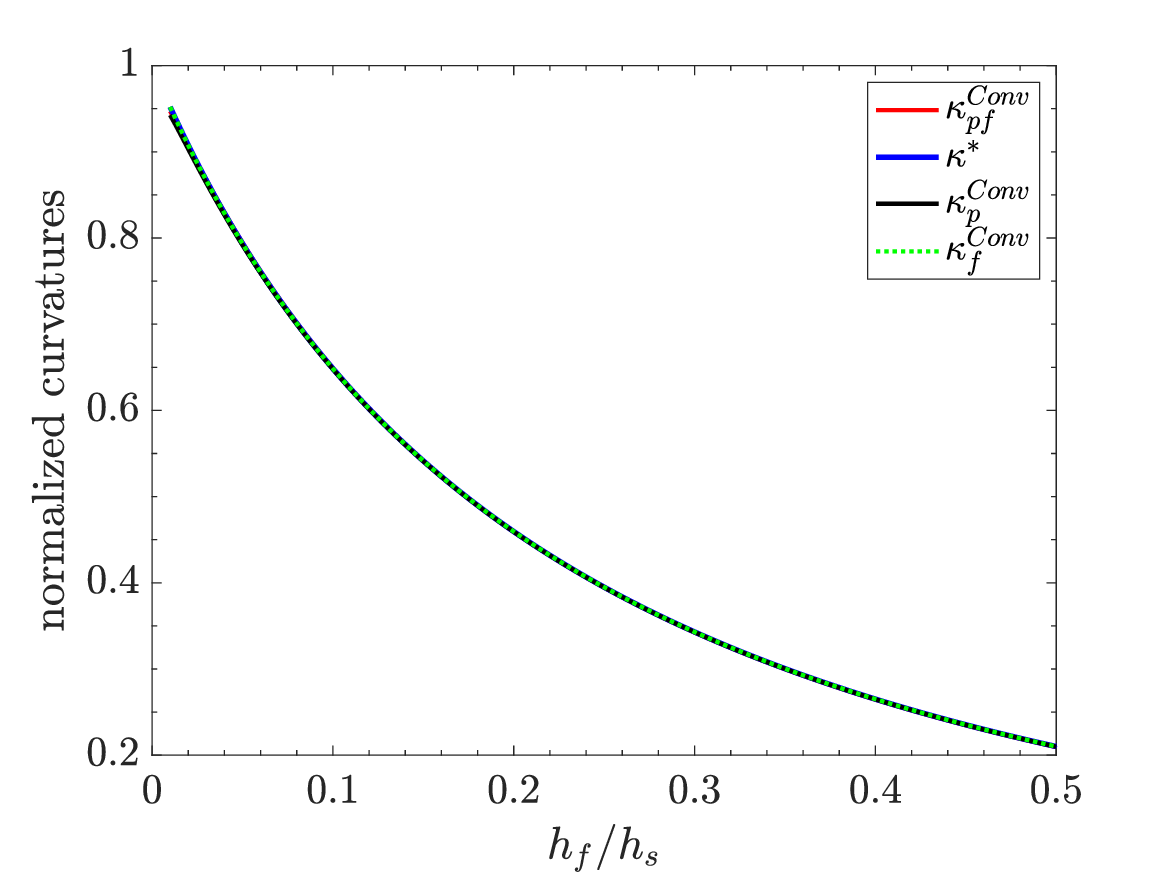}}
{\caption{Normalized stretching strains and curvatures for the converse case shown as a function of the ratio of the thickness of film and substrate for different cases of stiffness ratios $M_f/M_s$. Figures (a)-(c) show the stretching strains, and (d)-(f) show the curvatures.} \label{Fig:ConvStrainUniform}}
\end{figure}

\begin{figure}[h!]\centering
\subfigure[$M_f/M_s=0.1$]{\includegraphics[keepaspectratio=true,width=0.30\textwidth]{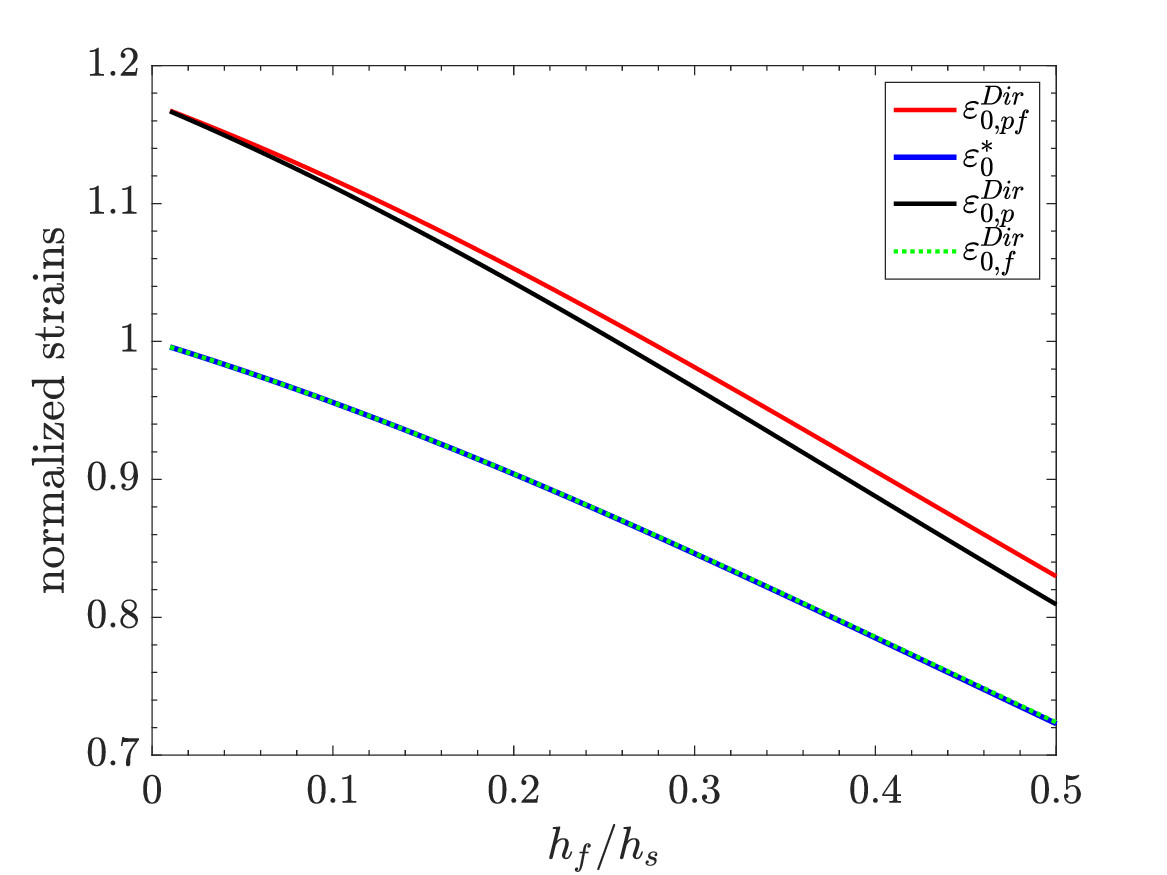}}
\subfigure[$M_f/M_s=1.0$]{\includegraphics[keepaspectratio=true,width=0.30\textwidth]{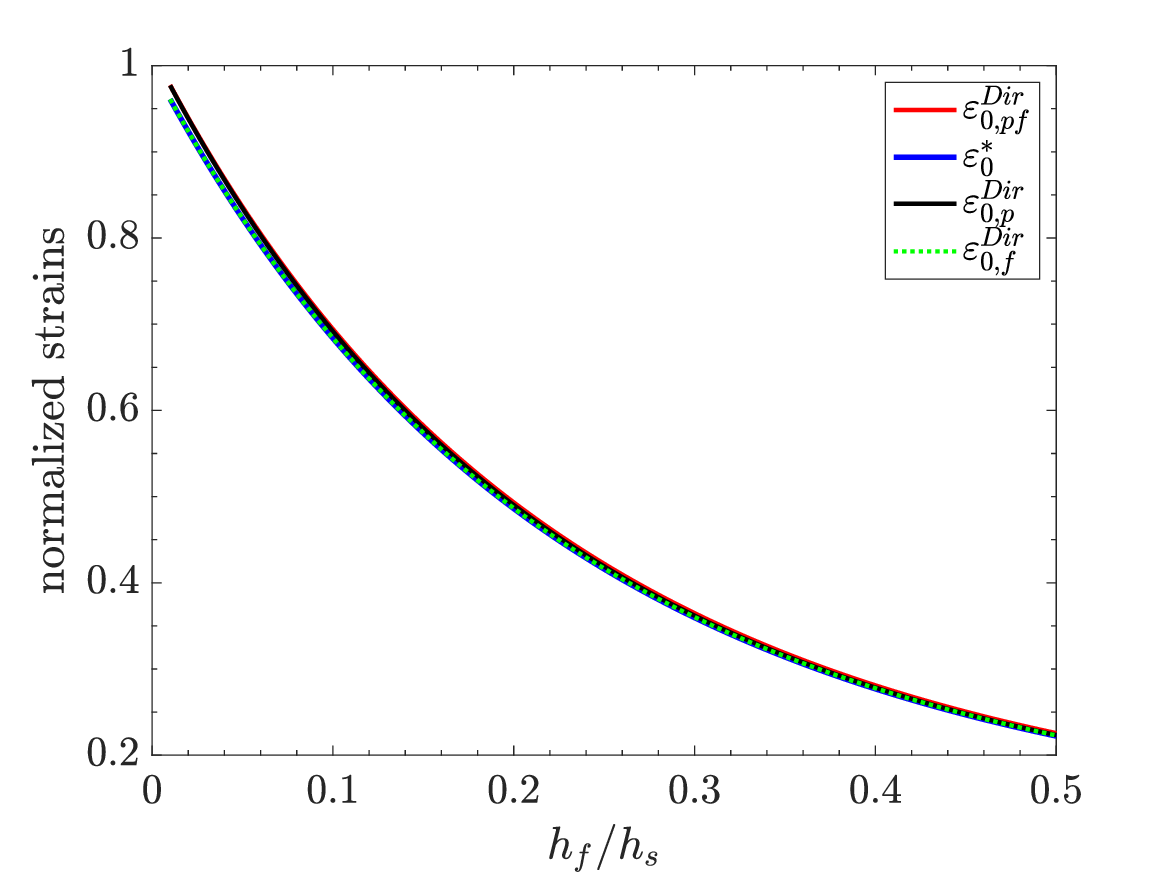}}
\subfigure[$M_f/M_s=1.5$]{\includegraphics[keepaspectratio=true,width=0.30\textwidth]{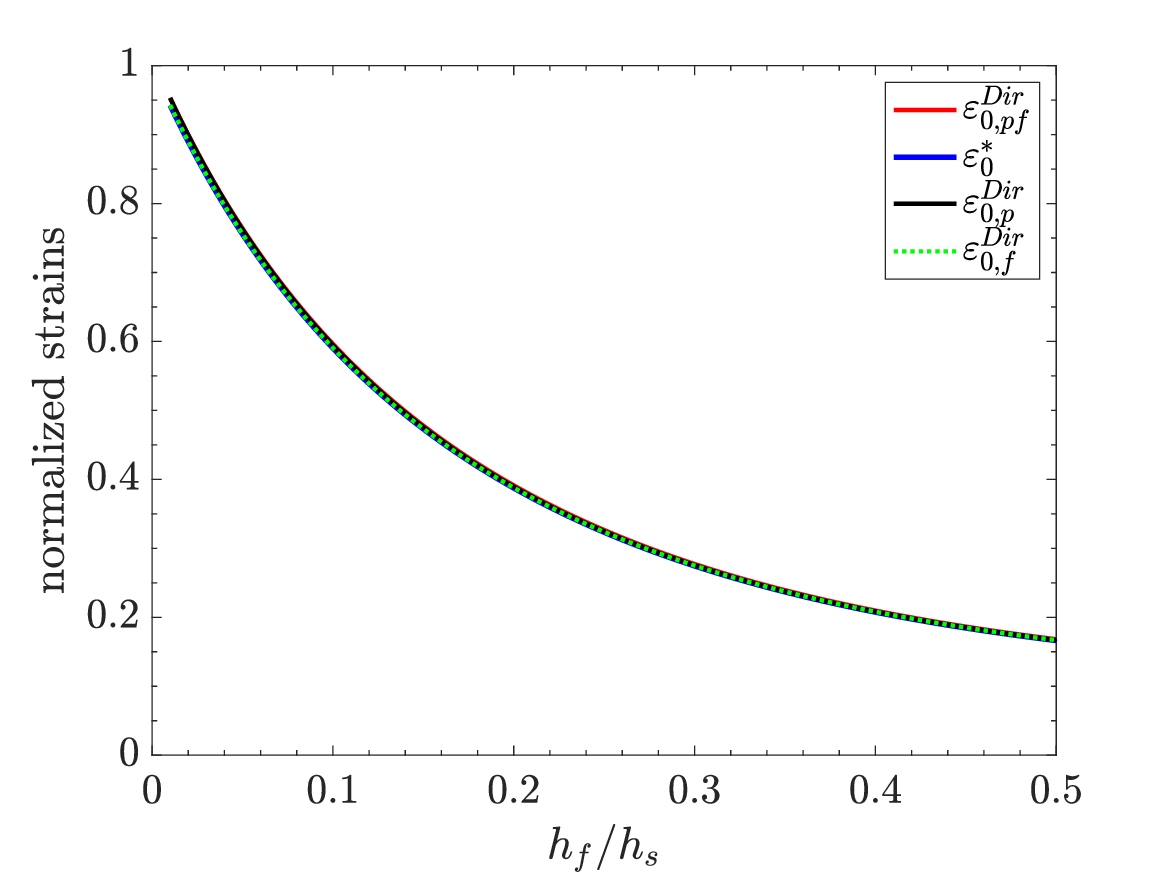}}
\subfigure[$M_f/M_s=0.1$]{\includegraphics[keepaspectratio=true,width=0.30\textwidth]{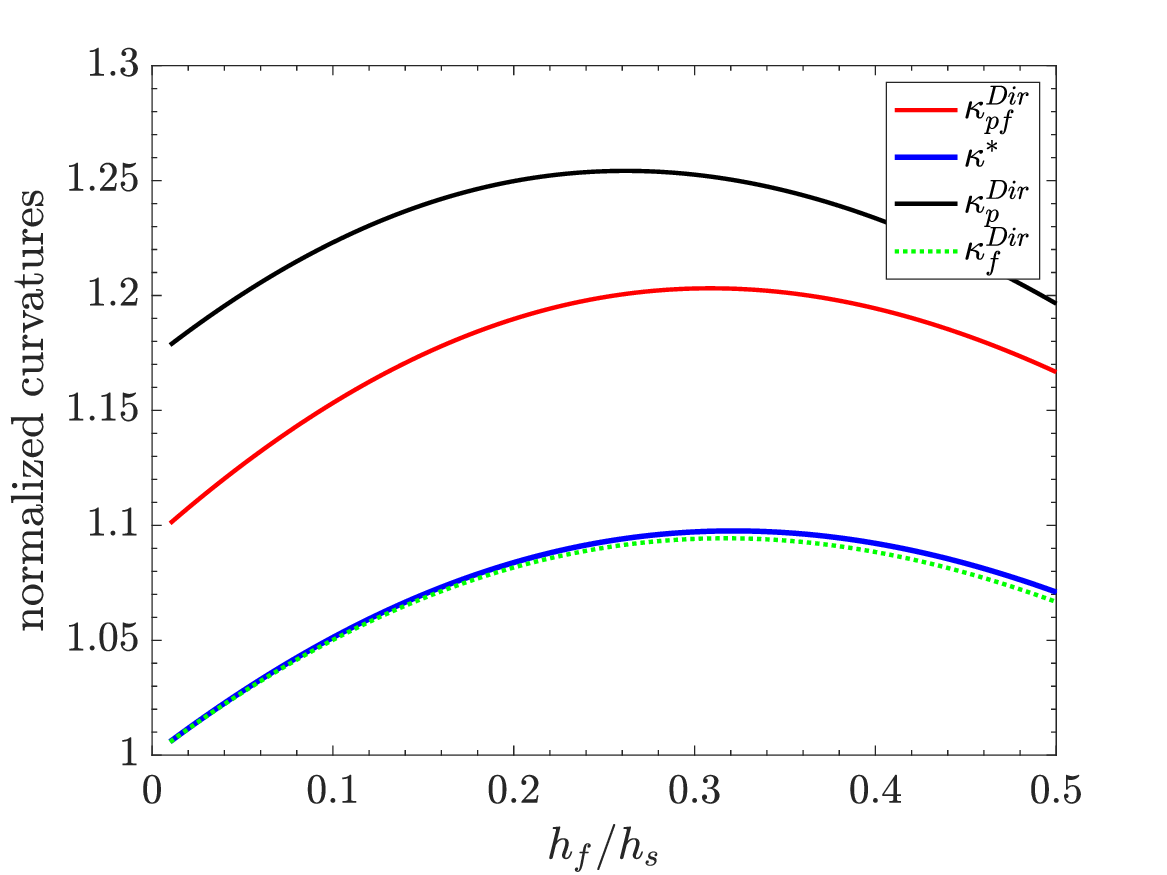}}
\subfigure[$M_f/M_s=1.0$]{\includegraphics[keepaspectratio=true,width=0.30\textwidth]{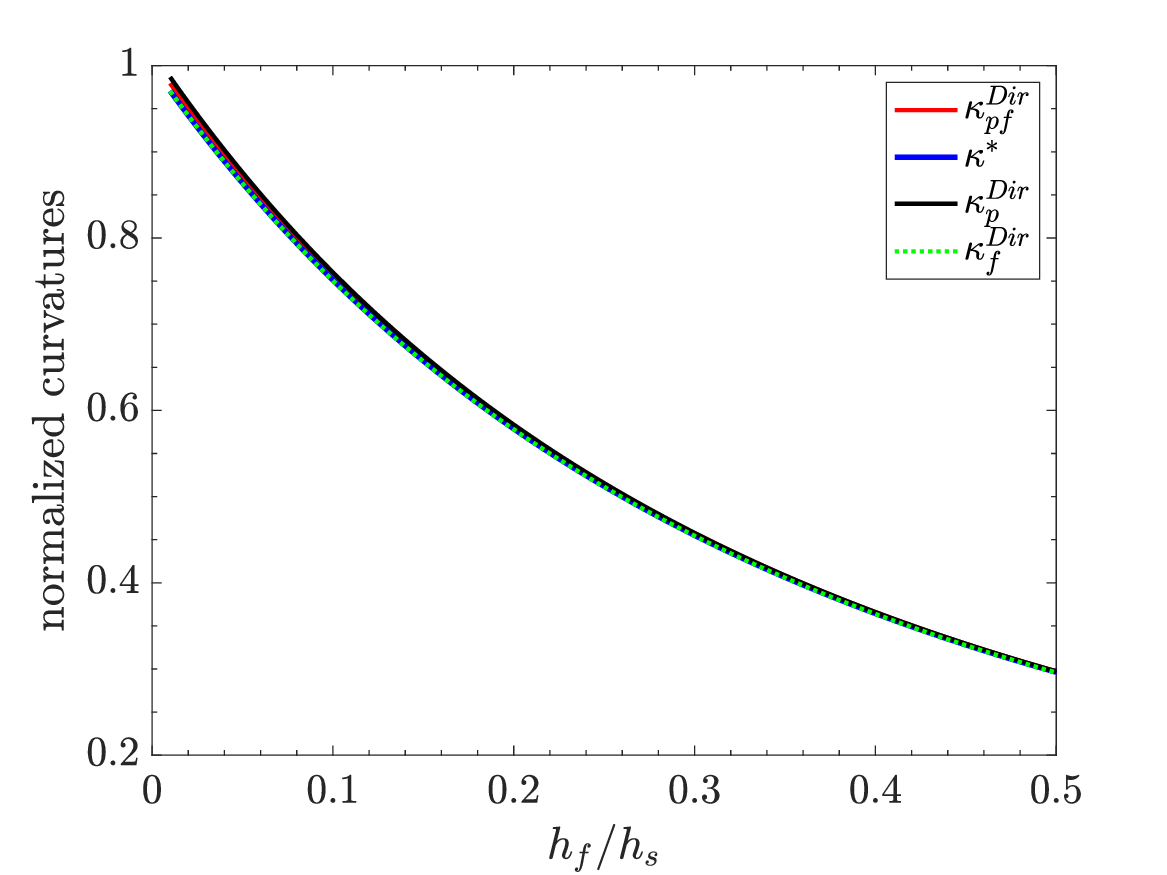}}
\subfigure[$M_f/M_s=1.5$]{\includegraphics[keepaspectratio=true,width=0.30\textwidth]{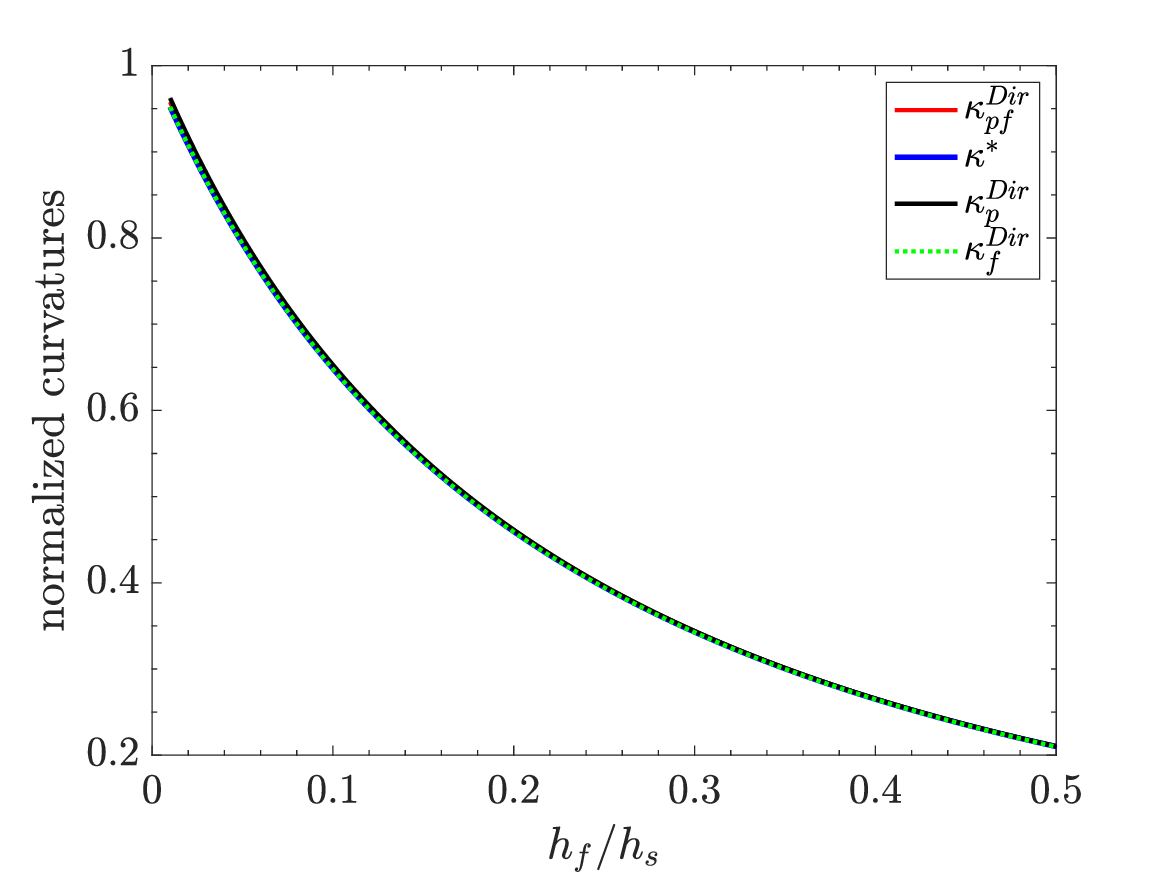}}
{\caption{Normalized stretching strains and curvatures for the direct case shown as a function of the ratio of the thickness of film and substrate for different cases of stiffness ratios $M_f/M_s$. Figures (a)-(c) shows the stretching strains, and (d)-(f) shows the curvatures.}\label{Fig:DirStrainUniform}}
\end{figure}

The dependence of stretching strains and curvatures on the ratio of thickness $h_f/h_s$ for stiffness ratios $M_f/M_s = 0.1$, $1.0$, and $1.5$, respectively, is shown in Figure \ref{Fig:ConvStrainUniform} for the converse case and in Figure \ref{Fig:DirStrainUniform} for the direct case. From these plots, we see that the trends in variation of curvature with thickness ratio depend strongly on stiffness ratio. In particular, when the stiffness ratio is small ($M_f/M_s=0.1$), the influence of the different electromechanical effects on the normalized stretching strains and curvatures is more prominent. With the increase in the relative stiffness of the film ($M_f/M_s=1.0$, to $1.5$), the normalized stretching strain and curvature follows a monotonically decreasing trend with increasing thickness ratio, and the distinction among the different electromechanical cases for the normalized stretching strains and curvatures vanishes.

\begin{figure}[h!]\centering
\subfigure[$M_f/M_s=0.1$]{\includegraphics[keepaspectratio=true,width=0.40\textwidth]{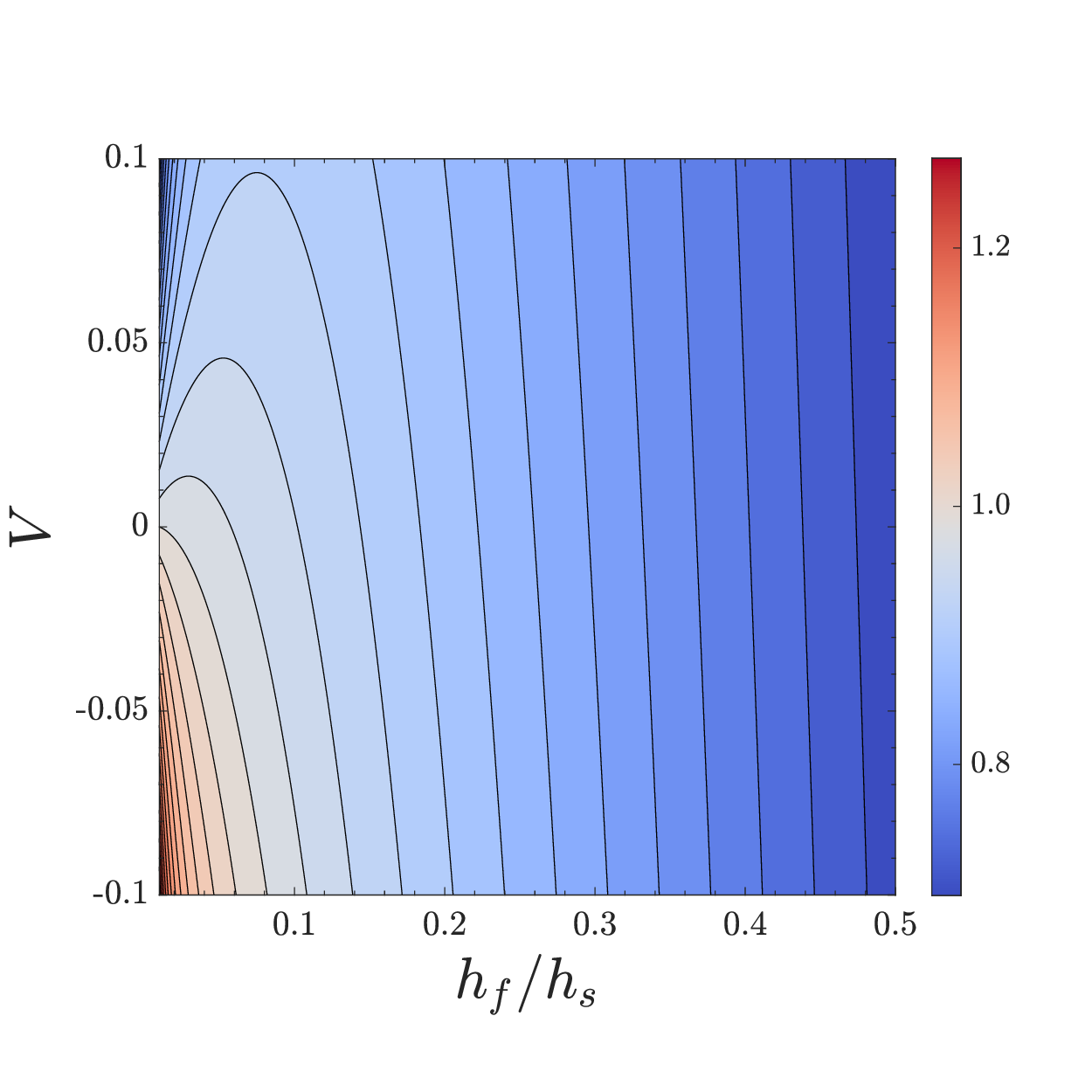}}
\subfigure[$M_f/M_s=1.0$]{\includegraphics[keepaspectratio=true,width=0.40\textwidth]{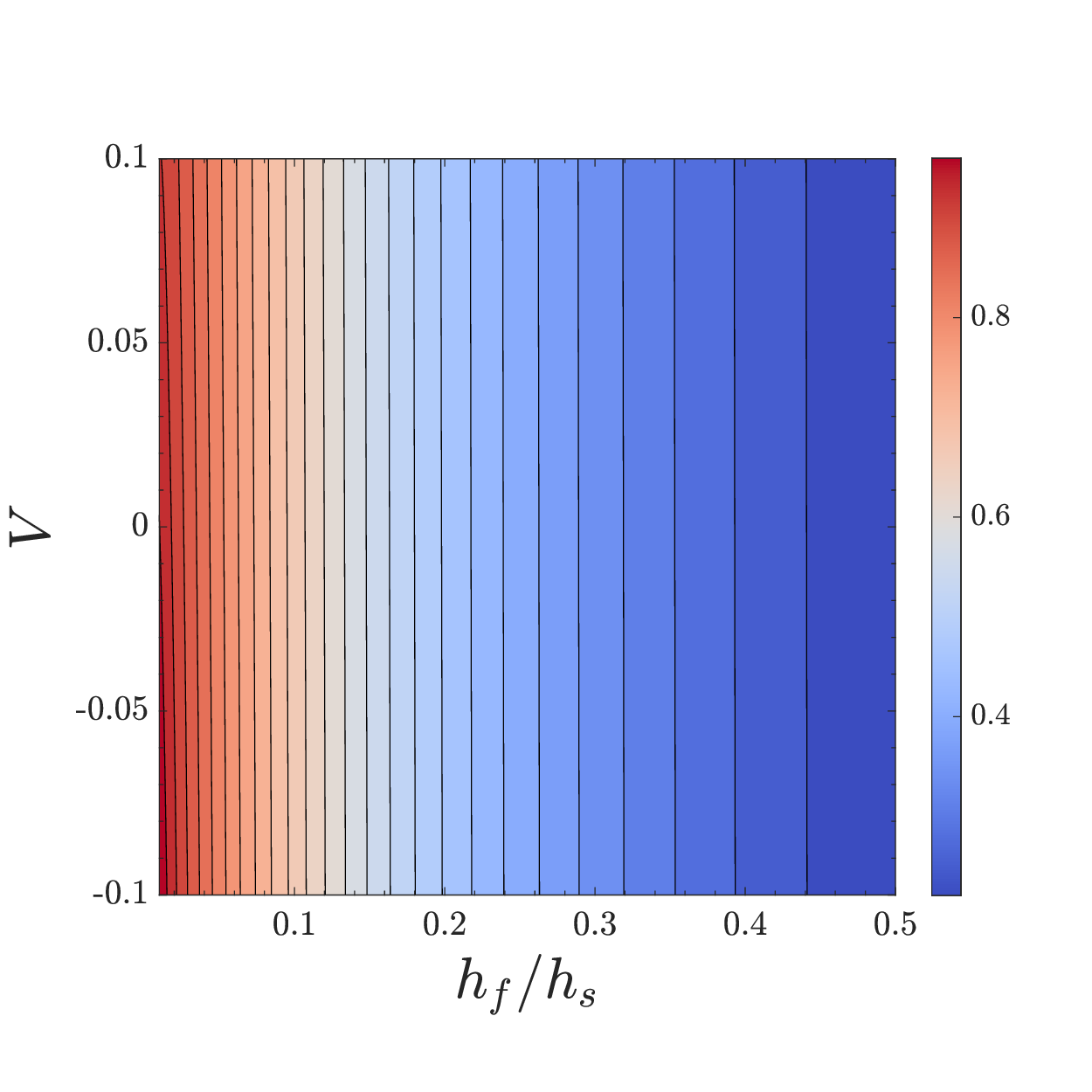}}\\
\subfigure[$M_f/M_s=0.1$]{\includegraphics[keepaspectratio=true,width=0.40\textwidth]{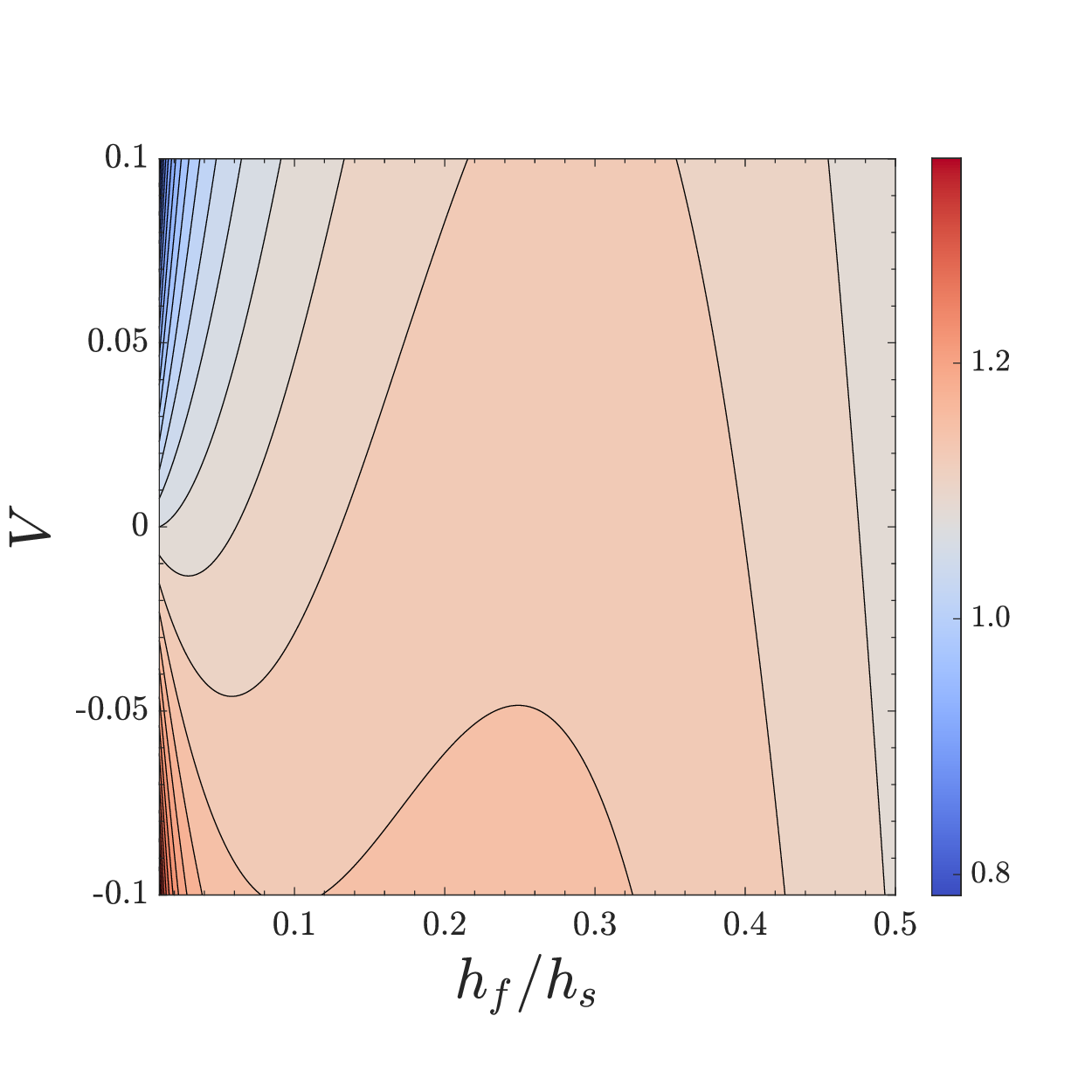}}
\subfigure[$M_f/M_s=1.0$]{\includegraphics[keepaspectratio=true,width=0.40\textwidth]{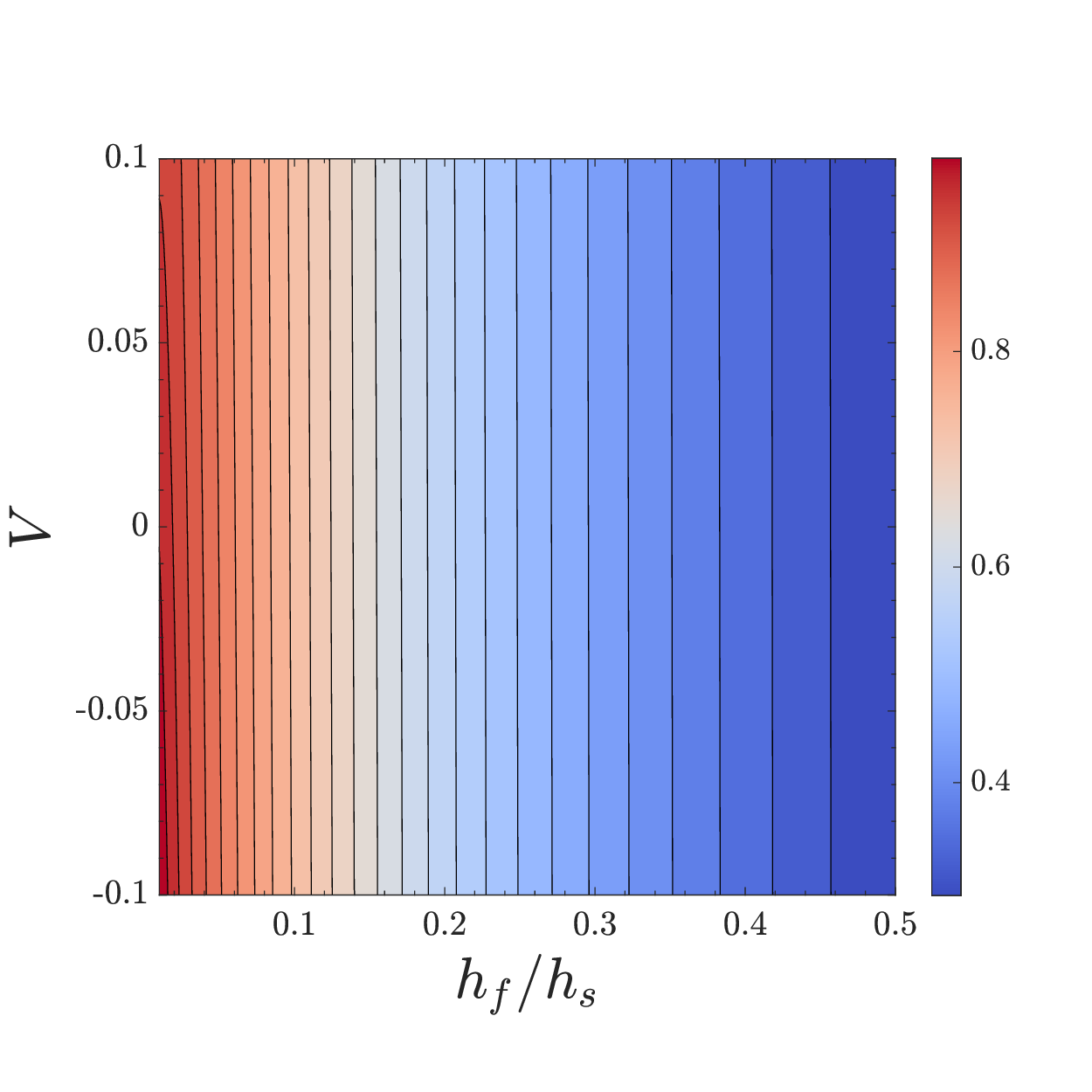}}
{\caption{Contour plot showing the effect of potential on the normalized stretching strains in (a) - (b) and normalized curvatures in (c) - (d) for the converse case shown for different ratios of the thickness of film and substrate.}\label{Fig:ConvStrainVeffect}}
\end{figure}

The effect of applied voltage $V$ on the normalized stretching strain (Equation \ref{Eq:strain:conv:pf}) and curvature (Equation \ref{Eq:kappa:conv:pf}) for the converse case is shown in Figure \ref{Fig:ConvStrainVeffect}. From these contour plots, it is clear that the effect of applied voltage on the normalized stretching strain and curvature is more pronounced when the relative film thickness and relative film stiffness are small.

\section{Non-uniform elastic mismatch strain and elastic properties}\label{Sec:Nonuniform}
\begin{figure}[h!]
\centering
\includegraphics[keepaspectratio=true,width=0.75\textwidth]{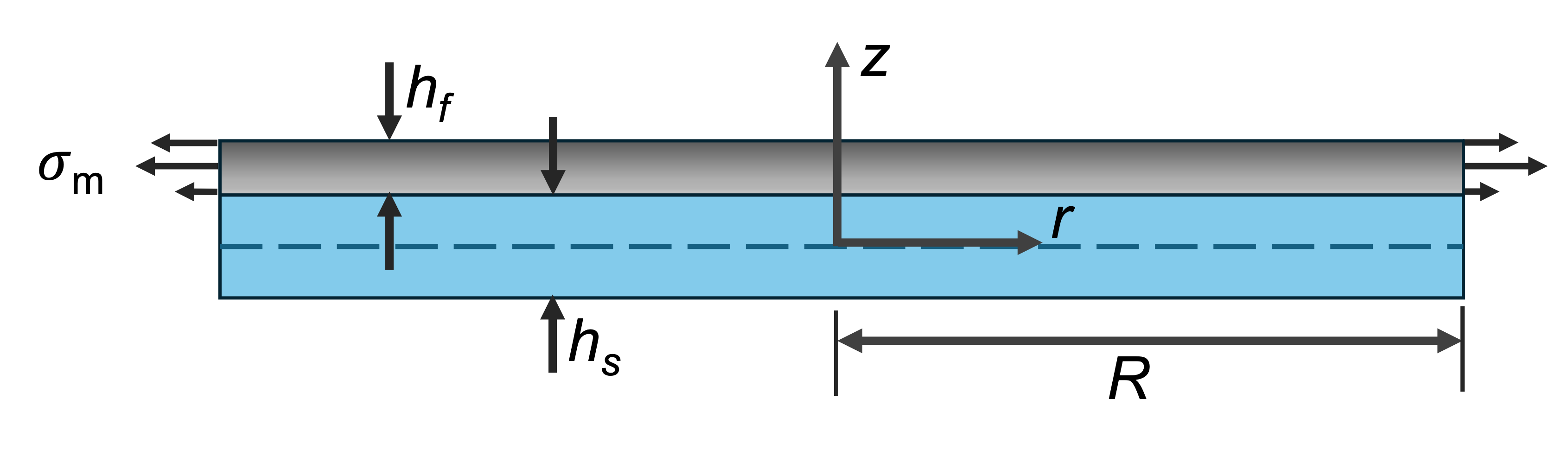}\label{Fig:NonUniformGeometry}
{\caption{A film substrate system with non-uniform film property and non-uniform elastic mismatch strain. A non-uniform stress is applied to the film to maintain the flat configuration. Upon removing the stress, the system deforms to a bent configuration.}\label{Fig:NonUniformGeometry}}
\end{figure}

In Section \ref{Sec:Uniform}, we derived the expressions of curvature and the stretching strain for uniform elastic properties and mismatch strain. We now proceed to the case where the mismatch strain and the film's elastic and electrical properties vary across the thickness of the film. The elastic properties of the substrate are assumed to be constant. In this case, the bending modulus of the film and substrate is
\begin{eqnarray}
    M(z) = \begin{cases} 
   M^f(z) = C_{11}^f(z) + C_{12}^f(z)  \,\, ,  \frac{h_s}{2} < z < \frac{h_s}{2}+ h_f\\
   M^s = C_{11}^s + C_{12}^s \,\,,  \,\,\,\,\,\,\,\,\,\,\, -\frac{h_s}{2} < z < \frac{h_s}{2} \\
    \end{cases}
\end{eqnarray}

Figure \ref{Fig:NonUniformGeometry} shows this film substrate system. Similar to our discussion in Section \ref{Sec:Uniform}, we consider two cases i.e. \emph{closed circuit} configuration corresponding to the converse effect and \emph{open circuit} configuration corresponding to the direct effect. Furthermore, the deformation is assumed to be small, the radial and the out-of-plane deformations are uncoupled, and therefore the radial and the transverse displacement fields are given by Equation \ref{Eq:displacementfield}. The elastic mismatch strain $\varepsilon_m(z)$ varies with thickness, and the strains in the bilayer are
\begin{equation}\label{Eq:bilayerstrainnonuniform}
    \varepsilon_{rr} =\varepsilon_{\theta\theta} = \begin{cases} 
    \varepsilon_0 - kz + \varepsilon_m(z) \,\, ,  \frac{h_s}{2} < z < \frac{h_s}{2}+ h_f\\
    \varepsilon_0 - kz \,\,,  \,\,\,\,\,\,\,\,\, -\frac{h_s}{2} < z < \frac{h_s}{2} \\
    \end{cases}
\end{equation}

We also assume that the piezoelectric constants $e_{15}(z)$ $e_{31}(z)$ and $e_{33}(z)$, the components of the electric permittivity tensor $k_{11}(z)$, $k_{22}(z)$, $k_{33}(z)$ and the components of the flexocoupling tensor $\mu_{11}(z)$, $\mu_{12}(z)$, $\mu_{44}(z)$ in the film to be varying across the thickness of the film. 

Using these assumptions and Equation \ref{Eqn:stress}, the non-zero components of the stress tensor in the film are
\begin{eqnarray}\label{Eq:stressnonuniformfilm}
& &\sigma_{rr}(z) = C_{11}^f(z) \varepsilon_{rr}(z) + C_{12}^f(z) \varepsilon_{\theta \theta}(z) - e_{31}(z) E_z(z) + \mu_{12}(z) \frac{\partial E_{z}}{\partial z} + \mu_{11}(z) \frac{\partial E_{r}}{\partial r} \,\,, \\
& &\sigma_{\theta \theta}(z) = C_{12}^f(z) \varepsilon_{rr}(z) + C_{11}^f(z) \varepsilon_{\theta \theta}(z) - e_{31}(z) E_z (z) + \mu_{12}(z) \frac{\partial E_{z}}{\partial z} + \mu_{12}(z) \frac{\partial E_{r}}{\partial r} \,\,, \\
& &\sigma_{r z}(z) = -e_{15}(z) E_r(z) \,\,,
\end{eqnarray}
and the non-zero components of the stress tensor in the substrate is
\begin{eqnarray}\label{Eq:stressnonuniformsubstrate}
& &\sigma_{rr}(z) = C_{11}^s \varepsilon_{rr}(z) + C_{12}^s \varepsilon_{\theta \theta}(z) \,\,,\\
& &\sigma_{\theta \theta}(z) = C_{12}^s \varepsilon_{rr}(z) + C_{11}^s \varepsilon_{\theta \theta}(z) \,\,.
\end{eqnarray}

Using Equation \ref{Eqn:displacement}, the non-zero components of the electric displacement in the film are
\begin{eqnarray}
& & D_r = k_{11}(z) E_r + \mu_{11}(z) \frac{\partial \varepsilon_{rr}}{\partial r} + \mu_{12}(z) \frac{\partial \varepsilon_{\theta \theta}}{\partial r}\,\,, \\
& & D_z = e_{31}(z) \varepsilon_{rr}(z) + e_{31}(z) \varepsilon_{\theta\theta}(z) + k_{33}(z) E_z(z) + \mu_{12}(z)\frac{\partial \varepsilon_{rr}}{\partial z} + \mu_{12}(z)\frac{\partial \varepsilon_{\theta \theta}}{\partial z} \,\,, \label{Eq:displacementnonuniformfilm}
\end{eqnarray} 
and the electric displacement in the substrate is zero, as piezoelectric and flexoelectric effects in the substrate are absent.

Using, Equations \ref{Eq:stressnonuniformfilm} - \ref{Eq:displacementnonuniformfilm} and \ref{Eq:ElectricFieldPotential}, the equilibrium equation for stress $\nabla\cdot \bm{\sigma} = 0$ and the electric displacement $\nabla \cdot {\bf{D}} = 0$ gives the following set of partial differential equations for the electrostatic potential $\phi$ in the film.

\begin{eqnarray}\label{Eq:EquilibriumNonuniform1}
 &&   \biggl(e_{31}(z) + e_{15}(z)\biggr) \frac{\partial ^2 \phi}{\partial r \partial z} - \frac{1}{r} \biggl(\mu_{11}(z) - \mu_{12}(z)\biggr) \frac{\partial^2 \phi}{ \partial r^2} \nonumber \\ &&+ \frac{\partial e_{15}}{\partial z} \frac{\partial \phi}{\partial r} - \mu_{12}(z) \frac{\partial^3 \phi}{\partial r \partial^2 z} - \mu_{12}(z) \frac{\partial^3 \phi}{\partial r^3} = 0 \\
   &&  \frac{\partial ^2 \phi}{\partial r^2} + \frac{1}{r} \frac{\partial \phi}{\partial r} = 0 \,\,, \\
   &&-k_{33} \frac{\partial ^2 \phi}{\partial z^2} - \frac{\partial k_{33}}{\partial z} \frac{\partial \phi}{\partial z} + 2e_{31}(z) \left(\frac{\partial \varepsilon_m(z)}{\partial z}-\kappa\right) + 2 \frac{\partial e_{31}}{\partial z} \varepsilon_{rr}^f(z) \nonumber \\ &&+ 2 \frac{\partial \mu_{12}}{\partial z} \left(\frac{\partial \varepsilon_m(z)}{\partial z}-\kappa\right) + 2 \mu_{12} (z) \frac{\partial^2 \varepsilon_{m}}{\partial z^2} = 0 \,\,.
\end{eqnarray}
For closed circuit configuration, these set of equations are first solved for the electrostatic potential with the boundary conditions $\phi(\frac{h_s}{2}) = V$ and $\phi(h_f+\frac{h_s}{2})=0$, and then we use  Equation \ref{Eq:ElectricFieldPotential} to obtain the corresponding electric field for the converse case. 


For open circuit configuration, the electric field corresponding to the direct effect is obtained by setting ${{\bf{D}} \cdot \bm{e}_z} = 0$, where $\bf{D}$ is given in Equation \ref{Eq:displacementnonuniformfilm}. In this case, the analytical expression for the electric field is
\begin{eqnarray} \label{Eq:ElectricFleldDirNonuniform}
    & & E_z^{Dir}(z) = \frac{2 e_{31}(z)}{k_{33}(z)} \left[ \kappa \left( z + \frac{\mu_{12}(z)}{e_{31}(z)} \right) - \biggl(\varepsilon_{0} + \varepsilon_{m}(z)\biggr) - \frac{\mu_{12}(z)}{e_{31}(z)} \frac{\partial \varepsilon_m}{\partial z}\right] \,\,, \nonumber \\
    & & E_r^{Dir}(r) = 0 \,\,.
\end{eqnarray}
We remark that the electric fields in both converse and direct cases depend on $\kappa$ and $\varepsilon_0$.
We use the electric field and strains to specify the enthalpy densities of the bilayer system with non-uniform film properties. The elastic enthalpy density is
\begin{eqnarray}\label{Eq:elasticenthalpydensity}
& &   \psi_{elastic} ({\bm{\varepsilon}}) = \begin{cases} 
    M^f (z) (\varepsilon_{rr}(z))^2 = M^f(z) (\varepsilon_0 - \kappa z + \varepsilon_m(z))^2 \,\, ,  \frac{h_s}{2} < z < \frac{h_s}{2}+ h_f\\
    M^s (z) (\varepsilon_{rr}(z))^2 = M^f(z) (\varepsilon_0 - \kappa z )^2 \,\,, -\frac{h_s}{2} < z < \frac{h_s}{2} \\
\end{cases} 
\end{eqnarray}
and the contribution of elasticity to the enthalpy of the system is
\begin{eqnarray}
\Pi_{elastic} (\varepsilon_0,\kappa) &=& \int_{0}^R \int_{-\frac{h_s}{2}}^{\frac{h_s}{2}+h_f} \psi_{elastic} ({\bm{\varepsilon}}) \,\, 2\pi r \mathrm{d}r \mathrm{d}z \,\,\nonumber \\
&=& \pi R^2 \left( I_{00} \varepsilon_{0}^2 + I_{01} + I_{20} \kappa^2 - 2 I_{10}\kappa \varepsilon_0  - 2 I_{11} \kappa  +2 I_{01} \varepsilon_0 \right) \,\,,
\end{eqnarray}
where the integrals $I_{pq}$ are given by
\begin{equation}\label{Eq:IntegralI}
    I_{pq} = \int_{-\frac{h_s}{2}}^{\frac{h_s}{2}+h_f} M(z)\,z^p \,\varepsilon_m^q(z) \,\mathrm{d}z \,\,.
\end{equation}

The dielectric enthalpy density in the film is
\begin{eqnarray}
& &\psi_{di}^{(\cdot)} ({\bf E}) = -\frac{1}{2} k_{33} (z)\left(E_z^{(\cdot)}\right)^2\,\,, 
\end{eqnarray}
where the superscript $^{(\cdot)}$ denotes converse or direct cases. The dielectric contribution to the total enthalpy is
\begin{eqnarray}
\Pi_{di}^{(\cdot)}(\varepsilon_0,\kappa) = \int_{0}^R \int_{\frac{h_s}{2}}^{\frac{h_s}{2}+h_f} \psi_{di}^{(\cdot)} ({\bf E}) \,\, 2\pi r \,\, \mathrm{d}r \mathrm{d}z 
= -\frac{1}{2}\pi R^2 K^{(\cdot)}(\varepsilon_0,\kappa) \,\,,
\end{eqnarray}
where the superscript $(\cdot)$ denotes the direct or converse cases and $K^{(\cdot)}$ is
\begin{equation}\label{Eq:IntegralK}
    K^{(\cdot)}(\varepsilon_0,\kappa) = \int_{\frac{h_s}{2}}^{\frac{h_s}{2}+h_f} k_{33} (z) \biggl(E^{(\cdot)}_z(z)\biggr)^2 \,\, \mathrm{d}z \,\,.
\end{equation}
As the electric field depends on  $\varepsilon_0$ and $\kappa$, $ K^{(\cdot)}$ also depends on these variables.

The piezoelectric enthalpy density in the film is
\begin{eqnarray}
& &\psi_{piezo}^{(\cdot)} ({\bm {\varepsilon}}, {\bf E}) = -2e_{31}(z)\,E_z^{(\cdot)}(z) \,\varepsilon_{rr}(z) \,\,,
\end{eqnarray}
and the piezoelectric contribution to the total enthalpy is given by
\begin{eqnarray}
\Pi_{piezo}^{(\cdot)}(\varepsilon_0,\kappa) &=& \int_{0}^R \int_{\frac{h_s}{2}}^{\frac{h_s}{2}+h_f} \psi_{piezo}^{(\cdot)} ({\bf {\varepsilon}}, {\bf E}) \,\, 2\pi r \, \mathrm{d}r \mathrm{d}z \,\, \nonumber \\
&=& -2 \pi R^2 \biggl[ J_{00}^{(\cdot)}(\varepsilon_0,\kappa) \varepsilon_0 - J_{10}^{(\cdot)}(\varepsilon_0,\kappa) \kappa + J_{01}^{(\cdot)}(\varepsilon_0,\kappa) \biggr]
\end{eqnarray}
where
\begin{equation}\label{Eq:IntegralJ}
J_{pq}^{(\cdot)}(\varepsilon_0,\kappa) = \int_{\frac{h_s}{2}}^{\frac{h_s}{2}+h_f} e_{31}(z) E_z^{(\cdot)}(z) z^p \varepsilon_m^q (z) \,\, \mathrm{d}z \,\,.
\end{equation}
Finally, the flexoelectric enthalpy density for the converse and direct cases is
\begin{eqnarray}
& & \psi_{flexo}^{Conv} ({\bm \varepsilon},  \nabla {\bf E}) = 2\mu_{12}  \varepsilon_{rr} 
 \frac{\partial  E_z^{Conv}}{\partial z} \,\,,\\
 & & \psi_{flexo}^{Dir} ( \nabla {\bm \varepsilon}, {\bf E}) = -2\mu_{12} \frac{\partial \varepsilon_{rr}}{\partial z} E_z^{Dir} \,\,.
\end{eqnarray}
The contribution of flexoelectricity to the total enthalpy for the converse case is
\begin{eqnarray}
    \Pi_{flexo}^{Conv}(\varepsilon_0,\kappa) &=& \int_{0}^R \int_{\frac{h_s}{2}}^{\frac{h_s}{2}+h_f} \psi_{flexo}^{Conv} ({\nabla \bm \varepsilon}, {\bf E}) \,\, 2\pi r \,\mathrm{d}r \mathrm{d}z \nonumber \\
    &=& 2 \pi R^2  \biggl[ M_{00}^{Conv}(\varepsilon_0,\kappa) \varepsilon_0  - M_{10}^{Conv} (\varepsilon_0,\kappa) \kappa + M_{01}^{Conv}(\varepsilon_0,\kappa) \biggr] \,\,
\end{eqnarray}
for the direct case is
\begin{eqnarray}
    \Pi_{flexo}^{Dir}(\varepsilon_0,\kappa) = \int_{0}^R \int_{\frac{h_s}{2}}^{\frac{h_s}{2}+h_f} \psi_{flexo}^{Dir} ({\nabla \bm \varepsilon}, {\bf E}) \,\, 2\pi r \,\mathrm{d}r \mathrm{d}z 
    = 2 \pi R^2  \biggl[L_0^{Dir}(\varepsilon_0,\kappa) \kappa - L_1^{Dir}(\varepsilon_0,\kappa) \biggr] \,\,,
\end{eqnarray}
where the integrals $L_p^{(\cdot)}$ and $M_{pq}^{(\cdot)}$ are given by
\begin{eqnarray}\label{Eq:IntegralL}
 & &  L_p^{(\cdot)} (\varepsilon_0,\kappa) =  \int_{\frac{h_s}{2}}^{\frac{h_s}{2}+h_f} \mu_{12} (z) E_z^{(\cdot)} (z) \left( \frac{\mathrm{d} \varepsilon_m}{\mathrm{d} z} \right)^p dz  \\
 & &  M_{pq}^{(\cdot)} (\varepsilon_0,\kappa) = \int_{\frac{h_s}{2}}^{\frac{h_s}{2}+h_f} \mu_{12}(z) \frac{\partial E_z^{(\cdot)}}{\partial z} z^p \varepsilon_m^q(z) \, dz \,\,. \label{Eq:IntegralM}
\end{eqnarray}
The total enthalpy of the system is given by the sum of elastic, dielectric, piezoelectric, and flexoelectric contributions (Equation \ref{Eq:TotalEnthalpy}). Setting the derivatives of the total enthalpy taken with respect to $\varepsilon_0$ and $\kappa$ to zero (Equation \ref{Eq:simultaneous}) and solving the two coupled equations gives us the values of $\varepsilon_0$ and $\kappa$. To demonstrate, we solve this for elastic mismatch strain with constant gradient and constant elastic and electrical properties in the next section.

\subsection{Linearly increasing elastic mismatch strain and constant material properties}
We consider the case where the elastic mismatch strain in the film increases linearly with thickness. This is given by
\begin{equation}\label{Eq:emspecialcase}
    \varepsilon_m(z) = \frac{\varepsilon_t}{h_f} \left(z - \frac{h_s}{2} \right) \,\,\,,
\end{equation}
where $\varepsilon_t$ is the mismatch strain on the film surface ($z=h_f+\frac{h_s}{2}$). The mismatch strain at the bottom of the film surface ($z=\frac{h_s}{2}$) is zero. We also assume the elastic, dielectric, piezoelectric, and flexoelectric properties to be constant. The average elastic mismatch strain is $<\varepsilon_m> = \frac{\varepsilon_t}{2}$

In this case, the equilibrium Equations \ref{Eq:EquilibriumNonuniform1} simplifies to the following 
\begin{eqnarray}\label{Eq:EquilibriumNonuniformSpecial}
     \frac{\partial ^2 \phi}{\partial r \partial z} = 0 \,,\,\,
    \frac{\partial ^2 \phi}{\partial r^2} + \frac{1}{r} \frac{\partial \phi}{\partial r} = 0 \,\,, \text{and} \,\,
   k_{33} \frac{\partial ^2 \phi}{\partial z^2} + 2e_{31} \left(\kappa - \frac{\varepsilon_t}{h_f} \right) = 0 \,\,.
\end{eqnarray}
Solving the above set of equations for the electrostatic potential $\phi$ with boundary conditions  $\phi(\frac{h_s}{2}) = V$ and $\phi(h_f+\frac{h_s}{2})=0$ for the closed circuit configuration and calculating the electric field for the converse case, we obtain the following
\begin{eqnarray}\label{Eq:ElectricFieldConverseSpecial}
 &&   E_z^{Conv}(z) = \frac{2 e_{31}}{k_{33}} \left( \kappa - \frac{\varepsilon_t}{h_f} \right) \left(z - \frac{h_s+h_f}{2}\right) + \frac{V}{h_f} \nonumber \\
 &&   E_r^{Conv}(r) = 0 \,\,. 
\end{eqnarray}

Again, from Equations \ref{Eq:ElectricFleldDirNonuniform} and \ref{Eq:emspecialcase}, the electric field for the \emph{direct} effect case is 
\begin{eqnarray}\label{Eq:ElectricFieldDirectSpecial}
    & & E_z^{Dir}(z) = \frac{2 e_{31}}{k_{33}} \left[ \kappa \left( z + \frac{\mu_{12}}{e_{31}} \right) - (\varepsilon_{0} + \varepsilon_{m}) - \frac{\mu_{12}}{e_{31}} \frac{\varepsilon_t}{h_f}\right] \,\,, \nonumber \\
    & & E_r^{Dir}(r) = 0 \,\,. 
\end{eqnarray}

We use the electric fields and strains to calculate the integrals (\ref{Eq:IntegralI}), (\ref{Eq:IntegralK}), (\ref{Eq:IntegralJ}), (\ref{Eq:IntegralL}), and (\ref{Eq:IntegralM}). These integrals are used to calculate the total enthalpy (\ref{Eq:TotalEnthalpy}) for the converse and direct cases. As the deformation of the mid-plane of the film-substrate system minimizes the total enthalpy of the system \cite{Freund:Book}, we set the derivative of the total enthalpy with stretching strain and curvature to zero as in Equation \ref{Eq:simultaneous} and solve this set of equations, giving us the stretching strain and the curvature. 

We remark that for this case of varying elastic mismatch strain with constant gradient, the stretching  strain $\varepsilon_0^{st}$ and curvature $\kappa^{st}$ with Stoney's approximation \cite{Freund:Book,Stoney} are given by
\begin{eqnarray} \label{stoney:nonuniform1}
    \varepsilon_0^{st} &=& -\frac{h_f M_f \varepsilon _t}{2 h_s M_s} \,\,,\\
    \kappa^{st} &=& \frac{3 h_f M_f \varepsilon _t}{h_s^2 M_s} \,\,.\label{stoney:nonuniform2}
\end{eqnarray}
The cases considering different electromechanical effects are discussed next. 

\paragraph{Case I: both piezoelectric and flexoelectric effects are present} We present expressions of the stretching strain and curvature for the film-substrate system when both piezoelectric and flexoelectric effects are present in the film. The stretching  strain and curvatures are respectively normalized by $\varepsilon_0^{st}$ in (\ref{stoney:nonuniform1}) and $\kappa^{st}$ in (\ref{stoney:nonuniform2}), respectively. 
For the converse case, the stretching  strain is
\begin{eqnarray}\label{Eq:strain:conv:pf:NU}
&& \frac{\varepsilon_{0,pf}^{Conv}}{\varepsilon_0^{st}} \nonumber \\ &&= \frac{\splitfrac{-h_s M_s \biggl\{4 V e_{31}^3 h_f^3 k_{33}+h_f h_s k_{33}^2 M_f (h_f^2 M_f-h_s^2 M_s) \varepsilon _t+2 e_{31}
k_{33} (V h_f^3 k_{33} M_f+h_s^3 M_s(V k_{33}+2 \varepsilon _t \mu _{12}))\mathstrut}{+2 e_{31}^2 h_f \biggl(h_f^2 h_s k_{33} M_f \varepsilon
_t-12 V h_f k_{33} \mu _{12}-12 h_s \mu _{12} (V k_{33}+2 \varepsilon _t \mu _{12}) \biggr) \biggr\}}}{\splitfrac{h_f M_f \varepsilon_t \biggl\{k_{33}^2
(h_f^4 M_f^2+4 h_f^3 h_s M_f M_s+6 h_f^2 h_s^2 M_f M_s+4 h_f h_s^3 M_f M_s+h_s^4 M_s^2)\mathstrut}{-24 e_{31} h_f h_s (h_f+h_s) k_{33}
M_s \mu _{12}+2 e_{31}^2 h_f^2 (h_f^2 k_{33} M_f+h_f h_s k_{33} M_s-24 \mu _{12}^2) \biggr\} }}\,\,, \nonumber \\
\end{eqnarray}
and the curvature is
\begin{eqnarray}\label{Eq:kappa:conv:pf:NU}
 \frac{\kappa_{pf}^{Conv}}{\kappa^{st}} \nonumber  = \frac{\splitfrac{h_s^2 M_s \biggl\{h_f k_{33}^2 M_f (h_f^2 M_f+4 h_f h_s M_s+3 h_s^2 M_s) \varepsilon _t+2 e_{31}^2 h_f (h_f^2
k_{33} M_f \varepsilon _t+h_f h_s k_{33} M_s \varepsilon _t\mathstrut}{-12 \mu _{12} (V k_{33}+2 \varepsilon _t \mu _{12}))-6 e_{31} h_s k_{33} M_s \biggl(h_s
(V k_{33}+2 \varepsilon _t \mu _{12})+h_f (V k_{33}+4 \varepsilon _t \mu _{12}) \biggr) \biggr\}}}{\splitfrac{3 h_f M_f \varepsilon_t
\biggl\{k_{33}^2 (h_f^4 M_f^2+4 h_f^3 h_s M_f M_s+6 h_f^2 h_s^2 M_f M_s+4 h_f h_s^3 M_f M_s+h_s^4 M_s^2)\mathstrut}{-24 e_{31} h_f h_s (h_f+h_s)
k_{33} M_s \mu _{12}+2 e_{31}^2 h_f^2 (h_f^2 k_{33} M_f+h_f h_s k_{33} M_s-24 \mu _{12}^2) \biggr\}}} \,\,. \nonumber \\
\end{eqnarray}
For the direct case, the stretching  strain is
\begin{eqnarray}\label{Eq:strain:dir:pf:NU}
&&\frac{\varepsilon_{0,pf}^{Dir}}{\varepsilon_0^{st}} \nonumber \\&& = \frac{\splitfrac{h_s^2 M_s \biggl\{-4 e_{31}^4 h_f^3+2 e_{31}^2 h_f k_{33} (-2 h_f^2 M_f+h_s^2 M_s )+4 e_{31} h_s^2 k_{33} M_s \mu
_{12}\mathstrut}{+h_f k_{33} M_f (-h_f^2 k_{33} M_f+h_s^2 k_{33} M_s-24 \mu_{12}^2) \biggr\}}}{\splitfrac{h_f M_f \biggl\{4 e_{31}^4 h_f^4+4 e_{31}^2
h_f k_{33} (h_f^3 M_f+2 h_f^2 h_s M_s+3 h_f h_s^2 M_s+2 h_s^3 M_s )+24 e_{31} h_f h_s (h_f+h_s ) k_{33} M_s \mu _{12}\mathstrut}{+k_{33}
\biggl(h_f^4 k_{33} M_f^2+4 h_f^3 h_s k_{33} M_f M_s+h_s^4 k_{33} M_s^2+6 h_f^2 M_f (h_s^2 k_{33} M_s+4 \mu _{12}^2)+4 h_f h_s M_s(h_s^2
k_{33} M_f+6 \mu _{12}^2) \biggr) \biggr\}}}\,\,, \nonumber \\
\end{eqnarray}
and the curvature
\begin{eqnarray}\label{Eq:kappa:dir:pf:NU}
&& \frac{\kappa_{pf}^{Dir}}{\kappa^{st}} \nonumber \\&& = \frac{\splitfrac{h_s^2 M_s \biggl\{4 e_{31}^4 h_f^3+2 e_{31}^2 h_f k_{33} (2 h_f^2 M_f+4 h_f h_s M_s+3 h_s^2 M_s)+12 e_{31} h_s (2
h_f+h_s) k_{33} M_s \mu _{12}\mathstrut}{+k_{33} \biggl(h_f^3 k_{33} M_f^2+4 h_f^2 h_s k_{33} M_f M_s+24 h_s M_s \mu _{12}^2+3 h_f M_f (h_s^2 k_{33}
M_s+8 \mu _{12}^2) \biggr) \biggr\}}}{\splitfrac{3 h_f M_f \biggl\{4 e_{31}^4 h_f^4+4 e_{31}^2 h_f k_{33} (h_f^3 M_f+2 h_f^2 h_s M_s+3 h_f
h_s^2 M_s+2 h_s^3 M_s)+24 e_{31} h_f h_s (h_f+h_s) k_{33} M_s \mu _{12} \mathstrut}{+k_{33} \biggl(h_f^4 k_{33} M_f^2 +4 h_f^3 h_s k_{33} M_f M_s+h_s^4
k_{33} M_s^2+6 h_f^2 M_f (h_s^2 k_{33} M_s+4 \mu _{12}^2)+4 h_f h_s M_s (h_s^2 k_{33} M_f+6 \mu _{12}^2) \biggr) \biggr\}}} \,\,.\nonumber \\
\end{eqnarray}

\paragraph{Case II: piezoelectric, dielectric and flexoelectric effects are absent} The expressions of stretching strain and curvature of the film-substrate system in the absence of piezoelectric and flexoelectric effects are obtained from Equations \ref{Eq:strain:conv:pf:NU} - \ref{Eq:kappa:conv:pf:NU} or Equations \ref{Eq:strain:dir:pf:NU} - \ref{Eq:kappa:dir:pf:NU} by setting $e_{31} = 0$, and $\mu_{12} = 0$. In this case, the stretching strain $\varepsilon_0^*$ and the curvature $\kappa^*$ are solely due to elastic effects of the film and substrate, and have been derived by Freund and co-workers \cite{Freund:Book}. These are
\begin{eqnarray}
    \frac{\varepsilon_0^*}{\varepsilon_0^{st}} = \frac{h_s^2 M_s (-h_f^2 M_f+h_s^2 M_s)}{h_f^4 M_f^2+4 h_f^3 h_s M_f M_s+6 h_f^2 h_s^2 M_f M_s+4 h_f h_s^3 M_f M_s+h_s^4
M_s^2}\,\,,
\end{eqnarray}
and 
\begin{eqnarray}
    \frac{\kappa^*}{\kappa_0^{st}} = \frac{h_s^2 M_s (h_f^2 M_f+4 h_f h_s M_s+3 h_s^2 M_s )}{3 (h_f^4 M_f^2+4 h_f^3 h_s M_f M_s+6 h_f^2 h_s^2 M_f M_s+4
h_f h_s^3 M_f M_s+h_s^4 M_s^2)}\,\,.
\end{eqnarray}

\paragraph{Case III: flexoelectric effect is absent} The expressions of stretching strain and curvature of the film-substrate system in the absence of flexoelectric effect are obtained from Equations \ref{Eq:strain:conv:pf:NU} - \ref{Eq:kappa:dir:pf:NU} by setting $\mu_{12} = 0$. Therefore, the stretching strain and the curvature arise due to piezoelectricity and dielectricity of the film, and the combined effects of elasticity in the film and substrate. 

\paragraph{Case IV: piezoelectric effect is absent} Finally, we discuss the case when the piezoelectric effect in the film is absent. In this case, stretching strain and curvature of the system are only influenced by the film and substrate's elasticity and flexoelectric effects in the film. The strains and curvatures can be obtained by setting $e_{31}=0$ in Equations \ref{Eq:strain:conv:pf:NU} - \ref{Eq:kappa:dir:pf:NU}. For the converse case, we see that the strains and the curvatures are the same as those obtained in Case II, 
\begin{eqnarray}
    \varepsilon_{0,f}^{Conv} = \varepsilon_0^* \,\,\,, \kappa_{f}^{Conv} = \kappa^* \,\,.
\end{eqnarray}

\paragraph{Electric polarization in the film with non-uniform elastic mismatch strain} Finally, we calculate the electric polarization in the film using the relation ${\bf{P}} = e_0 \bf{E} + \bf{D}$. The polarization in the $z$-direction for the converse case is 
\begin{eqnarray}\label{Eq:PolarizationConvNonuni}
    && P_z^{Conv} (z) = (e_0 + k_{33}) E_z^{Conv} (z) + 2 e_{31} \varepsilon_{rr}(z) + 2\mu_{12} \frac{\partial \varepsilon_{rr}}{\partial z} \,\,, \nonumber \\
     &=& (e_0 + k_{33})\Bigg[  \frac{2e_{31}}{k_{33}} \left(\kappa - \frac{\varepsilon_t}{h_f} \right)\left( z - \frac{h_f+h_s}{2} \right) + \frac{V}{h_f} \Bigg ] + 2 e_{31} \left[\varepsilon_0 - \kappa z + \frac{\varepsilon_t}{h_f} \left( z-\frac{h_s}{2} \right) \right] - 2\mu_{12} \left( \kappa - \frac{\varepsilon_t}{h_f} \right)\,\,, \nonumber \\
\end{eqnarray}
and for the direct case, we obtain 
\begin{eqnarray}\label{Eq:PolarizationDirNonuni}
    && P_z^{Dir} (z) = (e_0 + k_{33}) E_z^{Dir} (z) + 2 e_{31} \varepsilon_{rr}(z) + 2\mu_{12} \frac{\partial \varepsilon_{rr}}{\partial z} \,\,, \nonumber \\
     &=& (e_0 + k_{33})\left( \frac{2 e_{31}}{k_{33}}\right)\left[ \kappa \left( z + \frac{\mu_{12}}{e_{31}} \right) - \left\{ \varepsilon_{0} +\frac{\varepsilon_t}{h_f}\left(z-\frac{h_s}{2}\right) \right\} \right]  + 2 e_{31} \left[\varepsilon_0 - \kappa z + \frac{\varepsilon_t}{h_f} \left( z-\frac{h_s}{2} \right) \right] - 2\mu_{12} \left( \kappa - \frac{\varepsilon_t}{h_f} \right)\,\,, \nonumber \\
\end{eqnarray}
The polarization in the radial and angular directions is zero.

Similar to our observation in the uniform case, we find that the polarization depends on the electromechanical constants, including the stiffness $M_f$ and $M_s$, as well as the thicknesses $h_f$ and $h_s$, in a complex manner.


\subsubsection{Discussion}
We discuss the effect of the ratios of thickness and stiffness on the normalized stretching strain and curvature, and compare these for the four cases. We take $\varepsilon_t=-0.0164$, and the other parameters used are listed in Table \ref{parametertable}.

\begin{figure}[h!]\centering
\subfigure[$\frac{\varepsilon_0^*}{\varepsilon_0^{st}}$]{\includegraphics[keepaspectratio=true,width=0.30\textwidth]{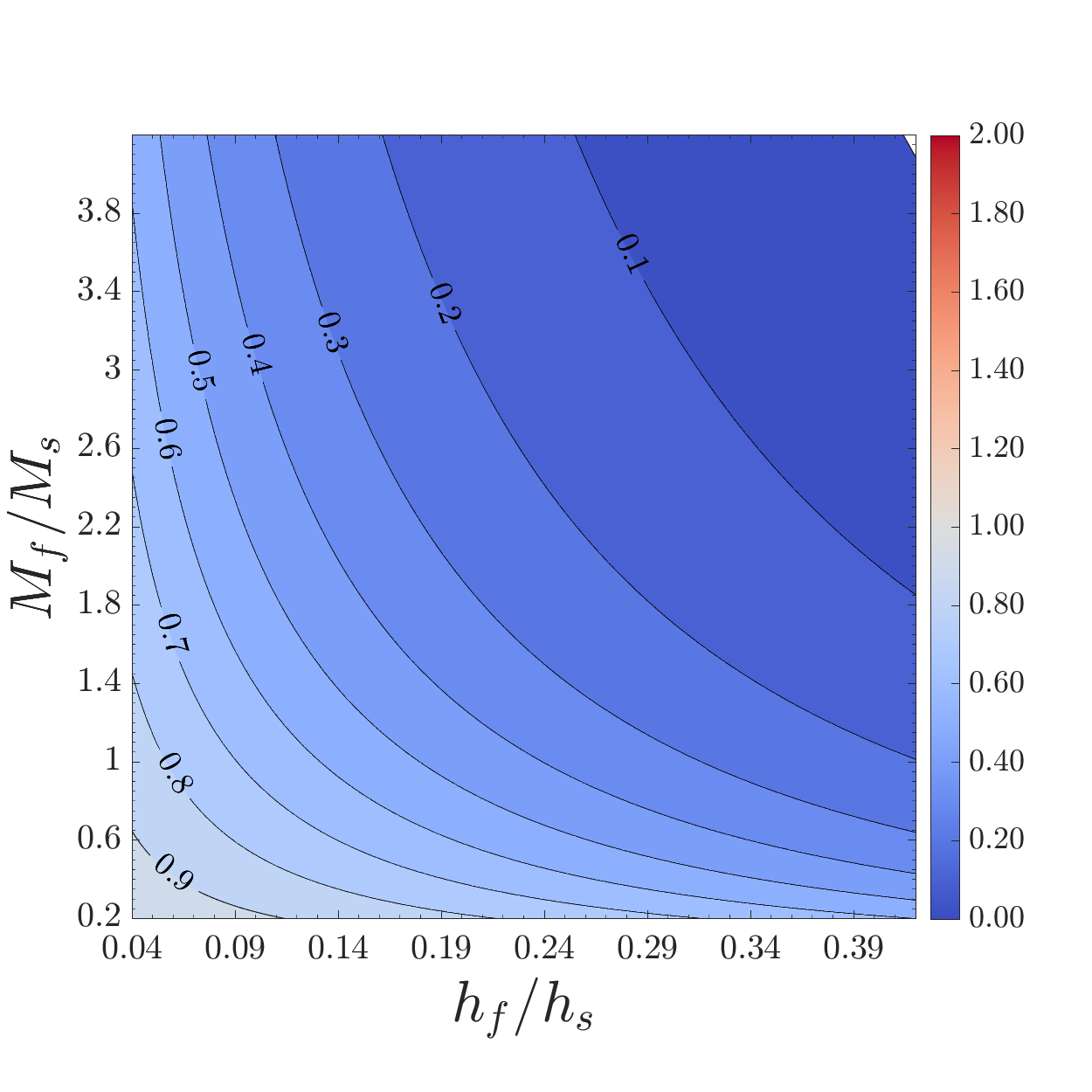}\label{Fig:Nonuni:e0:Star}}
\subfigure[$\frac{\varepsilon_{0,pf}^{Conv}}{\varepsilon_0^{st}}$]{\includegraphics[keepaspectratio=true,width=0.30\textwidth]{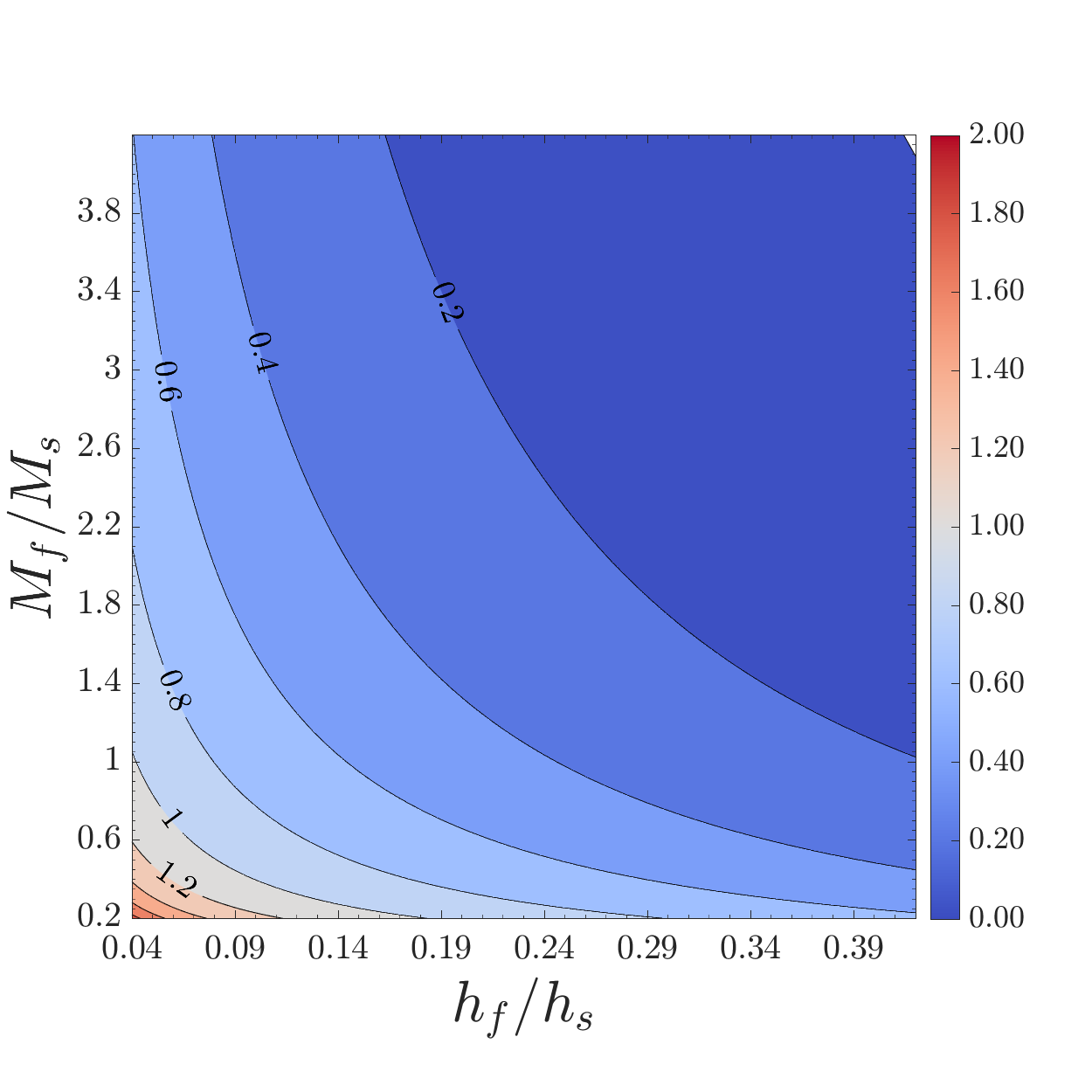}\label{Fig:Nonuni:e0:Conv}}
\subfigure[$\frac{\varepsilon_{0,pf}^{Dir}}{\varepsilon_0^{st}}$]{\includegraphics[keepaspectratio=true,width=0.30\textwidth]{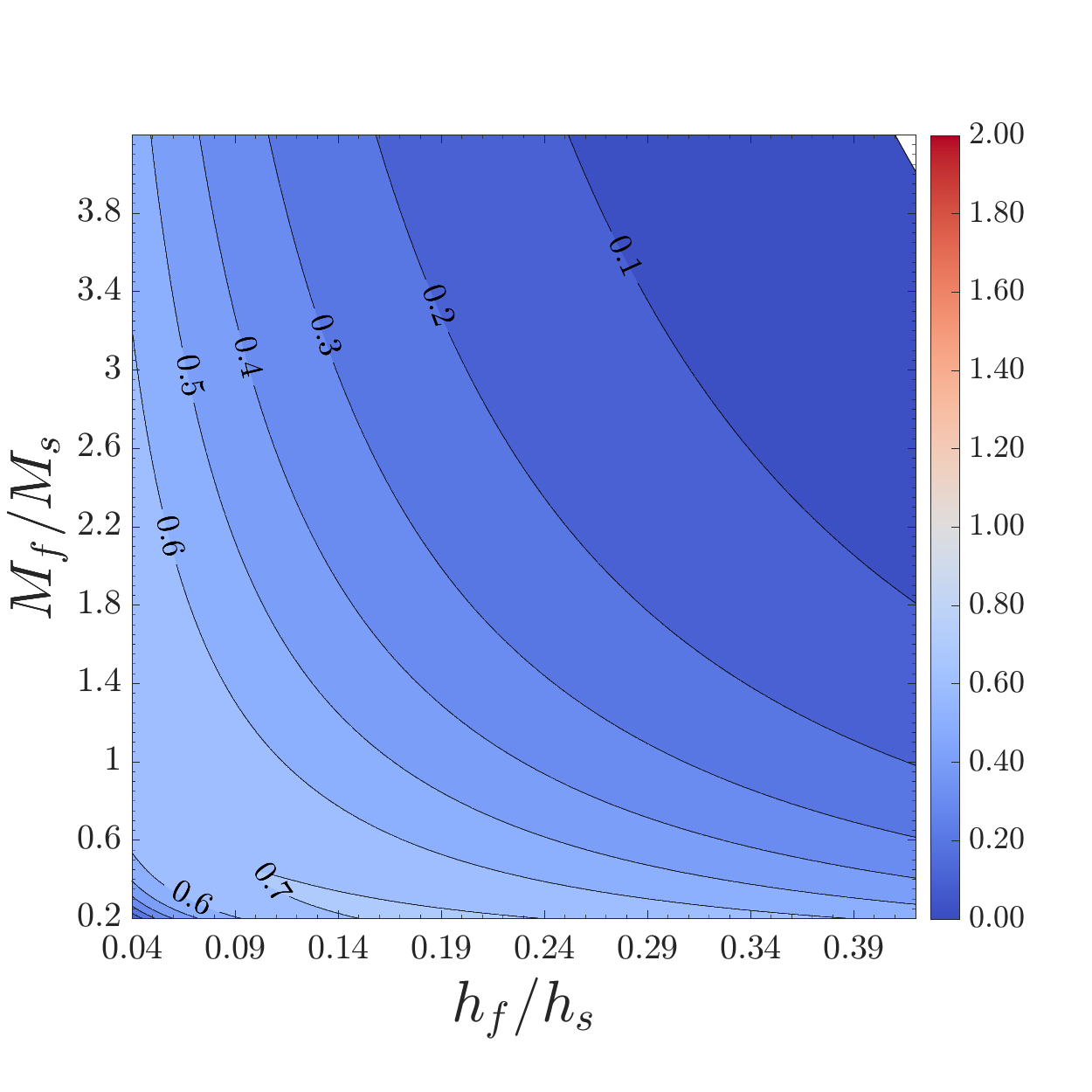}\label{Fig:Nonuni:e0:Dir}}\\
\subfigure[$\frac{\kappa^*}{\kappa^{st}}$]{\includegraphics[keepaspectratio=true,width=0.30\textwidth]{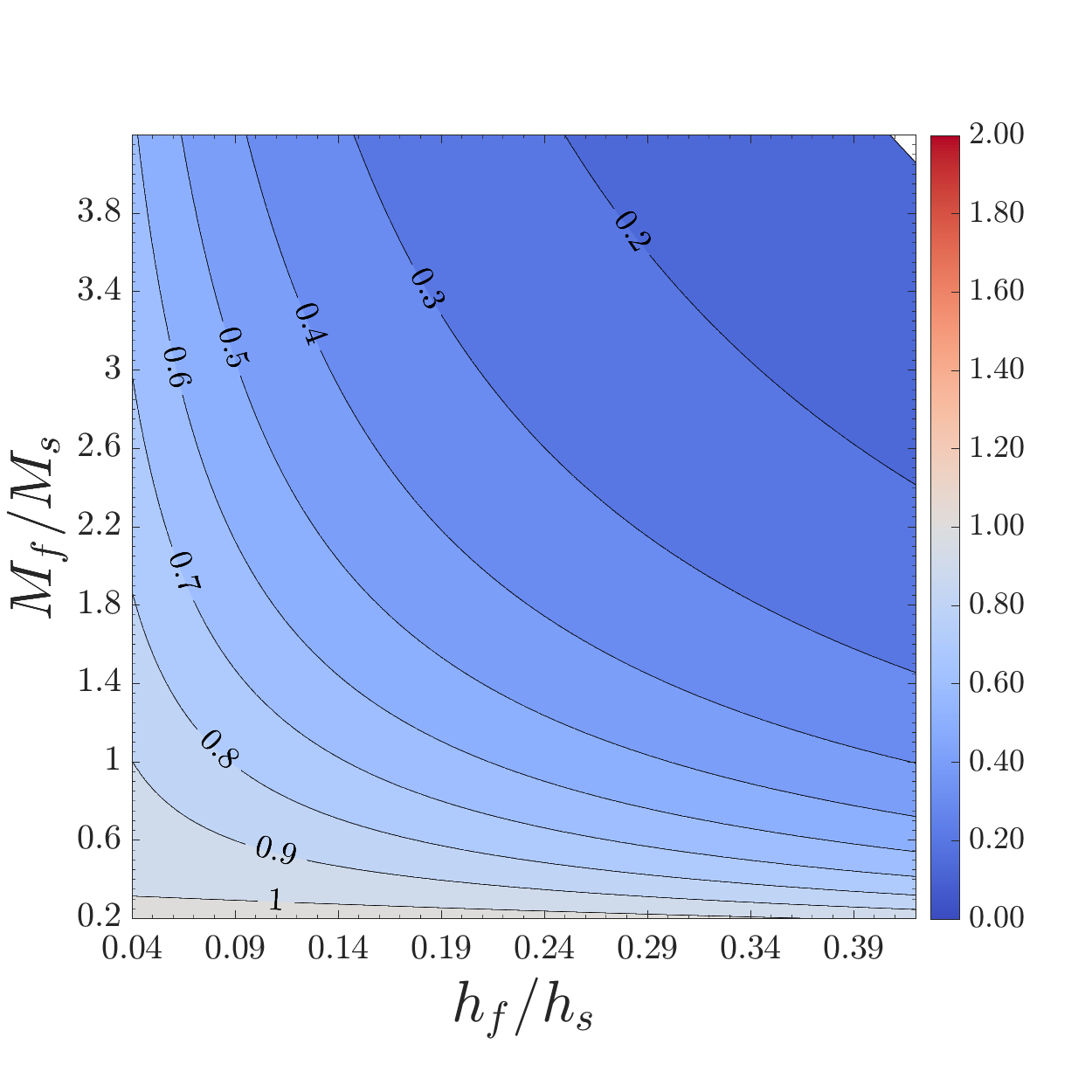}\label{Fig:Nonuni:Curv:Star}}
\subfigure[$\frac{\kappa_{pf}^{Conv}}{\kappa^{st}}$]{\includegraphics[keepaspectratio=true,width=0.30\textwidth]{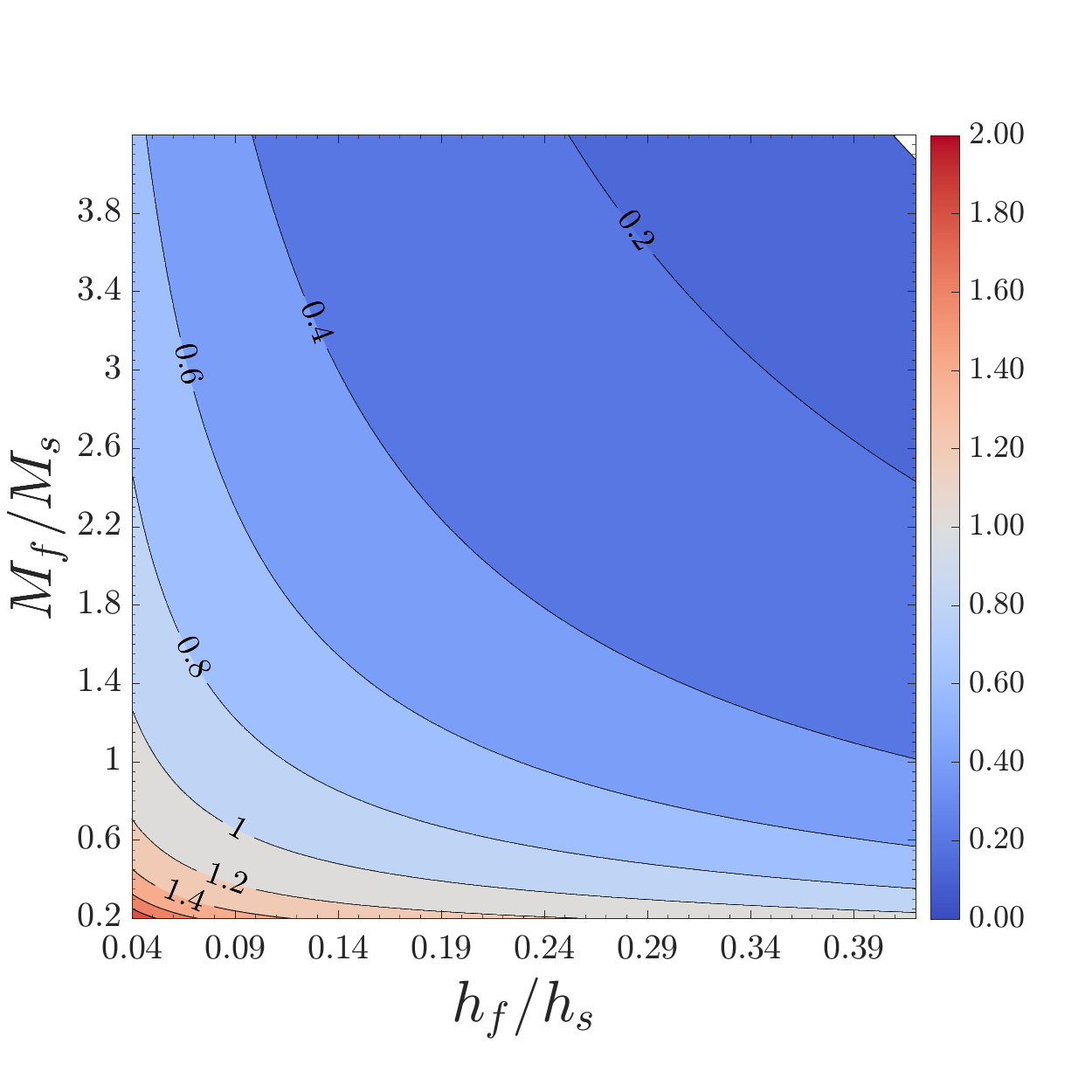}\label{Fig:Nonuni:Curv:Conv}}
\subfigure[$\frac{\kappa_{pf}^{Dir}}{\kappa^{st}}$]{\includegraphics[keepaspectratio=true,width=0.30\textwidth]{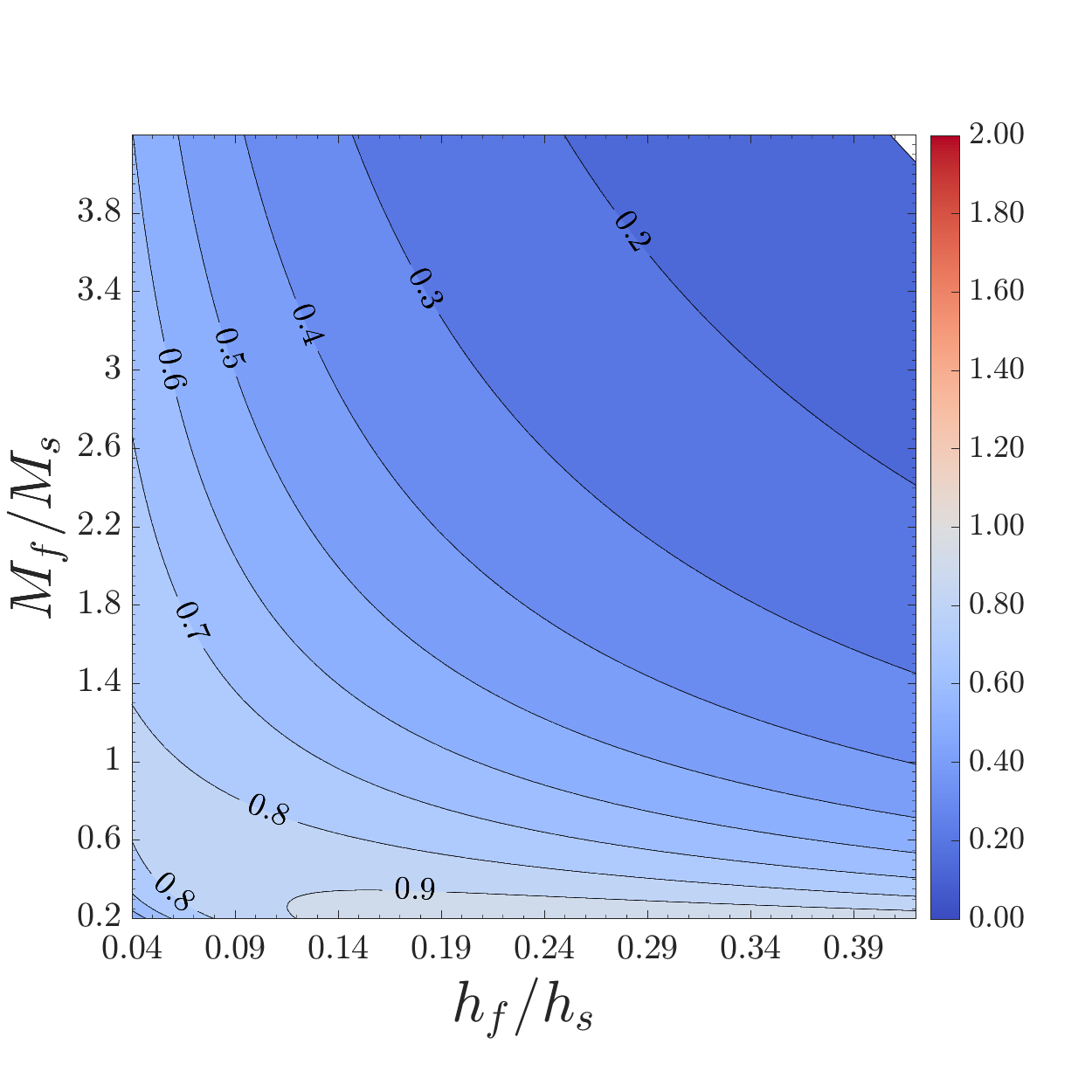}\label{Fig:Nonuni:Curv:Dir}}
{\caption{Normalized stretching strains and curvatures of the film-substrate system for various ratios of film-substrate thickness $h_f/h_s$ and stiffness $M_f/M_s$ with non-uniformly varying elastic mismatch strain with constant gradient and constant material properties. (a) depict the stretching strain and (d) the curvature when only elastic effects are accounted for (case II). (b), (c) show the stretching  strains and (e), (f)  the curvatures for converse and direct effects when both piezoelectric and flexoelectric effects are present (case I).}\label{Fig:Nonuni:e0kappa}}
\end{figure}

Figure \ref{Fig:Nonuni:e0kappa} shows the contour plots of normalized stretching strains and curvatures with varying ratios of film and substrate thickness $h_f/h_s$ and stiffness $M_f/M_s$. Figure \ref{Fig:Nonuni:e0:Star} shows the stretching strain, and Figure \ref{Fig:Nonuni:Curv:Star} shows the normalized curvature when only elastic effects are considered. When both piezo- and flexoelectric effects are considered, the normalized stretching strains are shown in Figure \ref{Fig:Nonuni:e0:Conv} for the converse case and Figure \ref{Fig:Nonuni:e0:Dir} for the direct case.  The normalized curvature is shown in Figure \ref{Fig:Nonuni:Curv:Conv} and Figure \ref{Fig:Nonuni:Curv:Dir} for the converse and direct cases, respectively. The contour line with the value $1$ gives the range of height and stiffness ratios for which Stoney's approximation (Equations \ref{stoney:nonuniform1} and \ref{stoney:nonuniform2}) is valid. From these figures, we also see that Stoney's formula overestimates stretching strains and curvatures when the relative thickness and the stiffness of the film increase. 

\begin{figure}[h!]\centering
\subfigure[$M_f/M_s=0.1$]{\includegraphics[keepaspectratio=true,width=0.30\textwidth]{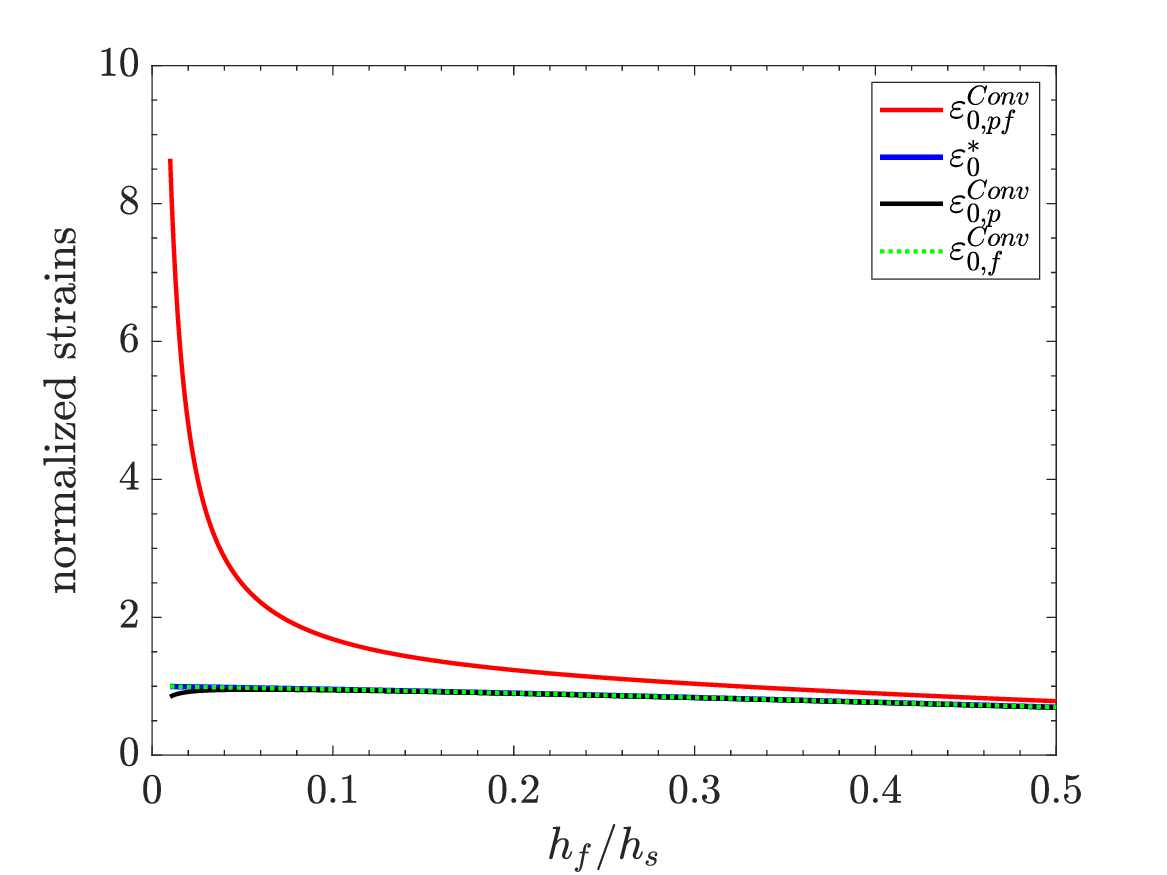}}
\subfigure[$M_f/M_s=1.0$]{\includegraphics[keepaspectratio=true,width=0.30\textwidth]{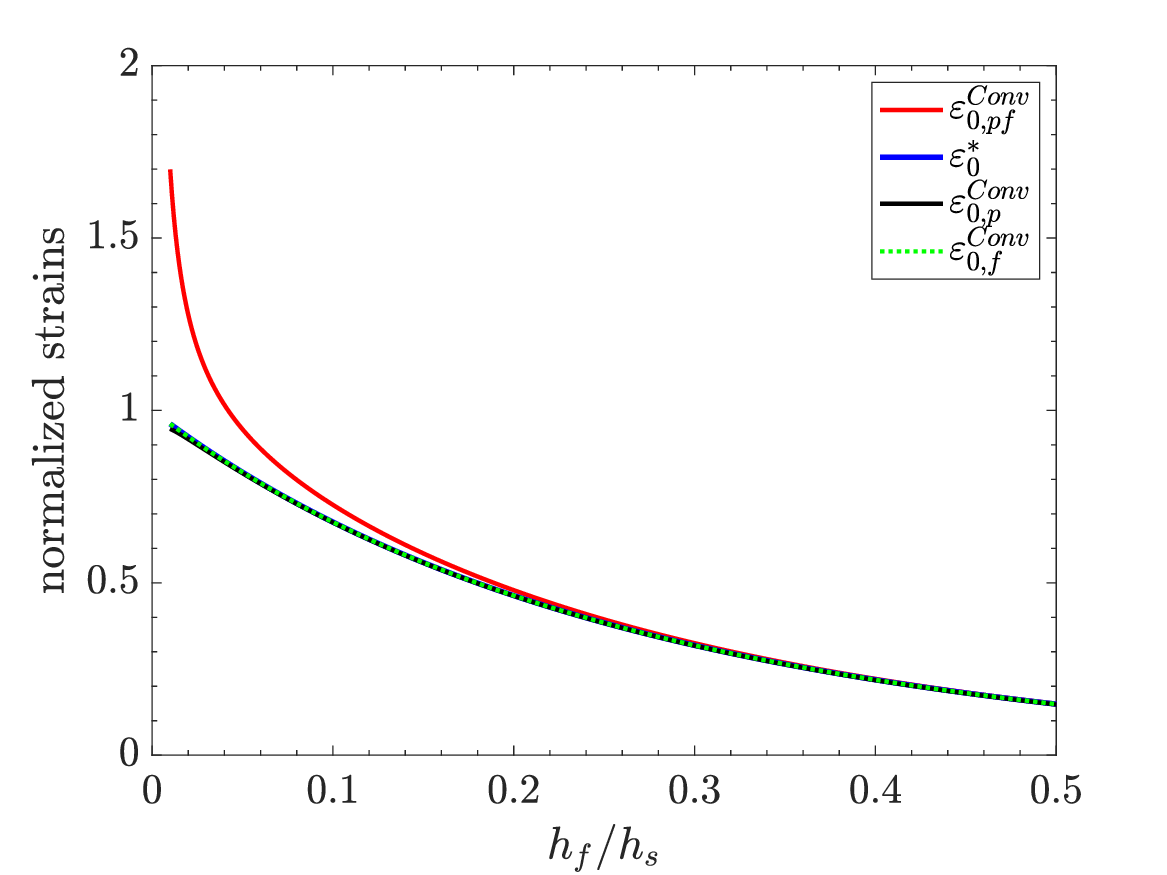}}
\subfigure[$M_f/M_s=1.5$]{\includegraphics[keepaspectratio=true,width=0.30\textwidth]{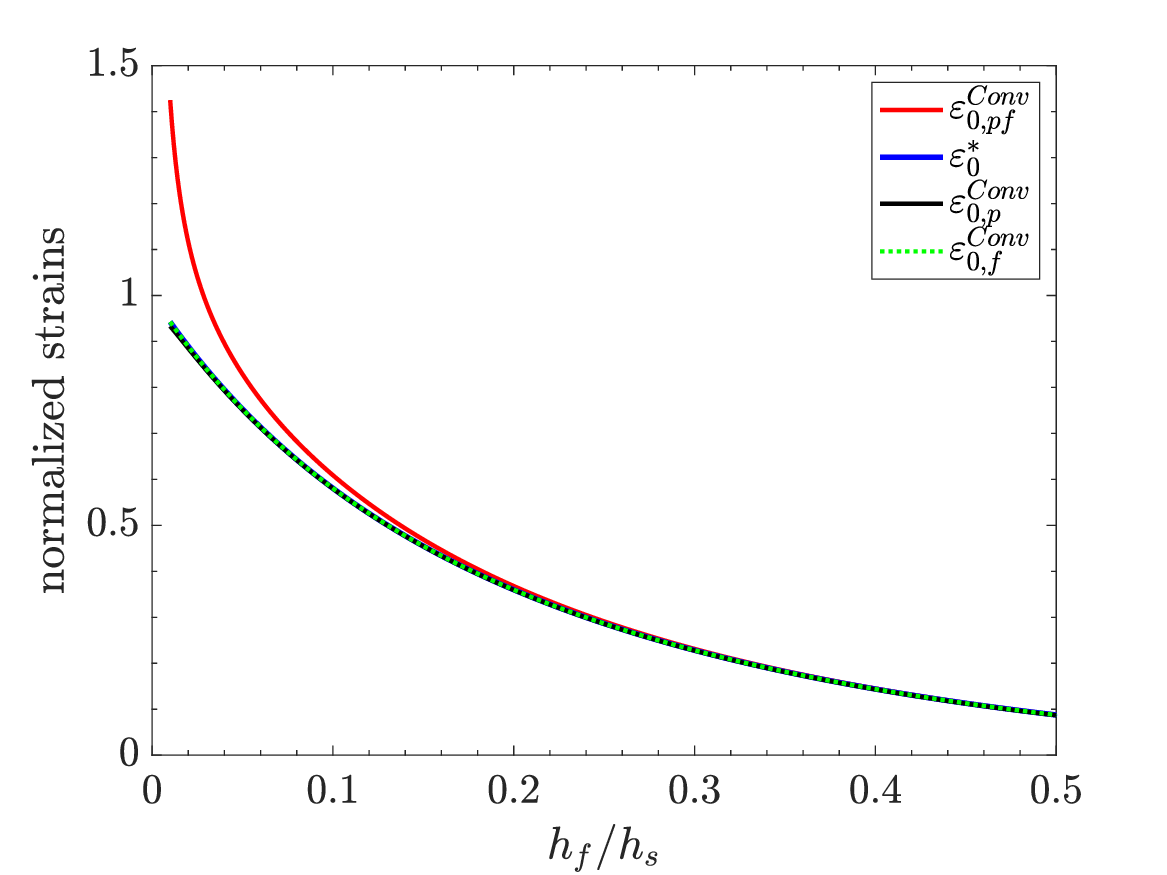}}
\subfigure[$M_f/M_s=0.1$]{\includegraphics[keepaspectratio=true,width=0.30\textwidth]{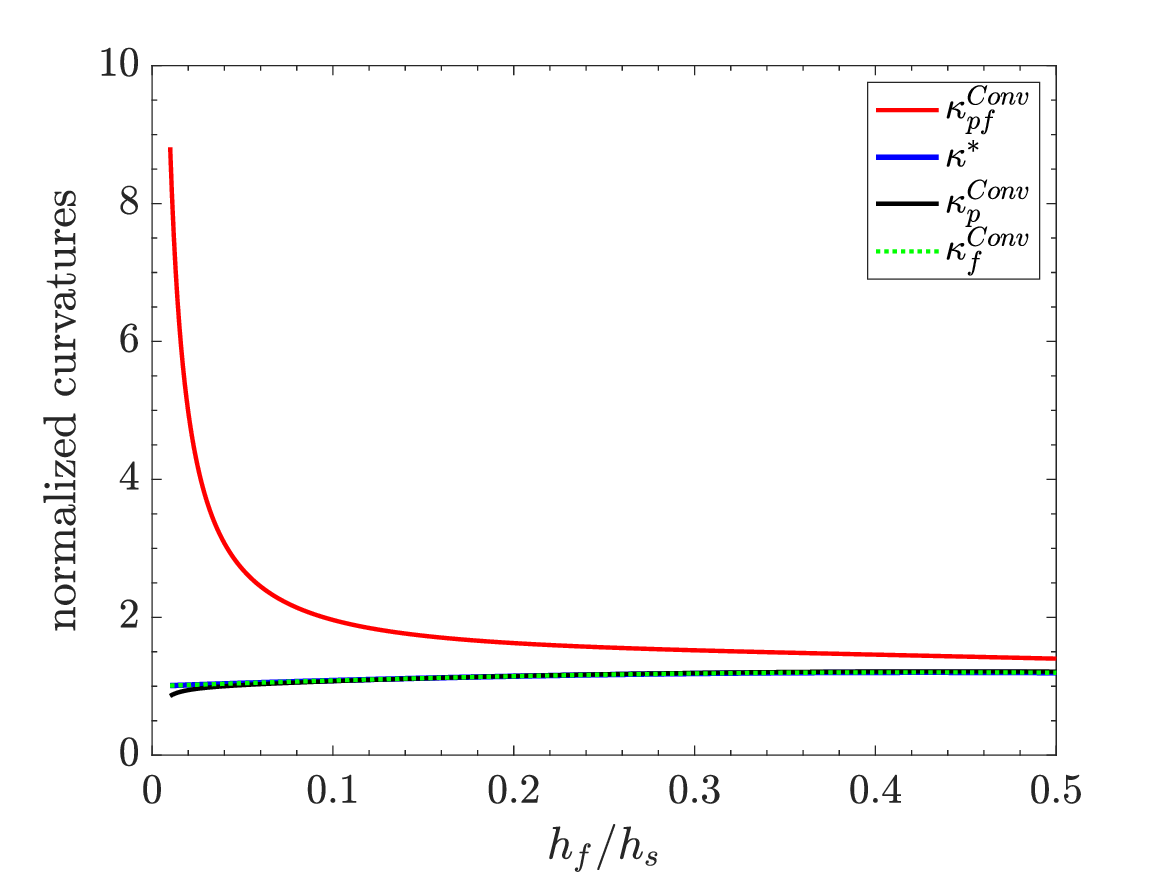}}
\subfigure[$M_f/M_s=1.0$]{\includegraphics[keepaspectratio=true,width=0.30\textwidth]{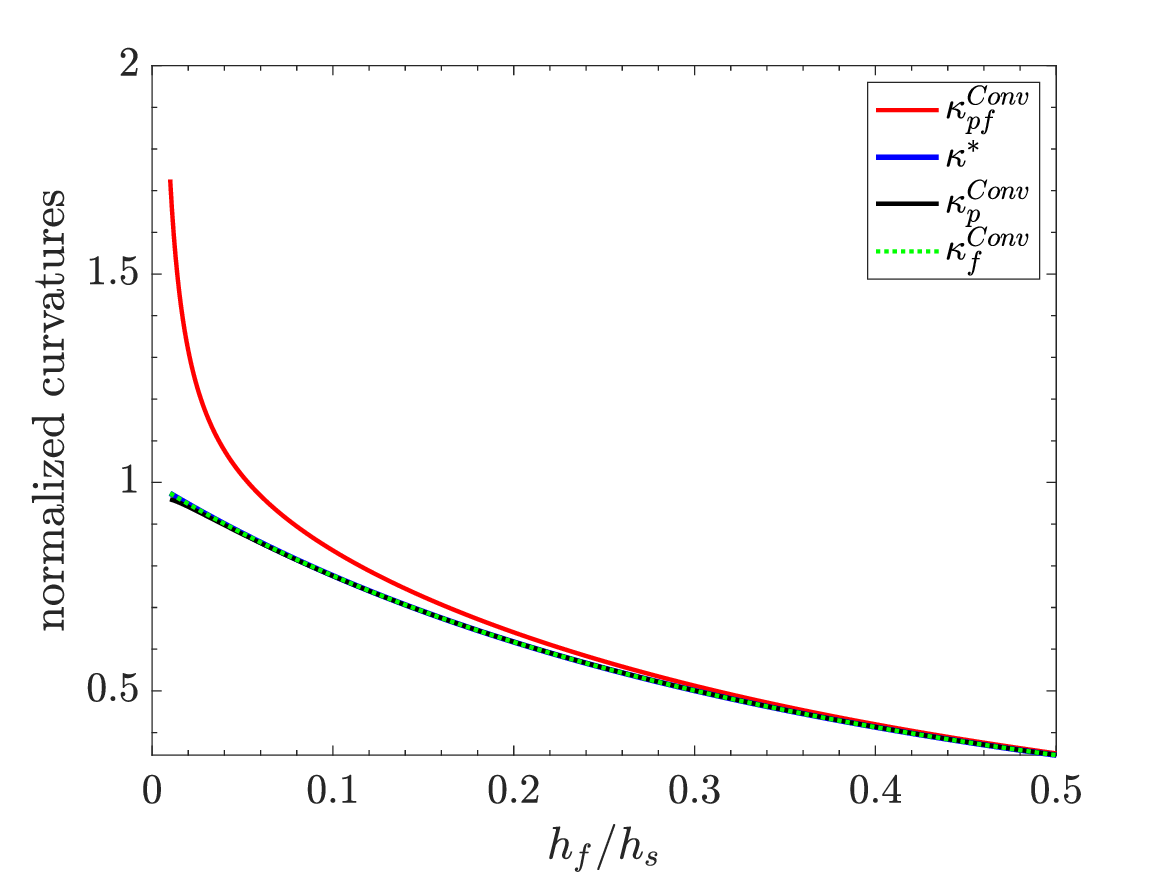}}
\subfigure[$M_f/M_s=1.5$]{\includegraphics[keepaspectratio=true,width=0.30\textwidth]{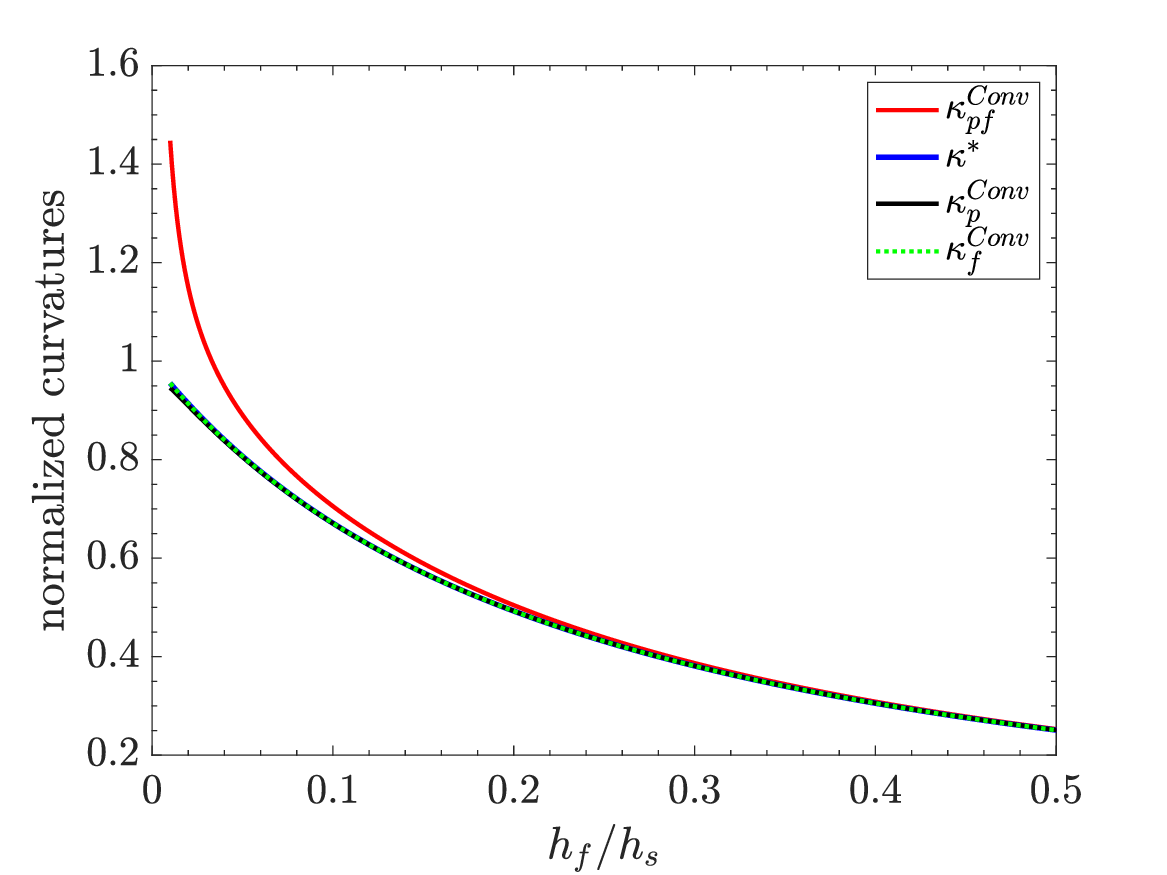}}
{\caption{Normalized stretching strains and curvatures for the converse case shown as a function of the ratio of the thickness of film and substrate, shown for different cases of stiffness ratios $M_f/M_s$ with nonuniform elastic mismatch strain. Figures (a)-(c) show the stretching strains, and (d)-(f) show the curvatures.}\label{Fig:ConvStrainNonuniform}}
\end{figure}

\begin{figure}[h!]\centering
\subfigure[$M_f/M_s=0.1$]{\includegraphics[keepaspectratio=true,width=0.30\textwidth]{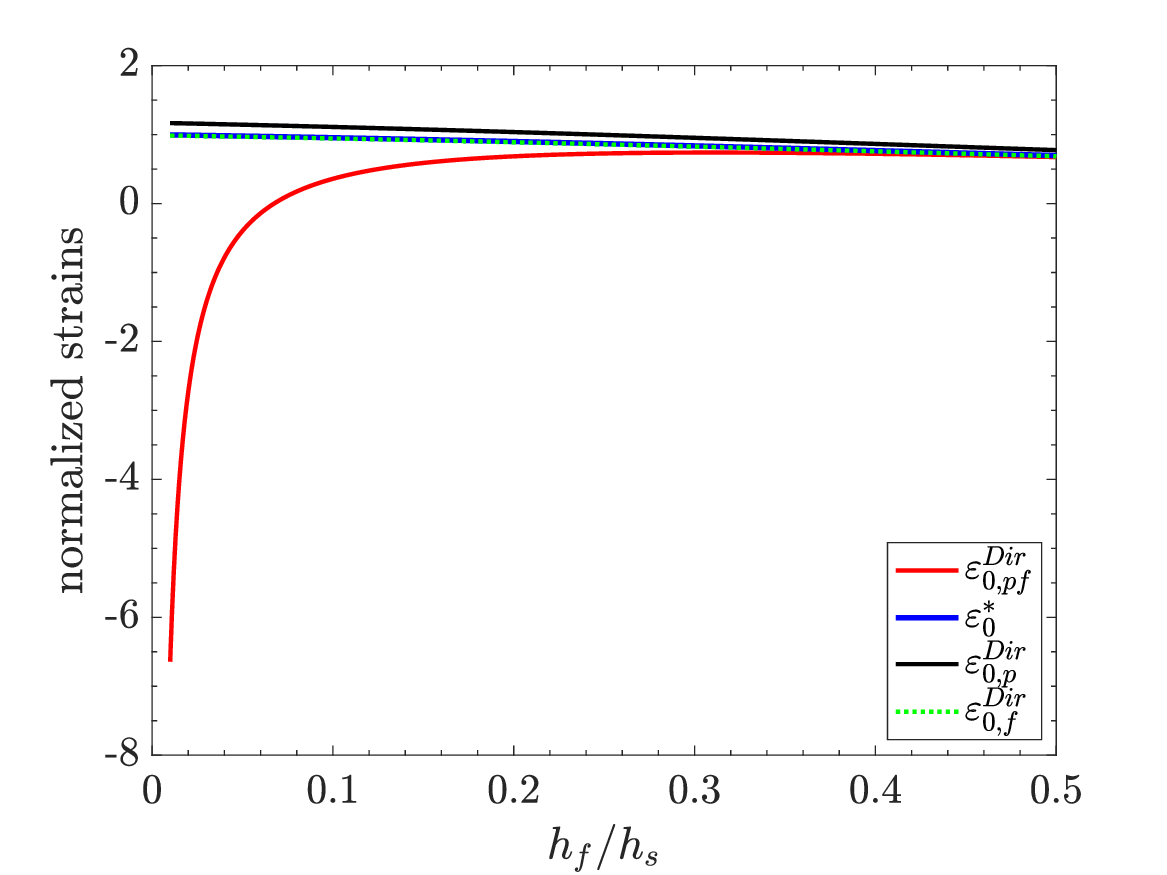}}
\subfigure[$M_f/M_s=1.0$]{\includegraphics[keepaspectratio=true,width=0.30\textwidth]{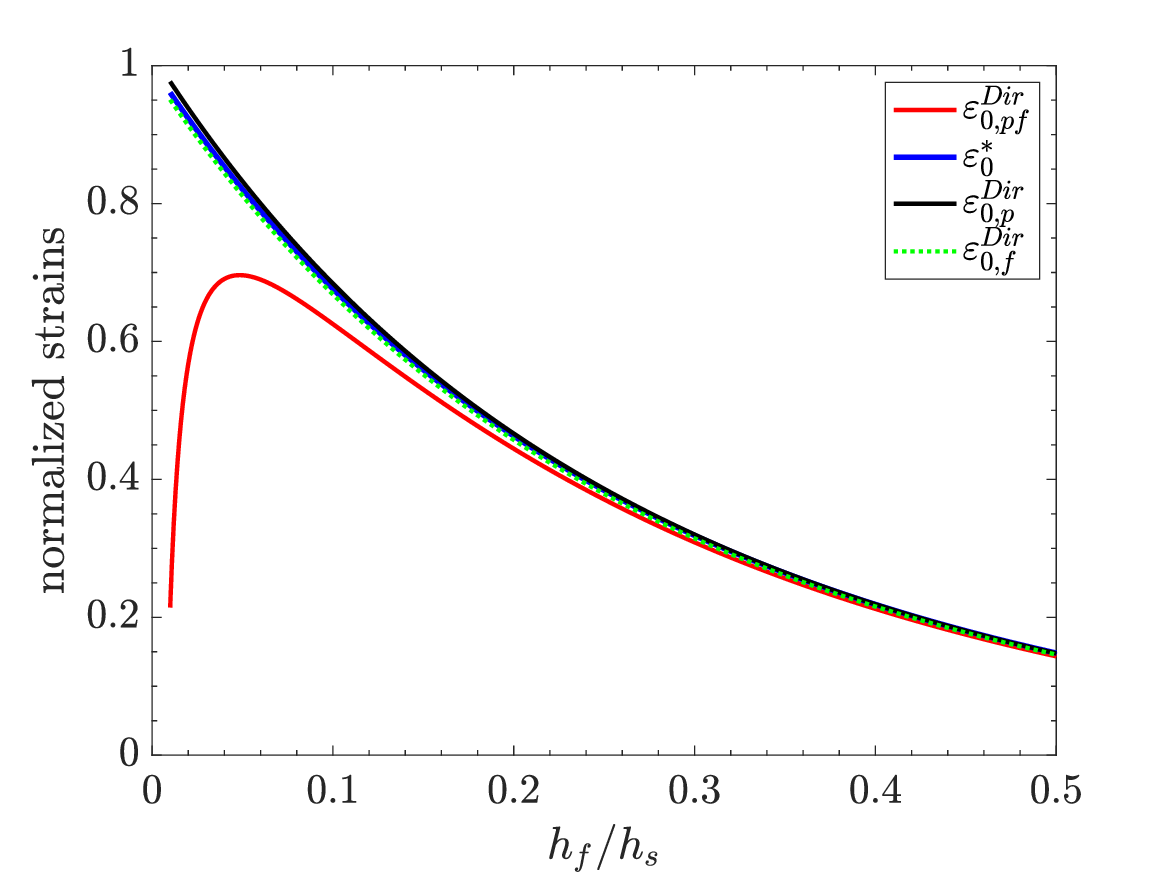}}
\subfigure[$M_f/M_s=1.5$]{\includegraphics[keepaspectratio=true,width=0.30\textwidth]{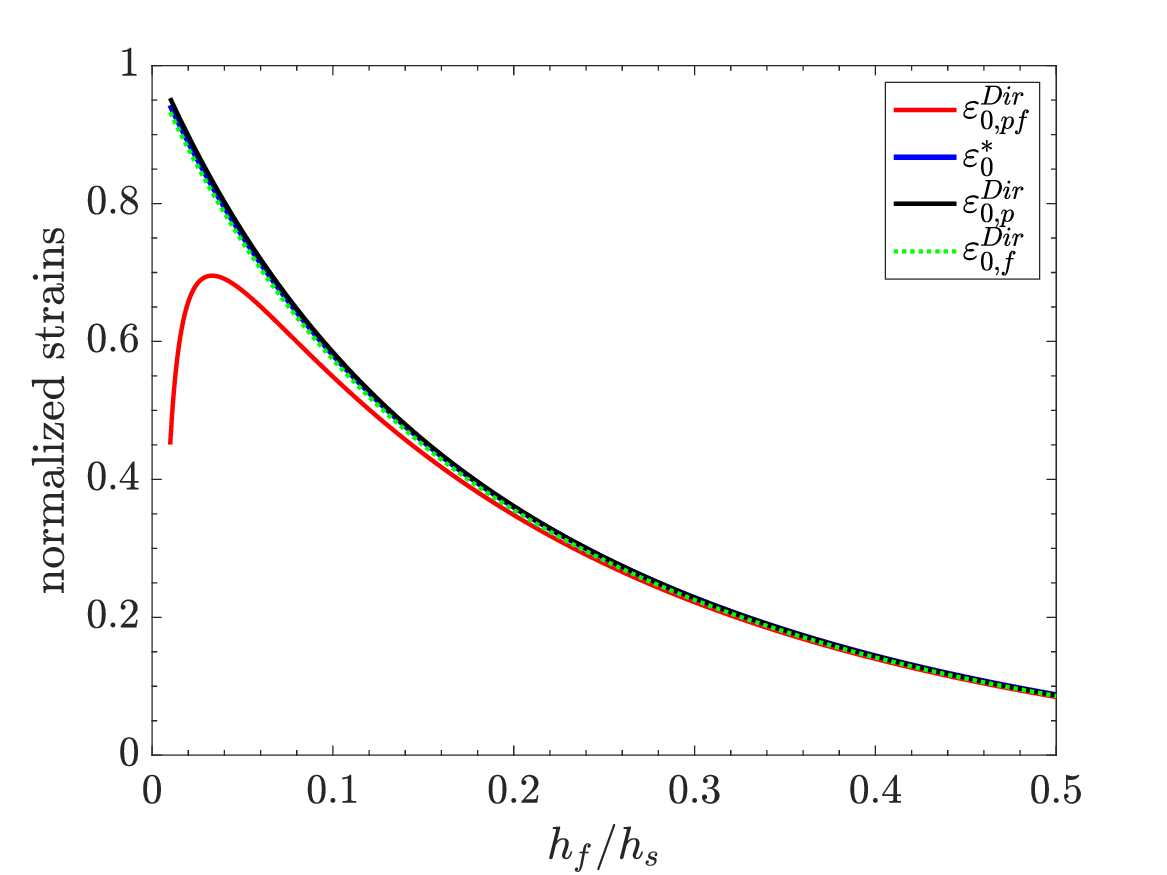}}
\subfigure[$M_f/M_s=0.1$]{\includegraphics[keepaspectratio=true,width=0.30\textwidth]{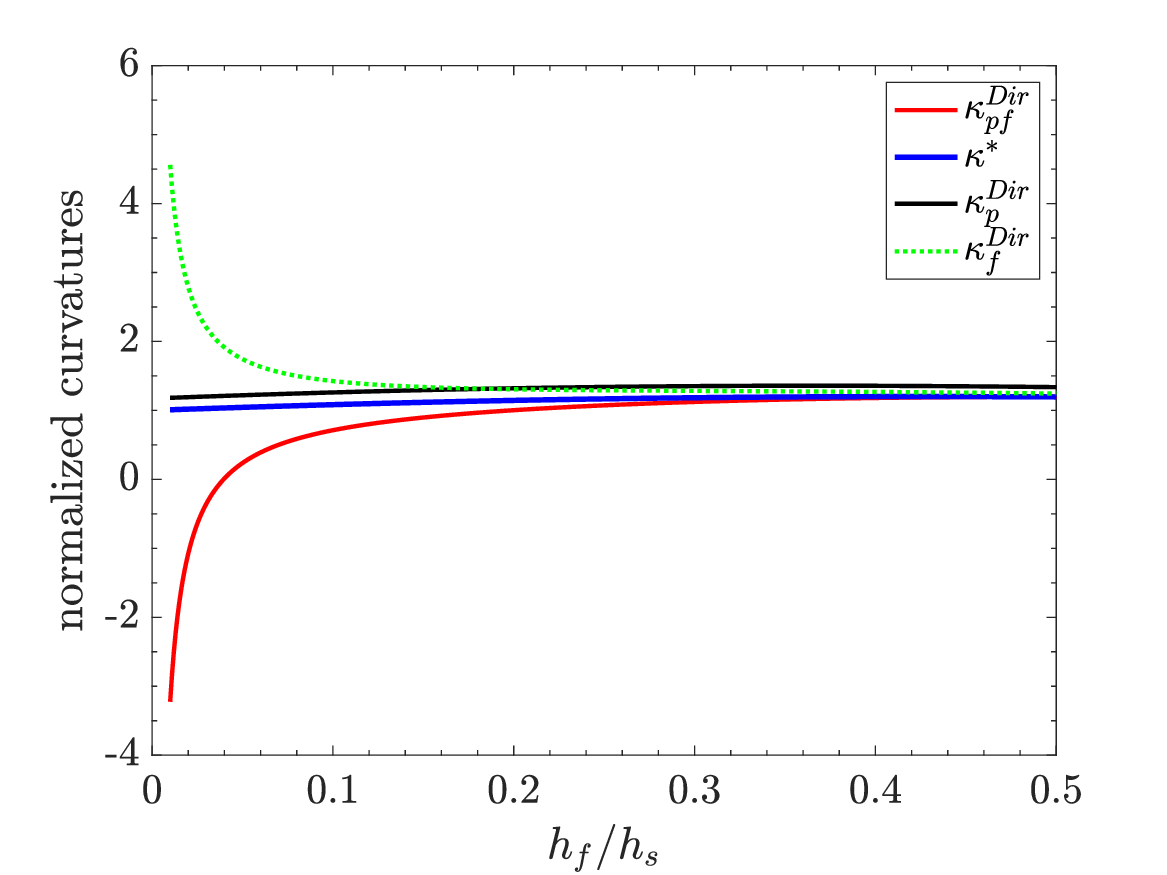}}
\subfigure[$M_f/M_s=1.0$]{\includegraphics[keepaspectratio=true,width=0.30\textwidth]{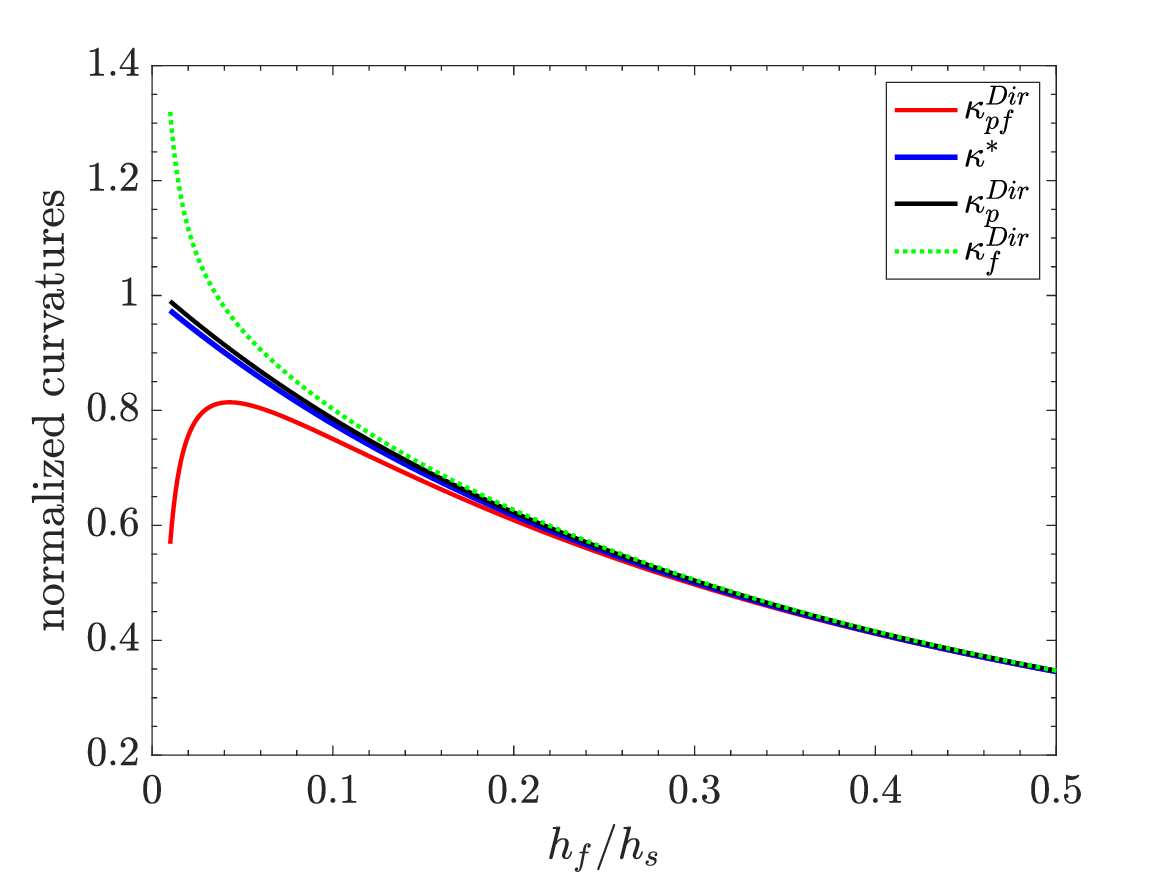}}
\subfigure[$M_f/M_s=1.5$]{\includegraphics[keepaspectratio=true,width=0.30\textwidth]{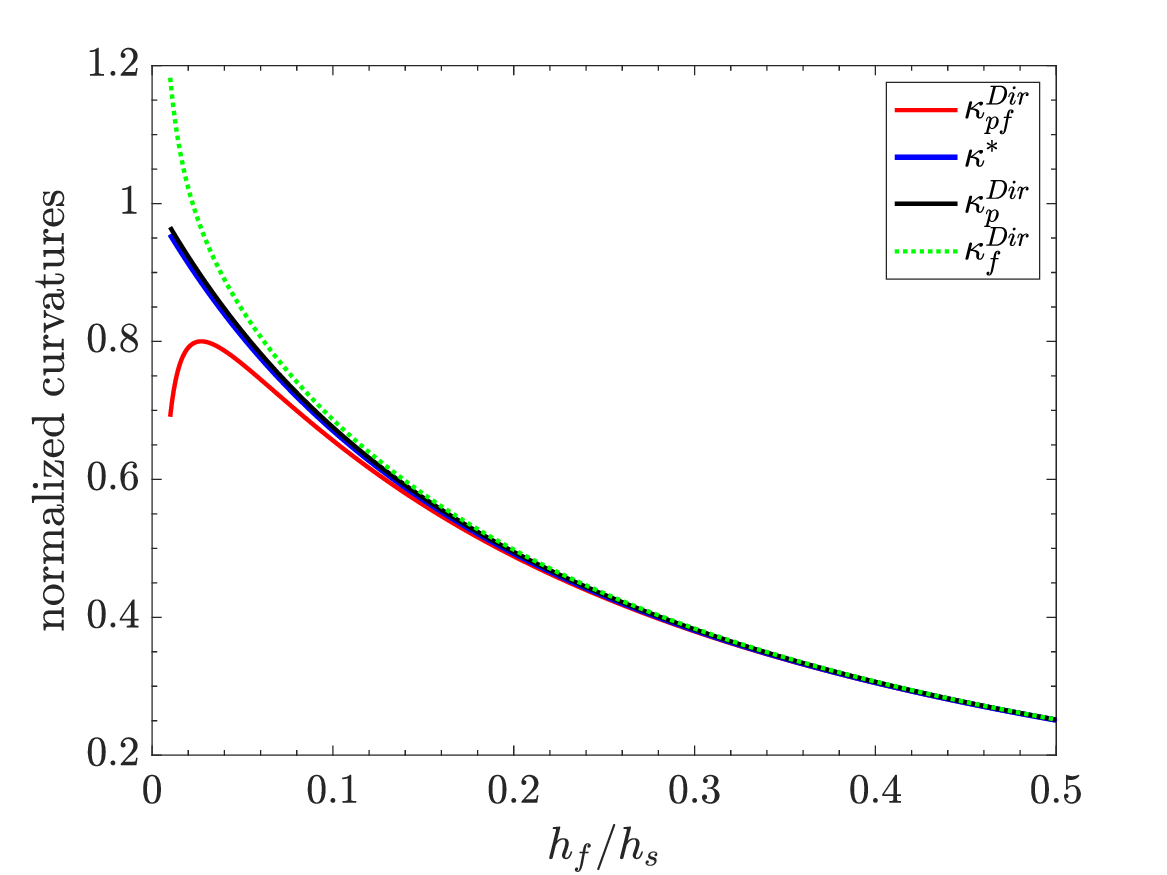}}
{\caption{Normalized stretching strains and curvatures for the direct case shown as a function of the ratio of the thickness of film and substrate, shown for different cases of stiffness ratios $M_f/M_s$ with nonuniform elastic mismatch strain. Figures (a)-(c) show the stretching strains, and (d)-(f) show the curvatures.}\label{Fig:DirStrainNonuniform}}
\end{figure}

Figure \ref{Fig:ConvStrainNonuniform} (converse case) and Figure \ref{Fig:DirStrainNonuniform} (direct case) show the dependence of normalized stretching strains and curvatures on the thickness ratio $h_f/h_s$ for stiffness ratios $M_f/M_s = 0.1$, $1.0$, and $1.5$, respectively. For the converse case, the normalized stretching strains and curvatures when both piezo- and flexoelectric effects are considered (Case I) are more than the stretching strains and curvatures for other cases (II-IV). However, for the direct case shown in Figure \ref{Fig:DirStrainNonuniform}, the normalized stretching strains and curvatures for Case I are smaller than the stretching strains and curvatures for other cases (II-IV). Additionally, when the thickness ratio is small, the influence of the different electromechanical effects on the normalized stretching strains and curvatures is more prominent, and with an increase in the relative film thickness, the distinction among the different electromechanical cases vanishes.

\section{Concluding remarks}\label{Sec:Conclusion}
In this work, we have derived explicit relations linking the properties of an electromechanically coupled thin film deposited on an elastic substrate to the resulting curvature and stretching strains, thereby generalizing Stoney's equation for piezoelectric and flexoelectric films deposited on elastic substrates. By considering films with both uniform and spatially varying properties, and open and closed circuit configurations, our analysis shows that the curvature, stretching strains, and electric polarization exhibit nonlinear dependence on the relative stiffness and thickness of the film and substrate, as well as on the nature of the electromechanical coupling. The formulas presented in this work can be employed for accurately characterizing the properties of electromechanically coupled thin films from substrate curvature measurements.

We remark that the scope of this work is limited to linear elastic isotropic materials. It is therefore pertinent to highlight some future research directions to extend its applicability. When a soft film is deposited on a substrate, it will show instabilities such as wrinkling, creasing, folding and localized ridges \cite{akerson2022stability}. It is natural to investigate the role of flexoelectricity in the soft thin film on these instabilities. As noted by Freund \cite{Freund:1999}, in the large-deformation regime, the equilibrium configuration of the film–substrate system renders the total potential energy stationary, though not necessarily minimum. In this case, if the deformation field is allowed to relax to include non-axisymmetric solutions, and sufficiently large elastic mismatch strains destabilizes the axisymmetric state, causing the system to adopt an ellipsoidal configuration rather than a spherical one. Other interesting avenues of investigation is to include the effects of electromechanical coupling and non-uniform properties of the substrate, and extension to anisotropic material properties. Overall, these developments will expand the purview of Stoney's equation for flexoelectric materials beyond the linear elastic isotropic regime.


\maketitle

\section*{Acknowledgements}
 This research used resources of the Oak Ridge Leadership Computing Facility, a DOE Office of Science User Facility operated by the Oak Ridge National Laboratory under contract DE-AC05-00OR22725. 
This manuscript has been authored in part by UT-Battelle, LLC, under contract DE-AC05- 00OR22725 with the US Department of Energy (DOE). The publisher acknowledges the US government license to provide public access under the DOE Public Access Plan (http://energy.gov/downloads/doe-public-access-plan).

During the preparation of this work the author used OpenAI (GPT-5) in order to search literature and revise language. After using this tool/service, the author reviewed and edited the content as needed and takes full responsibility for the content of the published article.

\section*{Author contributions}
S.G. conceived the project, performed derivations, calculations, analysis, and wrote the manuscript.
\section*{Conflict of interest declaration}
The author declares no competing interests.
\bibliographystyle{unsrt}
\bibliography{Flexoelectricity}

@article{mccartney2014,
  title={Methods for determining piezoelectric properties of thin epitaxial films: Theoretical foundations},
  author={McCartney, LN and Wright, L and Cain, MG and Crain, Jason and Martyna, Glenn J and Newns, Dennis M},
  journal={Journal of Applied Physics},
  volume={116},
  number={1},
  year={2014},
  publisher={AIP Publishing}
}

@article{zhou2017stoney,
  title={Stoney formula for piezoelectric film/elastic substrate system},
  author={Zhou, Wang-Min and Li, Wang-Jun and Hong, Sheng-Yun and Jin, Jie and Yin, Shu-Yuan},
  journal={Chinese Physics B},
  volume={26},
  number={3},
  pages={037701},
  year={2017},
  publisher={IOP Publishing}
}

@article{hohenberg1964,
  title={Inhomogeneous electron gas},
  author={Hohenberg, Pierre and Kohn, Walter},
  journal={Physical review},
  volume={136},
  number={3B},
  pages={B864},
  year={1964},
  publisher={APS}
}

@article{kohn1965,
  title={Self-consistent equations including exchange and correlation effects},
  author={Kohn, Walter and Sham, Lu Jeu},
  journal={Physical review},
  volume={140},
  number={4A},
  pages={A1133},
  year={1965},
  publisher={APS}
}

@article{ghosh2017a,
  title={SPARC: Accurate and efficient finite-difference formulation and parallel implementation of density functional theory: Isolated clusters},
  author={Ghosh, Swarnava and Suryanarayana, Phanish},
  journal={Computer Physics Communications},
  volume={212},
  pages={189--204},
  year={2017},
  publisher={Elsevier}
}

@article{ghosh2017b,
  title={SPARC: Accurate and efficient finite-difference formulation and parallel implementation of Density Functional Theory: Extended systems},
  author={Ghosh, Swarnava and Suryanarayana, Phanish},
  journal={Computer Physics Communications},
  volume={216},
  pages={109--125},
  year={2017},
  publisher={Elsevier}
}

@article{pureza2009enhancing,
  title={Enhancing accuracy to Stoney equation},
  author={Pureza, JM and Lacerda, MM and De Oliveira, AL and Fragalli, JF and Zanon, RAS},
  journal={Applied Surface Science},
  volume={255},
  number={12},
  pages={6426--6428},
  year={2009},
  publisher={Elsevier}
}

@article{feng2007stoney,
  title={On the Stoney formula for a thin film/substrate system with nonuniform substrate thickness},
  author={Feng, X and Huang, Y and Rosakis, AJ},
  journal={Journal of Applied Mechanics},
  volume={74},
  number={6},
  pages={1276--1281},
  year={2007},
  publisher={American Society of Mechanical Engineers Digital Collection}
}

@article{injeti2016extending,
  title={Extending Stoney's equation to thin, elastically anisotropic substrates and bilayer films},
  author={Injeti, Sai Sharan and Annabattula, Ratna Kumar},
  journal={Thin Solid Films},
  volume={598},
  pages={252--259},
  year={2016},
  publisher={Elsevier}
}

@article{liu2021modified,
  title={Modified Stoney formula for determining stress within thin films on large-deformation isotropic circular plates},
  author={Liu, Haijun and Dai, Minghui and Tian, Xiaoqing and Chen, Shan and Dong, Fangfang and Lu, Lei},
  journal={AIP Advances},
  volume={11},
  number={12},
  year={2021},
  publisher={AIP Publishing}
}

@article{janssen2009celebrating,
  title={Celebrating the 100th anniversary of the Stoney equation for film stress: Developments from polycrystalline steel strips to single crystal silicon wafers},
  author={Janssen, Guido CAM and Abdalla, MM and Van Keulen, F and Pujada, BR and Van Venrooy, B},
  journal={Thin Solid Films},
  volume={517},
  number={6},
  pages={1858--1867},
  year={2009},
  publisher={Elsevier}
}

@article{chen2015modified,
  title={Modified Stoney's Equation for Evaluation of Residual Stresses on Thin Film},
  author={Chen, Kuen Tsann and Chang, Jui Hsing and Wu, Jiun Yu},
  journal={Applied Mechanics and Materials},
  volume={789},
  pages={25--32},
  year={2015},
  publisher={Trans Tech Publ}
}

@article{qiang2021extension,
  title={Extension of the Stoney formula for the incremental stress of thin films},
  author={Qiang, Jun and Jiang, Bingyan and Dong, Yanzhuo and Roth, Benedikt and Jiang, Fengze},
  journal={Applied Physics Letters},
  volume={118},
  number={9},
  year={2021},
  publisher={AIP Publishing}
}

@article{finot1997large,
  title={Large deformation and geometric instability of substrates with thin-film deposits},
  author={Finot, M and Blech, IA and Suresh, S and Fujimoto, H},
  journal={Journal of applied physics},
  volume={81},
  number={8},
  pages={3457--3464},
  year={1997},
  publisher={American Institute of Physics}
}

@article{masters1993geometrically,
  title={Geometrically nonlinear stress-deflection relations for thin film/substrate systems},
  author={Masters, Christine B and Salamon, NJ},
  journal={International journal of engineering science},
  volume={31},
  number={6},
  pages={915--925},
  year={1993},
  publisher={Elsevier}
}

@article{injeti2021modified,
  title={Modified Stoney’s equation with anisotropic substrates undergoing large deformations},
  author={Injeti, Sai Sharan and Vyas, Nihit and Annabattula, Ratna Kumar},
  journal={Mechanics Research Communications},
  volume={113},
  pages={103685},
  year={2021},
  publisher={Elsevier}
}

@article{schicker2016stress,
  title={Stress-warping relation in thin film coated wafers},
  author={Schicker, J and Khan, WA and Arnold, T and Hirschl, C},
  journal={Modelling and Simulation in Materials Science and Engineering},
  volume={25},
  number={2},
  pages={025005},
  year={2016},
  publisher={IOP Publishing}
}

@article{Mindlin1965,
  title={Second gradient of strain and surface-tension in linear elasticity},
  author={Mindlin, Raymond David},
  journal={International journal of solids and structures},
  volume={1},
  number={4},
  pages={417--438},
  year={1965},
  publisher={Elsevier}
}

@article{Grasinger2021,
  title={Flexoelectricity in soft elastomers and the molecular mechanisms underpinning the design and emergence of giant flexoelectricity},
  author={Grasinger, Matthew and Mozaffari, Kosar and Sharma, Pradeep},
  journal={Proceedings of the National Academy of Sciences},
  volume={118},
  number={21},
  pages={e2102477118},
  year={2021},
  publisher={National Academy of Sciences}}

@article{Sharma2010,
  title={Piezoelectric thin-film superlattices without using piezoelectric materials},
  author={Sharma, ND and Landis, CM and Sharma, Pradeep},
  journal={Journal of Applied Physics},
  volume={108},
  number={2},
  year={2010},
  publisher={AIP Publishing}}

@article{Yan2013,
  title={Flexoelectric effect on the electroelastic responses of bending piezoelectric nanobeams},
  author={Yan, Z and Jiang, LY},
  journal={Journal of Applied Physics},
  volume={113},
  number={19},
  year={2013},
  publisher={AIP Publishing}}

@article{Abdollahi2015,
  title={Constructive and destructive interplay between piezoelectricity and flexoelectricity in flexural sensors and actuators},
  author={Abdollahi, Amir and Arias, Irene},
  journal={Journal of Applied Mechanics},
  volume={82},
  number={12},
  pages={121003},
  year={2015},
  publisher={American Society of Mechanical Engineers}}

@article{MaoPurohit2014,
  title={Insights into flexoelectric solids from strain-gradient elasticity},
  author={Mao, Sheng and Purohit, Prashant K},
  journal={Journal of Applied Mechanics},
  volume={81},
  number={8},
  pages={081004},
  year={2014},
  publisher={American Society of Mechanical Engineers}
}

@article{Mao2015,
  title={Defects in flexoelectric solids},
  author={Mao, Sheng and Purohit, Prashant K},
  journal={Journal of the Mechanics and Physics of Solids},
  volume={84},
  pages={95--115},
  year={2015},
  publisher={Elsevier}}

@article{Kothari2018,
  title={Critical curvature localization in graphene. I. Quantum-flexoelectricity effect},
  author={Kothari, Mrityunjay and Cha, Moon-Hyun and Kim, Kyung-Suk},
  journal={Proceedings of the Royal Society A: Mathematical, Physical and Engineering Sciences},
  volume={474},
  number={2214},
  pages={20180054},
  year={2018},
  publisher={The Royal Society Publishing}}

@article{Kothari2019,
  title={Critical curvature localization in graphene. II. Non-local flexoelectricity--dielectricity coupling},
  author={Kothari, Mrityunjay and Cha, Moon-Hyun and Lefevre, Victor and Kim, Kyung-Suk},
  journal={Proceedings of the Royal Society A},
  volume={475},
  number={2221},
  pages={20180671},
  year={2019},
  publisher={The Royal Society Publishing}}

@article{Lee2012,
  title={Giant flexoelectric effect through interfacial strain relaxation},
  author={Lee, Daesu and Noh, Tae Won},
  journal={Philosophical Transactions of the Royal Society A: Mathematical, Physical and Engineering Sciences},
  volume={370},
  number={1977},
  pages={4944--4957},
  year={2012},
  publisher={The Royal Society Publishing}}

@article{Le2011,
  title={The number and types of all possible rotational symmetries for flexoelectric tensors},
  author={Le Quang, H and He, Q-C},
  journal={Proceedings of the royal society a: mathematical, physical and engineering sciences},
  volume={467},
  number={2132},
  pages={2369--2386},
  year={2011},
  publisher={The Royal Society Publishing}}

@article{Rahmati2025,
  title={A static and dynamic theory for photo-flexoelectric liquid crystal elastomers and the coupling of light, deformation and electricity},
  author={Rahmati, Amir Hossein and Mozaffari, Kosar and Liu, Liping and Sharma, Pradeep},
  journal={Journal of the Mechanics and Physics of Solids},
  volume={195},
  pages={105949},
  year={2025},
  publisher={Elsevier}}

@article{Mishra2025,
  title={Modeling direct and converse flexoelectricity in soft dielectric rods with application to the follower load},
  author={Mishra, Pushkar and Gupta, Prakhar},
  journal={Journal of the Mechanics and Physics of Solids},
  volume={195},
  pages={105956},
  year={2025},
  publisher={Elsevier}}

@article{Song2024,
  title={Analyzing flexoelectric polarization of suspended membrane by nonlinear bending theory of plate},
  author={Song, Chunlin and Zhang, Mei and Ming, Wenjie and Fan, Xuhui and Huang, Boyuan and Li, Jiangyu},
  journal={Journal of the Mechanics and Physics of Solids},
  volume={193},
  pages={105898},
  year={2024},
  publisher={Elsevier}
}

@article{Witt2023,
  title={Modelling and numerical simulation of remodelling processes in cortical bone: An IGA approach to flexoelectricity-induced osteocyte apoptosis and subsequent bone cell diffusion},
  author={Witt, Carina and Kaiser, Tobias and Menzel, Andreas},
  journal={Journal of the Mechanics and Physics of Solids},
  volume={173},
  pages={105194},
  year={2023},
  publisher={Elsevier}
}

@article{Deng2014,
  title={Flexoelectricity in soft materials and biological membranes},
  author={Deng, Qian and Liu, Liping and Sharma, Pradeep},
  journal={Journal of the Mechanics and Physics of Solids},
  volume={62},
  pages={209--227},
  year={2014},
  publisher={Elsevier}
}

@article{Chen2015,
  title={Utilizing mechanical loads and flexoelectricity to induce and control complicated evolution of domain patterns in ferroelectric nanofilms},
  author={Chen, Weijin and Zheng, Yue and Feng, Xue and Wang, Biao},
  journal={Journal of the Mechanics and Physics of Solids},
  volume={79},
  pages={108--133},
  year={2015},
  publisher={Elsevier}
}

@article{Deng2019,
  title={The collusion of flexoelectricity and Hopf bifurcation in the hearing mechanism},
  author={Deng, Qian and Ahmadpoor, Fatemeh and Brownell, William E and Sharma, Pradeep},
  journal={Journal of the Mechanics and Physics of Solids},
  volume={130},
  pages={245--261},
  year={2019},
  publisher={Elsevier}
}

@article{Chen2018,
  title={Mechanical switching of ferroelectric domains beyond flexoelectricity},
  author={Chen, Weijin and Liu, Jianyi and Ma, Lele and Liu, Linjie and Jiang, GL and Zheng, Yue},
  journal={Journal of the Mechanics and Physics of Solids},
  volume={111},
  pages={43--66},
  year={2018},
  publisher={Elsevier}
}

@article{Shen2010,
  title={A theory of flexoelectricity with surface effect for elastic dielectrics},
  author={Shen, Shengping and Hu, Shuling},
  journal={Journal of the Mechanics and Physics of Solids},
  volume={58},
  number={5},
  pages={665--677},
  year={2010},
  publisher={Elsevier}
}

@article{Lun2022,
  title={Asymmetric mechanical properties in ferroelectrics driven by flexo-deformation effect},
  author={Lun, Yingzhuo and Hong, Jiawang and Fang, Daining},
  journal={Journal of the Mechanics and Physics of Solids},
  volume={164},
  pages={104891},
  year={2022},
  publisher={Elsevier}
}

@article{Sharma2007,
  title={On the possibility of piezoelectric nanocomposites without using piezoelectric materials},
  author={Sharma, Nikhil D and Maranganti, Ravi and Sharma, Pradeep},
  journal={Journal of the Mechanics and Physics of Solids},
  volume={55},
  number={11},
  pages={2328--2350},
  year={2007},
  publisher={Elsevier}
}

@article{Gao2008,
  title={An electromechanical liquid crystal model of vesicles},
  author={Gao, Ling-Tian and Feng, Xi-Qiao and Yin, Ya-Jun and Gao, Huajian},
  journal={Journal of the Mechanics and Physics of Solids},
  volume={56},
  number={9},
  pages={2844--2862},
  year={2008},
  publisher={Elsevier}
}

@article{Banerjee2016,
  title={Cyclic density functional theory: A route to the first principles simulation of bending in nanostructures},
  author={Banerjee, Amartya S and Suryanarayana, Phanish},
  journal={Journal of the Mechanics and Physics of Solids},
  volume={96},
  pages={605--631},
  year={2016},
  publisher={Elsevier}
}

@article{Ghosh2019,
  title={Symmetry-adapted real-space density functional theory for cylindrical geometries: Application to large group-IV nanotubes},
  author={Ghosh, Swarnava and Banerjee, Amartya S and Suryanarayana, Phanish},
  journal={Physical Review B},
  volume={100},
  number={12},
  pages={125143},
  year={2019},
  publisher={APS}
}

@article{kumar2025,
  title={Ab initio study of flexoelectricity in MXene monolayers},
  author={Kumar, Shashikant and Zhang, Zixi and Suryanarayana, Phanish},
  journal={Nanotechnology},
  year={2025}
}

@article{codony2021,
  title={Transversal flexoelectric coefficient for nanostructures at finite deformations from first principles},
  author={Codony, David and Arias, Irene and Suryanarayana, Phanish},
  journal={Physical Review Materials},
  volume={5},
  number={3},
  pages={L030801},
  year={2021},
  publisher={APS}
}

@article{kumar2021,
  title={Flexoelectricity in atomic monolayers from first principles},
  author={Kumar, Shashikant and Codony, David and Arias, Irene and Suryanarayana, Phanish},
  journal={Nanoscale},
  volume={13},
  number={3},
  pages={1600--1607},
  year={2021},
  publisher={Royal Society of Chemistry}
}

@article{Yudin2013Review,
  title={Fundamentals of flexoelectricity in solids},
  author={Yudin, Peter V and Tagantsev, Alexander K},
  journal={Nanotechnology},
  volume={24},
  number={43},
  pages={432001},
  year={2013},
  publisher={IOP Publishing}
}

@article{Tripathy2021Review,
  title={Comprehensive review on flexoelectric energy harvesting technology: Mechanisms, device configurations, and potential applications},
  author={Tripathy, Alekhika and Saravanakumar, Balasubramaniam and Mohanty, Smita and Nayak, Sanjay K and Ramadoss, Ananthakumar},
  journal={ACS Applied Electronic Materials},
  volume={3},
  number={7},
  pages={2898--2924},
  year={2021},
  publisher={ACS Publications}
}

@article{Codony2021Review,
  title={Mathematical and computational modeling of flexoelectricity},
  author={Codony, David and Mocci, Alice and Barcel{\'o}-Mercader, Jordi and Arias, Irene},
  journal={Journal of Applied Physics},
  volume={130},
  number={23},
  year={2021},
  publisher={AIP Publishing}
}

@article{Nguyen2013Review,
  title={Nanoscale flexoelectricity},
  author={Nguyen, Thanh D and Mao, Sheng and Yeh, Yao-Wen and Purohit, Prashant K and McAlpine, Michael C},
  journal={Advanced Materials},
  volume={25},
  number={7},
  pages={946--974},
  year={2013},
  publisher={Wiley Online Library}
}

@article{Wang2016Review,
  title={Electroactive polymers for sensing},
  author={Wang, Tiesheng and Farajollahi, Meisam and Choi, Yeon Sik and Lin, I-Ting and Marshall, Jean E and Thompson, Noel M and Kar-Narayan, Sohini and Madden, John DW and Smoukov, Stoyan K},
  journal={Interface focus},
  volume={6},
  number={4},
  pages={20160026},
  year={2016},
  publisher={The Royal Society}
}

@article{Stoney,
  title={The tension of metallic films deposited by electrolysis},
  author={Stoney, George Gerald},
  journal={Proceedings of the Royal Society of London. Series A, Containing Papers of a Mathematical and Physical Character},
  volume={82},
  number={553},
  pages={172--175},
  year={1909},
  publisher={The Royal Society London}
}

@book{Book:Anand&Govindjee,
  title={Continuum mechanics of solids},
  author={Anand, Lallit and Govindjee, Sanjay},
  year={2020},
  publisher={Oxford University Press}
}

@article{Abdollahi:2014,
  title={Computational evaluation of the flexoelectric effect in dielectric solids},
  author={Abdollahi, Amir and Peco, Christian and Millan, Daniel and Arroyo, Marino and Arias, Irene},
  journal={Journal of Applied Physics},
  volume={116},
  number={9},
  year={2014},
  publisher={AIP Publishing}
}

@article{codony:JAP,
  title={Mathematical and computational modeling of flexoelectricity},
  author={Codony, David and Mocci, Alice and Barcel{\'o}-Mercader, Jordi and Arias, Irene},
  journal={Journal of Applied Physics},
  volume={130},
  number={23},
  year={2021},
  publisher={AIP Publishing}
}

@article{codony:Finite,
  title={Modeling flexoelectricity in soft dielectrics at finite deformation},
  author={Codony, David and Gupta, Prakhar and Marco, Onofre and Arias, Irene},
  journal={Journal of the Mechanics and Physics of Solids},
  volume={146},
  pages={104182},
  year={2021},
  publisher={Elsevier}
}

@book{Freund:Book,
  title={Thin film materials: stress, defect formation and surface evolution},
  author={Freund, Lambert Ben and Suresh, Subra},
  year={2004},
  publisher={Cambridge university press}
}

@article{Freund:1999,
  title={Extensions of the Stoney formula for substrate curvature to configurations with thin substrates or large deformations},
  author={Freund, LB and Floro, JA and Chason, E},
  journal={Applied Physics Letters},
  volume={74},
  number={14},
  pages={1987--1989},
  year={1999},
  publisher={American Institute of Physics}
}

@article{akerson2022stability,
  title={Stability and post-bifurcation of film-substrate systems},
  author={Akerson, Andrew and Elliott, Ryan S},
  journal={Proceedings of the Royal Society A},
  volume={478},
  number={2264},
  pages={20220181},
  year={2022},
  publisher={The Royal Society}
}

\end{document}